\documentclass[a4paper,fleqn,longmktitle]{cas-dc}
\usepackage[authoryear]{natbib}

\usepackage{appendix}
\usepackage{booktabs}
\usepackage{microtype}
\usepackage{subcaption}
\usepackage{supertabular}
\usepackage[autostyle,english=american]{csquotes}

\usepackage{tcolorbox}
\usepackage{pdfpages}

\usepackage{url}

\usepackage{multibib}
\newcites{appendix}{Appendix References}
\def\@mb@citenamelist{cite,citep,citet,citealp,citealt}

\definecolor{newcolor}{rgb}{.8,.349,.1}

\newcounter{appendix}

\newcommand{\appendixtitle}{}
\newcommand{\appendixtitleFull}{%
  \space\Alph{appendix}.\quad\appendixtitle}
\AtBeginEnvironment{appendices}{%
  \captionsetup{labelsep=period}
  \counterwithin{figure}{appendix}
  \counterwithin{table}{appendix}
  \counterwithin{equation}{appendix}
  \counterwithin{section}{appendix}  
  \preto{\section}{%
    \clearpage
    \refstepcounter{appendix}
    \phantomsection 
    \addcontentsline{toc}{section}{\appendixtitleFull}%
  }
}

\begin{document}
\let\WriteBookmarks\relax
\def\floatpagepagefraction{1}
\def\textpagefraction{.001}
\shorttitle{Surgical Data Science – from Concepts toward Clinical Translation}
\shortauthors{Maier-Hein/Eisenmann et~al.}

\title [mode = title]{Surgical Data Science – from Concepts \textcolor{black}{toward} Clinical Translation}                      

\author[1,2,3]{Lena Maier-Hein}[orcid=0000-0003-4910-9368]
\cormark[1]
\fnmark[1]
\ead{l.maier-hein@dkfz-heidelberg.de}
\credit{Designed the poll for the SDS 2019 workshop, organized and chaired the SDS 2019 workshop including the interactive discussions, structured and drafted the document, wrote sections 1 and 7, contributed to all other sections, edited the whole document.}

\author[1]{Matthias Eisenmann}[orcid=0000-0002-0713-8761]
\fnmark[1]
\credit{Conceptualization, Methodology, Software, Formal analysis, Data curation, Writing - Original Draft, Writing - Review \& Editing, Visualization, Supervision, Project administration}

\author[4,5]{Duygu Sarikaya}[orcid=0000-0002-2083-4999]
\credit{Co-organized the SDS 2019 workshop as part of the executive board, chaired sessions, moderated interactive discussions, took part in panel discussion, presented AI4OR initiative, co-wrote section 5, and contributed to other sections.}

\author[1]{Keno März}[orcid=0000-0003-3503-1918]
\credit{}

\author[6]{Toby Collins}[orcid=0000-0002-9441-8306]
\credit{}

\author[7]{Anand Malpani}[orcid=0000-0001-9477-9403]
\credit{}

\author[8]{Johannes Fallert}
\credit{}

\author[9]{Hubertus Feussner}[orcid=0000-0003-0964-4497]
\credit{}

\author[10]{Stamatia Giannarou}[orcid=0000-0002-8745-1343]
\credit{}

\author[11,12]{Pietro Mascagni}[orcid=0000-0001-7288-3023]
\credit{Participated in the poll for the SDS 2019 workshop, presented the CONDOR project at the SDS 2019 workshop and participated in  the interactive discussions, revised the whole manuscript.}

\author[13]{Hirenkumar Nakawala}[orcid=0000-0002-0121-2828]
\credit{Co-wrote the clinical translation section and revised the whole manuscript, participated in the poll for the SDS 2019 workshop and in the interactive discussions.}

\author[14,15]{Adrian Park}
\credit{}

\author[16]{Carla Pugh}[orcid=0000-0001-9139-8082]
\credit{Participated in the SDS 2019 workshop as a speaker and workshop moderator. Contributed section on surgical metrics and revised the whole manuscript.}

\author[17]{Danail Stoyanov}[orcid=0000-0002-0980-3227]
\credit{}

\author[7]{Swaroop S. Vedula}
\credit{}

\author[19]{Kevin Cleary}[orcid=0000-0001-9809-4823]
\credit{}

\author[20]{Gabor Fichtinger}[orcid=0000-0002-6354-262X]
\credit{}

\author[21,22]{Germain Forestier}[orcid=0000-0002-4960-7554]
\credit{}

\author[5]{Bernard Gibaud}[orcid=0000-0002-1772-4887]
\credit{}

\author[23,24]{Teodor Grantcharov}
\credit{}

\author[25,26]{Makoto Hashizume}[orcid=0000-0002-5140-0982]
\credit{}

\author[1]{\textcolor{black}{Doreen Heckmann-Nötzel}}
\credit{}

\author[18]{Hannes G. Kenngott}
\credit{}

\author[27]{Ron Kikinis}
\credit{}

\author[8]{Lars Mündermann}
\credit{}

\author[28,29]{Nassir Navab}
\credit{}

\author[1]{Sinan Onogur}[orcid=0000-0002-8371-760X]
\credit{Co-wrote section 3 and revised other parts of the manuscript.}

\author[30]{Raphael Sznitman}
\credit{Participated in the SDS 2016 and 2019 workshops. Revised and reviewed parts of the manuscript.}

\author[29]{Russell H. Taylor}[orcid=0000-0001-6272-1100]
\credit{}

\author[1]{Minu D. Tizabi}
\credit{Wrote sections of the manuscript and revised the entire manuscript.}

\author[18]{Martin Wagner}[orcid=0000-0002-9831-9110]
\credit{Co-Wrote Section 5 and revised other parts of the manuscript.}

\author[7,29]{Gregory D. Hager}
\credit{}

\author[31]{Thomas Neumuth}
\credit{}

\author[11,12]{Nicolas Padoy}[orcid=0000-0002-5010-4137]
\credit{Contributed to the two SDS workshops and reviewed and revised the whole manuscript.}

\author[32]{\textcolor{black}{Justin Collins}}[orcid=0000-0003-2228-1213]
\credit{}

\author[33]{\textcolor{black}{Ines Gockel}}[orcid=0000-0001-7423-713X]
\credit{}

\author[34]{\textcolor{black}{Jan Goedeke}}[orcid=0000-0002-1587-5956]
\credit{}

\author[35,36]{\textcolor{black}{Daniel A. Hashimoto}}[orcid=0000-0003-4725-3104]
\credit{}

\author[37,38,39,40]{\textcolor{black}{Luc Joyeux}}[orcid=0000-0001-7331-6354]
\credit{}

\author[41]{\textcolor{black}{Kyle Lam}}[orcid=0000-0001-6407-4912]
\credit{}

\author[42,43,44]{\textcolor{black}{Daniel R. Leff}}[orcid=0000-0002-5310-1046]
\credit{}

\author[45]{\textcolor{black}{Amin Madani}}[orcid=0000-0003-0901-9851]
\credit{}

\author[46]{\textcolor{black}{Hani J. Marcus}}[orcid=0000-0001-8000-392X]
\credit{}

\author[47]{\textcolor{black}{Ozanan Meireles}}[orcid=0000-0003-3876-5802]
\credit{}

\author[1]{\textcolor{black}{Alexander Seitel}}[orcid=0000-0002-5919-9646]
\credit{}

\author[48]{\textcolor{black}{Dogu Teber}}
\credit{}

\author[49]{\textcolor{black}{Frank Ückert}}[orcid=0000-0001-8362-5636]
\credit{}

\author[18]{Beat P. Müller-Stich}
\credit{}

\author[5]{Pierre Jannin}[orcid=0000-0002-7415-071X]
\fnmark[1]
\credit{}

\author[50,51]{Stefanie Speidel}[orcid=0000-0002-4590-1908]
\fnmark[1]
\credit{}

\address[1]{Division of Computer Assisted Medical Interventions (CAMI), German Cancer Research Center (DKFZ), Heidelberg, Germany}
\address[2]{Faculty of Mathematics and Computer Science, Heidelberg University, Heidelberg, Germany}
\address[3]{Medical Faculty, Heidelberg University, Heidelberg, Germany }
\address[4]{Department of Computer Engineering, Faculty of Engineering, Gazi University, Ankara, Turkey}
\address[5]{LTSI, Inserm UMR 1099, University of Rennes 1, Rennes, France}
\address[6]{IRCAD, Strasbourg, France}
\address[7]{The Malone Center for Engineering in Healthcare, The Johns Hopkins University, Baltimore, Maryland, USA}
\address[8]{KARL STORZ SE \& Co. KG, Tuttlingen, Germany}
\address[9]{Department of Surgery, Klinikum rechts der Isar, Technical University of Munich, Munich, Germany}
\address[10]{The Hamlyn Centre for Robotic Surgery, Imperial College London, London, United Kingdom}
\address[11]{ICube, University of Strasbourg, CNRS, France}
\address[12]{\textcolor{black}{IHU Strasbourg, Strasbourg, France}}
\address[13]{Altair Robotics Lab, University of Verona, Verona, Italy}
\address[14]{Department of Surgery, Anne Arundel Health System, Annapolis, Maryland, USA}
\address[15]{Johns Hopkins University School of Medicine, Baltimore, Maryland, USA}
\address[16]{Department of Surgery, Stanford University School of Medicine, Stanford, California, USA}
\address[17]{Wellcome/EPSRC Centre for Interventional and Surgical Sciences, University College London, London, United Kingdom}
\address[18]{Department for General, Visceral and Transplantation Surgery, Heidelberg University Hospital, Heidelberg, Germany}
\address[19]{The Sheikh Zayed Institute for Pediatric Surgical Innovation, Children's National Hospital, Washington, D.C., USA}
\address[20]{The Perk Lab, Queen’s University, Kingston, Ontario, Canada}
\address[21]{L'Institut de Recherche en Informatique, Mathématiques, Automatique et Signal (IRIMAS), University of Haute-Alsace, Mulhouse, France}
\address[22]{Faculty of Information Technology, Monash University, Clayton, Victoria, Australia}
\address[23]{University of Toronto, Toronto, Ontario, Canada}
\address[24]{The Li Ka Shing Knowledge Institute of St. Michael’s Hospital, Toronto, Ontario, Canada}
\address[25]{Kyushu University, Fukuoka, Japan}
\address[26]{Kitakyushu Koga Hospital, Fukuoka, Japan}
\address[27]{Department of Radiology, Brigham and Women’s Hospital, and Harvard Medical School, Boston, Massachusetts, USA}
\address[28]{Computer Aided Medical Procedures, Technical University of Munich, Munich, Germany}
\address[29]{Department of Computer Science, The Johns Hopkins University, Baltimore, Maryland, USA}
\address[30]{ARTORG Center for Biomedical Engineering Research, University of Bern, Bern, Switzerland}
\address[31]{Innovation Center Computer Assisted Surgery (ICCAS), University of Leipzig, Leipzig, Germany}
\address[32]{\textcolor{black}{Division of Surgery and Interventional Science, University College London, London, United Kingdom}}
\address[33]{\textcolor{black}{Department of Visceral, Transplant, Thoracic and Vascular Surgery, Leipzig University Hospital, Leipzig, Germany}}
\address[34]{\textcolor{black}{Pediatric Surgery, Dr. von Hauner Children's Hospital, Ludwig-Maximilians-University, Munich, Germany}}
\address[35]{\textcolor{black}{University Hospitals Cleveland Medical Center, Case Western Reserve University, Cleveland, Ohio, USA}}
\address[36]{\textcolor{black}{Surgical AI and Innovation Laboratory, Massachusetts General Hospital, Harvard Medical School, Boston, Massachusetts, USA}}
\address[37]{\textcolor{black}{MyFetUZ Fetal Research Center, Department of Development and Regeneration, Biomedical Sciences, KU Leuven, Leuven, Belgium}}
\address[38]{\textcolor{black}{Center for Surgical Technologies, Faculty of Medicine, KU Leuven, Leuven, Belgium}}
\address[39]{\textcolor{black}{Department of Obstetrics and Gynecology, Division Woman and Child, Fetal Medicine Unit, University Hospitals Leuven, Leuven, Belgium}}
\address[40]{\textcolor{black}{Michael E. DeBakey Department of Surgery, Texas Children’s Hospital and Baylor College of Medicine, Houston, Texas, USA}}
\address[41]{\textcolor{black}{Department of Surgery and Cancer, Imperial College London, London, United Kingdom}}
\address[42]{\textcolor{black}{Department of BioSurgery and Surgical Technology, Imperial College London, London, United Kingdom}}
\address[43]{\textcolor{black}{Hamlyn Centre for Robotic Surgery, Imperial College London, London, United Kingdom}}
\address[44]{\textcolor{black}{Breast Unit, Imperial Healthcare NHS Trust, London, United Kingdom}}
\address[45]{\textcolor{black}{Department of Surgery, University Health Network, Toronto, Ontario, Canada}}
\address[46]{\textcolor{black}{National Hospital for Neurology and Neurosurgery, and UCL Queen Square Institute of Neurology, London, United Kingdom}}
\address[47]{\textcolor{black}{Massachusetts General Hospital, and Harvard Medical School, Boston, Massachusetts, USA}}
\address[48]{\textcolor{black}{Department of Urology, City Hospital Karlsruhe, Karlsruhe, Germany}}
\address[49]{\textcolor{black}{Institute for Applied Medical Informatics, Hamburg University Hospital, Hamburg, Germany}}
\address[50]{Division of Translational Surgical Oncology, National Center for Tumor Diseases (NCT/UCC) Dresden, Dresden, Germany}
\address[51]{Centre for Tactile Internet with Human-in-the-Loop (CeTI), TU Dresden, Dresden, Germany}

\cortext[cor1]{Corresponding author}
\fntext[fn1]{Shared first/senior author}

\begin{abstract}
Recent developments in data science in general and machine learning in particular have transformed the way experts envision the future of surgery. Surgical Data Science \textcolor{black}{(SDS)} is a new research field that aims to improve the quality of interventional healthcare through the capture, organization, analysis and modeling of data. While an increasing number of data-driven approaches and clinical applications have been studied in the fields of radiological and clinical data science, translational success stories are still lacking in surgery. In this publication, we shed light on the underlying reasons and provide a roadmap for future advances in the field. Based on an international workshop involving leading researchers in the field of SDS, we review current practice, key achievements and initiatives as well as available standards and tools for a number of topics relevant to the field, namely (1) infrastructure for data acquisition, storage and access in the presence of regulatory constraints, (2) data annotation and sharing and (3) data analytics. \textcolor{black}{We further complement this technical perspective with (4) a review of currently available SDS products and the translational progress from academia and (5) a roadmap for faster clinical translation and exploitation of the full potential of SDS, based on an international multi-round Delphi process}.
\end{abstract}




\maketitle

\section{Introduction}
\label{sec:introduction}

More than 15 years ago, in 2004, leading researchers in the field of computer aided surgery (CAS) organized the workshop \enquote{OR2020: Operating Room of the Future}. Around 100 invited experts including physicians, engineers, and operating room (OR) personnel attended the workshop \citep{cleary_or2020_2004} to define the OR of the future, with 2020 serving as target time frame. Interestingly, many of the problems and challenges identified back in 2004 do not differ substantially from those we are facing today. Already then, researchers articulated the need for \enquote{integration of technologies and a common set of standards}, \enquote{improvements in electronic medical records and access to information in the operating room}, as well as \enquote{interoperability of equipment}. In the context of data-driven approaches, they criticized the lack of an \enquote{ontology or standard} for \enquote{high-quality surgical informatics systems} and underlined the need for \enquote{clear understanding of surgical workflow and modeling tools}. Broadly speaking, the field has not made progress as quickly as researchers had hoped for at the time.

\begin{figure*}
\centering
\includegraphics[width=\textwidth]{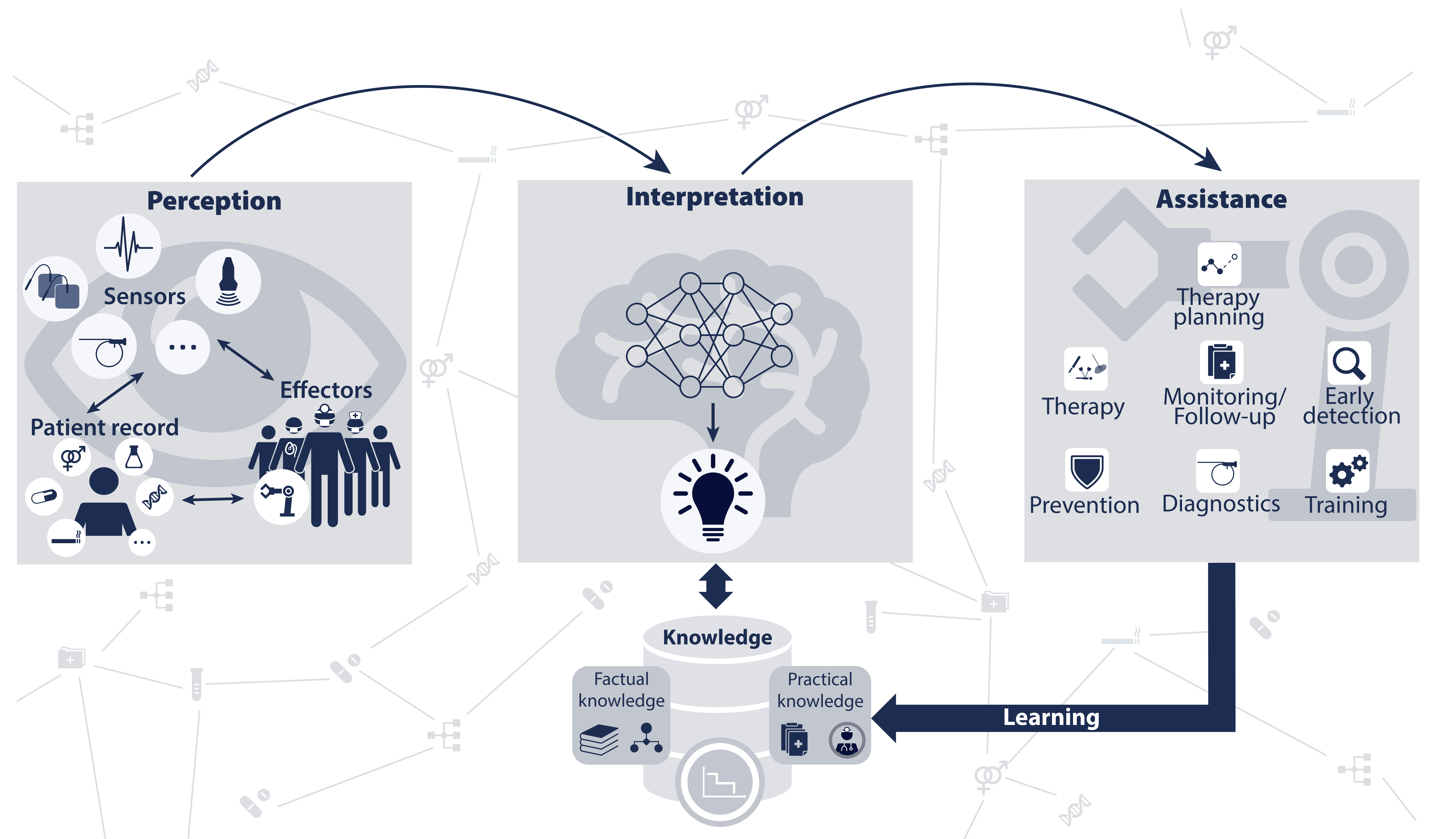}
\caption{Building blocks of a surgical data science (SDS) system. \textit{Perception}: Relevant data is perceived by the system (Sec.~\ref{sec:technicalInfrastructure}). In this context, \textit{effectors} include humans and/or devices that manipulate the patient including surgeons, operating room (OR) team, anesthesia team, nurses and robots. \textit{Sensors} are devices for perceiving patient- and procedure-related data such as images, vital signals and motion data from effectors. Data about the \textit{patient} includes preoperative images and laboratory data, for example. Domain \textit{knowledge} serves as the basis for data interpretation (Sec.~\ref{sec:dataAnnotation}). It comprises \textit{factual knowledge}, such as previous findings from studies, clinical guidelines or hospital-specific standards related to the clinical workflow as well as \textit{practical knowledge} from previous procedures. \textit{Interpretation}: The perceived data is interpreted in a context-aware manner (Sec.~\ref{sec:dataAnalytics}) to provide real-time \textit{assistance} (Sec.~\ref{sec:clinicalTranslation}). Applications of SDS are manifold, ranging from surgical education to various clinical tasks, such as early detection, diagnosis, and therapy assistance.}
\label{fig:sdsComponents}
\end{figure*}

More recently, the renaissance of data science techniques in general and deep learning (DL) in particular has given new momentum to the field of CAS. In response to the general artificial intelligence (AI) hype, a consortium of international experts joined forces to discuss the role of data-driven methods for the OR of the future. Based on a workshop held in 2016 in Heidelberg, Germany, the consortium defined Surgical Data Science (SDS) as a scientific discipline with the objective of improving \enquote{the quality of interventional healthcare and its value through capture, organization, analysis, and modelling of data} \citep{maier-hein_surgical_2017}. In this context, \enquote{data may pertain to any part of the patient care process (from initial presentation to long-term outcomes), may concern the patient, caregivers, and/or technology used to deliver care, and are analyzed in the context of generic domain-specific knowledge derived from existing evidence, clinical guidelines, current practice patterns, caregiver experience, and patient preferences}. Importantly, SDS involves the physical \enquote{manipulation of a target anatomical structure to achieve a specified clinical objective during patient care} \citep{maier-hein_surgical_2018}. In contrast to general biomedical data science, it also includes procedural data as depicted in Fig.~\ref{fig:sdsComponents}.

Three years later, in 2019, an international poll revealed that no commonly recognized surgical data science success stories exist to date, while success stories in other fields have been dominating media reports for years, as detailed in Sec.~\ref{sec:lackOfSuccessStories}. The purpose of this paper was therefore to go beyond the broad discussion of the \textit{potential} of SDS by providing an extensive review of the field and identifying concrete measures to pave the way for clinical success stories. The paper is based on an international workshop that took place in June 2019 in Rennes, France, and structured according to core topics discussed at the workshop. In Section~\ref{sec:lackOfSuccessStories}, we will review the questionnaire that served as the basis for the workshop as well as the international 4-round Delphi process \citep{hsu_delphi_2007} we conducted with clinical and technical stakeholders from more than 50 institutions to present concrete goals for the future. In the ensuing sections, we will present the current practice, key initiatives and achievements, standards, platforms and tools as well as current challenges and next steps for the main building blocks of SDS, namely technical infrastructure for data acquisition, storage and access (Sec.~\ref{sec:technicalInfrastructure}), methods for data annotation and sharing (Sec.~\ref{sec:dataAnnotation}) as well as data analytics (Sec.~\ref{sec:dataAnalytics}). A section about achievements, pitfalls and current challenges related to clinical translation of SDS (Sec.~\ref{sec:clinicalTranslation}) and a discussion of our findings (Sec.~\ref{sec:discussion}) will close the manuscript. While, by definition, SDS encompasses multiple interventional disciplines, such as interventional radiology and gastroenterology, the present paper puts a strong focus on surgery.

\section{Lack of success stories in surgical data science}
\label{sec:lackOfSuccessStories}

Machine learning (ML) has begun to revolutionize almost all areas of healthcare. Success stories cover a wide variety of application fields ranging from radiology and dermatology to gastroenterology and mental health applications \citep{miotto_deep_2018, topol_high-performance_2019}. Strikingly, such success stories appear to be lacking in surgery.

The international \citeauthor{surgical_data_science_initiative_surgical_2015} \citep{maier-hein_surgical_2017} was founded in 2015 with the mission to pave the way for AI success stories in surgery. Key result of the first workshop, which was inspired by current open space and think tank formats, was a common definition of SDS~\citep{maier-hein_surgical_2017} and a thorough description of the challenges in applying AI in interventional healthcare. The second edition of the workshop in 2019 focused on a comprehensive overview of the field including key research initiatives, industrial perspectives and first success stories. Prior to the workshop, the registered participants were asked to fill out a questionnaire, covering various aspects related to SDS. 43\% of the 77~participants were professors/academic group leaders (clinical or engineering), while the remaining were mostly either from industry (14\%) or PhD students / Postdocs (36\%). The majority of participants (61\%) agreed that the most important developments since the last workshop in 2016 were related to advances in AI. Notably, however, when participants were asked about the most impressive SDS paper, only a single paper \citep{maier-hein_surgical_2017} (the position paper from the first workshop) was mentioned more than twice (primarily by non-co-authors). The majority of participants agreed that the lack of representative annotated data is the main obstacle in the field and the main reason for the failure of previous SDS projects. Also, when referring to their personal experience, 33\% associated the main reason of failure of an SDS project with lack of data, followed by underestimation of the problem complexity (29\%). \citeauthor{endovis_endovis_2015} (28\%), Cholec80 \citep{twinanda_endonet_2017} (21\%) and JIGSAWS \citep{gao_jhu-isi_2014} (17\%) were mentioned as the most useful publicly available data sets but the small size/limited representativeness of the data set was identified as a core issue~(45\%).

Based on the replies to the questionnaire and the subsequent workshop discussions, we identified four areas that are essential for moving the field forward: (1) Technical infrastructure for data acquisition, storage and access, (2) data annotation and sharing, (3) data analytics, and (4) aspects related to clinical translation. These are reflected in the four main sections of this paper. We then conducted a Delphi process involving a consortium of medical and technical experts from more than 50 institutions (see list of co-authors) to formulate a mission statement along with a set of goals that are necessary to accomplish the respective mission (see Tab.~\ref{tab:goalsMission1}, ~\ref{tab:goalsMission2}, ~\ref{tab:goalsMission3} and ~\ref{tab:goalsMission4}) for each of the four areas.
More specifically, the coordinating team of the Delphi process put forth an initial mission statement and an initial set of goals for each of the four missions based on the workshop discussions. In a 4-round Delphi process, the remaining consortium members then iteratively refined the phrasing of the missions statements and goals and added further proposals for goals. This process yielded a set of 6-9 goals per mission that received support by at least two thirds of the voting members. Finally, the consortium collaboratively compiled a list of relevant stakeholders (Tab.~\ref{tab:stakeholders}) and then rated their importance for the four missions (Appendix~\ref{app:stakeholderImportance}). To avoid redundancy, the consortium further agreed on the following:

   Context statement: \textit{Unless otherwise specified, in all of the following text, a) surgical data science (SDS) represents the general context of the suggested phrases and b) “data” may pertain to any part of the patient care process (from initial presentation to long-term outcomes), may concern the patient, caregivers and/or technology used to deliver care and must be acquired, stored, and shared in accordance with both local and international regulatory constraints. In general, c) data handling should comply with the FAIR (\textbf{F}indability, \textbf{A}ccessibility, \textbf{I}nteroperability, and \textbf{R}euse) principles and d) user-friendliness should be a guiding principle in all processes related to data handling. Finally, e) the term SDS stakeholders refers to clinical, research, industrial, regulatory, public and private stakeholders.}\\


\begin{table}[pos=htp]

\caption{List of relevant SDS stakeholders.}
\begin{tcolorbox}[title= Surgical Data Science Stakeholders, colback=white, halign=left, subtitle style={boxrule=0.4pt,colback=gray}]
    %
       \tcbsubtitle{Clinical stakeholders}
    \begin{itemize}
    \item Hospital administration
    \item Hospital information technology (IT)
    \item Data-generating units in healthcare institutions (e.g. imaging departments, laboratories, centers of clinical studies)
    \item Surgical teams (e.g., surgeons, nurses, anesthesiologists)
    \item Medical professional bodies (e.g. the Society of American Gastrointestinal and Endoscopic Surgeons (SAGES) and the European Association of Endoscopic Surgery
(EAES))
    \end{itemize}
    \tcbsubtitle{Research stakeholders}
    \begin{itemize}
    \item Researchers (including clinician scientists)
    \item Research institutions such as university hospitals
    \item Scientific societies, (e.g. the international Society for Medical Image Computing and Computer Assisted Intervention (MICCAI))
    \item Journals/editors
    \item Funding agencies / institutions (e.g. the European Research Council (ERC))
    \end{itemize}
    \tcbsubtitle{Industrial stakeholders}
    \begin{itemize}
        \item Medtech companies - large
        \item Medtech companies - medium-sized
        \item Medtech companies - small-sized
        \item Industry federations
        \item Investors
    \end{itemize}
    \tcbsubtitle{Regulatory stakeholders}
    \begin{itemize}
        \item Lawmakers
        \item Regulatory agencies (e.g. the U.S. Food and Drug Association (FDA))
        \item Institutional review boards
        \item Insurance companies
    \end{itemize}
    \tcbsubtitle{Public and private stakeholders}
    \begin{itemize}
        \item Patients and/or their legal guardians/family
        \item Charities and donors
        \item Public health organizations, such as the World Health Organization (WHO)
        \item Media
        \item Citizens
    \end{itemize}
\end{tcolorbox}
\label{tab:stakeholders}
\end{table}

Based on the international questionnaire, the on-site workshop and the subsequent Delphi process, the following sections present the perspective of the members of the international data science initiative on the identified key aspects for generating SDS success stories. 

\section{Technical infrastructure for data acquisition, storage and access}
\label{sec:technicalInfrastructure}

To date, the application of data science in interventional medicine (e.g. surgery, interventional radiology, endoscopy, radiation therapy) has found comparatively limited attention in the literature. This can partly be attributed to the fact that only a fraction of patient-related data and information is being digitized and stored in a structured manner \citep{hager_chapter_2020} and that doing so is often an infeasible challenge in modern ORs. This section focuses on current hurdles in creating an environment that can record and structure highly heterogeneous surgical data for long-term usage. 

\subsection{Current practice}

Different types of data pose different types of challenges:\\

\textbf{Not all data can currently be acquired:} The OR is a highly dynamic environment where a team of health workers with varying specializations (e.g. surgeons, anesthesia team) continuously makes decisions based on device data, observation of the patient, and the outcome of previous actions. However, a lot of information that the healthcare workers perceive by interacting with the patient and each other is currently not at all acquired although it crucially affects decision making. This information relates to different human senses including vision, touch (e.g. palpation and tactile feedback from tissue) and hearing (e.g. acoustic signals resulting from instrument-tissue interaction \citep{ostler_acoustic_2020}, communication in the OR, etc.). First initiatives have begun addressing these issues (see Sec.~\ref{sec:technicalInfrastructure_keyInitiatives}) but the infrastructure is not yet widely available.\\

\textbf{Not all data that can be acquired is recorded and permanently stored:} Surgical data in minimally invasive surgery (MIS) routinely involves live image data of high resolution and frame rate. 
Modern stereoscopic endoscopes create two Full High Definition (HD) video streams at 60~Hz. If this data is to be stored uncompressed, it can quickly exceed 50~GB per video, with much larger file sizes possible depending on the situation and additional sensory input, and even larger again considering 4K resolutions.
Healthcare information technology (HIT) is currently not designed to prospectively record and store such large data files. \\


\textbf{Not all acquired data is digitized and stored in a structured manner:} A large proportion of documentation in the hospital is still unstructured. Reports, doctors’ letters, transcripts from examinations, treatment strategy plans and many more need to be documented in their original form for legal reasons \citep{kilian_method_2015}. When creating such documents, it is not uncommon to use printouts or Portable Document Format (PDF) documents that then form the basis of discussions between healthcare personnel or with patients. Resulting decisions are subsequently entered into the most relevant information systems as scans, unstructured, or semi-structured documents. As a result, all processes are documented in a manner satisfactory for legal aspects, but largely inaccessible to computation. This is especially true for information related to the surgical procedure, where the decision process leading up to the final operation strategy may not be stored at all (in simple cases) or only in the form of handwritten plans (in complex cases). 
Additionally, the exact parameters recorded for a specific intervention may differ between hospitals, leading to missing values if such data sets are merged. 
A host of information is potentially available from the actual surgery, including the exact steps taken, instruments used, information exchanged between personnel, haptic feedback, distractions, adaptations of the strategy plan, etc., many of which are not documented at all in OR reports, or documented incompletely. Evidence of this are e.g. similarly sized reports of the same procedures while the corresponding surgeries have radically different lengths. Additionally, problems during surgery may systematically be underreported \citep{hamilton_are_2018}. \\

\textbf{Not all data that is stored can be exchanged between systems:} Perioperative data is distributed over varying information systems. For example, Picture Archiving and Communication Systems (PACS) contain image data and videos, Radiology Information Systems (RIS) contain reports, findings and radiotherapy plans, and Laboratory Information Systems (LIS) contain laboratory data. Information systems that focus on a single aspect, e.g. laboratory data, can implement efficient storage, manipulation and retrieval methods specific to the given data types. At the same time, user interaction can be kept as simple as possible, with a large degree of workflow optimization for relevant personnel interfacing with the information systems. Linking data from several systems effectively complicates these models. The more data types are incorporated in a model, the more special cases need to be considered, making the model less accessible and harder to query. However, a strict semantic annotation is a prerequisite for guaranteeing retrievability and interoperability \citep{lehne_why_2019}. As a result, data exchange between information systems is rare. A positive example has been set in radiology, where the \citeauthor{digital_imaging_and_communications_in_medicine_dicom_nema_nodate} standard has enabled the structured exchange of imaging data. OR data recording systems have also started to offer connection to other hospital infrastructure systems like electronic medical records (EMR), e.g. NUCLeUS\texttrademark{} (Sony Corporation, Tokyo, Japan).
At present, however, this connectivity is typically not utilized widely or effectively. Also, stored OR data is generally not labelled and hence has limited utilization for SDS projects without significant efforts to restructure it.\\
\newpage
\textbf{Regulatory constraints make data acquisition, storage and access challenging:} SDS data collection, management and use must comply with standards in security and fidelity which typically vary depending on the data type and level of patient-specific information. Data governance in healthcare and specifically in surgery is still challenging and less mature compared to other domains \citep{tse_challenges_2018}. In the European Union (EU), the \textit{General Data Protection Regulation (GDPR)} covers issues pertaining to personal data both within the EU and its entry to or exit out of the EU since 2018 \citep{general_data_protection_regulation_gdpr_regulation_2016}. Similarly, in the United States of America (USA) the healthcare-specific \textit{Health Insurance Portability and Accountability Act of 1996 (HIPAA)} protects the confidentiality and integrity of patient data. In the United Kingdom (UK), the \textit{Data Protection Act (2018)} was put in place for the National Health Service (NHS). In other countries, equivalents for data protection exist and are related to the legal frameworks of the respective healthcare system. 

From an ethico-legal perspective, it is worth noting that companies commonly obtain surgical data either through contracts with individual consulting surgeons, licensing agreements with hospitals or in exchange for discounted pricing of their products. This current practice raises important issues regarding power imbalances and the democratization of data access \citep{august_value_2021}.

\subsection{Key initiatives and achievements}
\label{sec:technicalInfrastructure_keyInitiatives}

This section presents prominent SDS initiatives with a specific focus on data acquisition, access and exchange.

\textbf{Data acquisition:} Several industrial and academic initiatives have been proposed to overcome the bottleneck of prospective surgical data acquisition.

The \textit{DataLogger} (KARL STORZ SE \& Co. KG, Tuttlingen, Germany) is a technical platform for synchronously capturing endoscopic video and device data from surgical devices, such as the endoscopic camera, light source, and insufflator \citep{wagner_big_2017}.
The DataLogger has served as a basis for the development of a Smart Data Platform as part of the \textit{InnOPlan} project \citep{roedder_digital_2016} and has been continuously expanded to support an increasing number of medical devices and clinical information systems. It has also been used to collect data for Endoscopic Vision challenges (e.g. \citeauthor{endovis-_endovissub-workflow_nodate, endovis-workflowandskill_endovissub-workflowandskill_nodate, endovis-robust-mis_endovis_nodate}).

The OR Black Box\textsuperscript{\textregistered} \citep{goldenberg_using_2017} is a platform that allows healthcare professionals to identify, understand, and mitigate risks that impact patient safety. It combines input from video cameras, microphones, and other sensors with human and automated processing to produce insights that lead to improved efficiency and reduced adverse events. 
The OR Black Box has been in operation in Canada since 2014, in Europe since 2017 and in the USA since 2019. An early analysis of the OR Black Box use in laparoscopic procedures of over 100 patients has demonstrated that errors and distractions as annotated by experts viewing the procedures took place in every case, and often went unnoticed  or were at least not recalled by the surgeon at the time \citep{jung_first-year_2020}. 

In Strasbourg, France, the Nouvel Hôpital Civil (NHC), the Institut de Recherche contre les Cancers de l'Appareil Digéstif (IRCAD) and the Institut hospitalo-universitaire (IHU) record surgery videos for education purposes and research. These are curated and used mainly for IRCAD's WebSurg \citep{mutter_websurg_2011}, a free online reference for video-based surgery training with over 370,000 members.

The \textit{Surgical Metrics Project} began in October 2019 at the Annual Clinical Congress meeting of the American College of Surgeons (ACS). Over 200 board certified surgeons were equipped with wearable technology while they performed a simulated open bowel repair on porcine intestines. Multi-modal data, including electroencephalography (EEG) audio and video data were acquired to quantify efficient and successful operative approaches \citep{pugh_how_2020}.

The CDEGenerator is an online platform that addresses the need to create and share definitions of joint Core Data Elements (CDE) \citep{Varghese2018-gb}. These definitions combine a list of recorded parameters together with an exact semantic description. By agreeing on a common CDE, two hospitals can guarantee that the collected data is compatible to the degree of the described acquisition processes.


\textbf{Data access and exchange:} In the perioperative environment, the nonprofit organization Integrating the Healthcare Enterprise (IHE, Oak Brook, Illinois, USA) has been a driving force in forming a set of standards that facilitate data exchange \citep{grimes_challenge_2005}. It identifies clinical use cases, their requirements and relevant standards, and publishes guidelines (called \enquote{profiles}) on how to fulfill such use cases. IHE does not publish standards by itself, but rather identifies sets of standards (e.g. DICOM for image exchange and Logical Observation Identifiers Names and Codes (LOINC) \citep{forrey_logical_1996} for nomenclature) that are best suited to solve specific aspects of healthcare interoperability. Additionally, IHE regularly hosts \enquote{Connectathons}, where vendors present services with IHE profile implementations and test their systems against those of other vendors, verifying correct data exchange.

Inside the OR, efforts for transmitting and centralizing data have been explored for some time, particularly with integrated OR solutions provided by endoscopic device manufacturers and medical technology providers (KARL STORZ: OR1\texttrademark; Olympus Medical Systems (Tokyo, Japan): ENDOALPHA; Stryker (Michigan, USA): iSuite; Getinge AB (Getinge, Sweden): Tegris\textsuperscript{\textregistered}; Richard Wolf GmbH (Knittlingen, Germany): core nova; STERIS plc (Derby, UK): Harmony iQ\textsuperscript{\textregistered}; Brainlab AG (Munich, Germany): Digital O.R.; caresyntax GmbH (Berlin, Germany): PRIME365; Medtronic plc (Dublin, Ireland): Touch Surgery\texttrademark{} Enterprise; Sony: NUCLeUS\texttrademark; General Electric Company (Boston, USA): Edison\texttrademark; EIZO Corporation (Hakusan, Japan): \newline CuratOR\textsuperscript{\textregistered}). The wide availability of such systems should be an enabling technology for SDS efforts, not only allowing capturing of data from the OR but also setting a precedent on data management, security, storage and transmission.

Frequently, integrated ORs only provide technical interoperability for connecting image sources with displays (sinks) by using video and broadcasting standards such as Video Graphics Array (VGA), Digital Visual Interface (DVI), High-Definition Multimedia Interface (HDMI) or DisplayPort (DP). 
Higher levels of interoperability are easier to achieve with Internet Protocol (IP)-based data exchange standards (see Sec.~\ref{sec:technicalInfrastructure_standardsAndTools}).

Additionally to video routing and capturing, the integration of data from further devices in the OR is relevant. The German Federal Ministry of Education and Research (BMBF) lighthouse project OR.NET \citep{rockstroh_ornet_2017}, now continued as a nonprofit organization OR.NET e.V., worked on cross-manufacturer concepts and standards for the dynamic and secure networking of medical devices and information technology (IT) systems in the operating room and clinics \citep{kricka_history_2019, miladinovic_nfv_2018}. Initial results laid important foundations in the shape of a service-oriented communication protocol for the dynamic cross-vendor networking of medical devices and resulted in the International Organization for Standardization (ISO)/Institute of Electrical and Electronics Engineers (IEEE) 11073 Service-oriented Device Connectivity (SDC) series of standards (see Sec.~\ref{sec:technicalInfrastructure_standardsAndTools}). The projects InnOPlan \citep{roedder_digital_2016} (see paragraph "Data acquisition") and \citeauthor{op_41_op_nodate} also used SDC as the basis for device communication. InnOPlan's Smart Data platform enables real-time provision and analysis of medical device data to enable data-driven services in the operating room. The project \textit{OP 4.1} aimed at developing a platform for the OR - in analogy to an operating system for smartphones - that allows for integration of new technical solutions via apps.

The project \citeauthor{connected_optimized_network__data_in_operating_rooms_condor_project_nodate} is another collaborative endeavor that aims to build a video-driven Surgical Control Tower \citep{padoy_machine_2019, mascagni2021_ORblackTower} within the new surgical facilities of the IRCAD and IHU Strasbourg hospital by developing a novel video standard and new surgical data analytics tools. A similar initiative is The Operating Room of the Future (ORF) that researches device integration in the OR, workflow process improvement, as well as decision support by combining patient data and OR devices for MIS \citep{stahl_introducing_2005}.


\subsection{Standards, platforms and tools}
\label{sec:technicalInfrastructure_standardsAndTools}
Standards, platforms and tools have focused on the topics of interoperability as well as data storage and exchange.
\subsubsection{Interoperability}

Interoperability is defined by IEEE as \enquote{the ability of two or more systems or components to exchange information and to use the information that has been exchanged} \citep{ieee_ieee_1991} or by the Association for the Advancement of Medical Instrumentation (AAMI) as \enquote{the ability of medical devices, clinical systems, or their components to communicate in order to safely fulfill an intended purpose} \citep{aami_medical_2020}.

Numerous standards have been introduced to provide interoperability including Health Level 7 (HL7), IEEE 11073, American Society for Testing
and Materials (ASTM) F2761 (Integrated Clinical Environment (ICE)), DICOM, ISO TC215, European Committee for Standardization (CEN) TC251 and International Electrotechnical Commission (IEC) 62A. Different levels of interoperability can be distinguished, for example through the 7 Level Conceptual Interoperability Model (LCIM) from \citet{tolk_applying_2007}, which is defined as follows \citep{wang_levels_2009}:

\begin{itemize}
\item Level 0 -- No interoperability: \textit{Two systems cannot interoperate.}
\item Level 1 -- Technical interoperability: Two systems have the means to communicate, but neither has a shared understanding of the structure nor meaning of the data communicated. \textit{The systems have common physical and transport layers.}
\item Level 2 -- Syntactic interoperability: Two systems communicate using an agreed-upon protocol with structure but without any meaning. \textit{The systems exchange data using a common format.}
\item Level 3 -- Semantic interoperability: Two systems communicate with structure and have agreed on the meaning of the exchanged terms. \textit{The meaning of only the exchanged data is understood.}
\item Level 4 -- Pragmatic interoperability: Two systems communicate with a shared understanding of data, the relationships between elements of the data, and the context of the data but these systems do not support changing relationships or context over time. \textit{The meaning of the exchanged data and the relationships between pieces of information is understood.}
\item Level 5 -- Dynamic interoperability: Two systems are able to adapt their information models based on changing meaning and context of data over time. \textit{Evolving semantics are understood.}
\item Level 6 -- Conceptual interoperability: Includes the understanding and exchange of complex concepts. \textit{Systems are aware of each other’s underlying assumptions, models and processes.}
\end{itemize}

The number of interoperability levels varies from model to model and depends on the goal of the intended classification. For example, \citet{lehne_why_2019} use only four levels, the first two being identical to those listed above; the third, also called \enquote{semantic interoperability} addresses the complexities mentioned in levels 3 to 5 here, and the fourth puts forth the concept of \enquote{Organisational Interoperability}, which includes aspects of level 5 and 6.
The following paragraphs use the LCIM to classify the standards of interest to this paper.

\textbf{(1) Technical interoperability:} Modern hospitals typically have sophisticated networks, which makes technical interoperability the most achievable level \citep{lehne_why_2019}. The main challenge inside the OR, where real-time capability is often critical, is the available bandwidth. An uncompressed Full HD video stream at 60~fps in a color depth of 24~bit requires a bandwidth of 2.98~Gigabit per second (Gbps, not to be confused with Gigabyte per second (GBps), which is eight times larger). Available Ethernet ports typically have a data transfer rate of 1~Gbps. While more modern installations may reach Ethernet data transfer rates of 10~Gbps, this technology is still expensive and typically reserved for networks in data centers. Wireless networks are even slower: Modern devices often support theoretical speeds between 0.45~Gbps and 1.3~Gbps, which results in an effective bandwidth of around 50\% of the theoretical limit. The newest Wi-Fi-6 Standard, released late 2019, increases this theoretical limit to over 10~Gbps under laboratory conditions, but the effective speeds and adoption rate remain to be seen. In general, Wi-Fi (Wireless Fidelity) suffers from a higher rate of associated uncertainties as well as latency, depending on a number of environment factors. Critically, Wi-Fi packets may get lost if interference between networks is too high, causing latency spikes of potentially several hundreds of milliseconds, which may negatively affect real-time applications. The new 5G standard for wireless communication can potentially ease some of these problems by reaching theoretical speeds of 20~Gbps and avoiding conflicts with other networks since the relevant frequencies are licensed for specific areas. Additionally, 5G as a method of internet access could enable the transfer of large amounts of data to and from the hospital in relatively short time, something which previously required not readily available fast physical connections like glass fibre. While limitations of available bandwidth can be mitigated by using data compression, importantly, \enquote{losses imperceptible to humans} can still impede algorithm performance. 

It is worth noting that, especially inside the OR, devices still exist that are entirely unable to connect to networks (from basic technical infrastructure like doors or lights to routine medical equipment like certain anesthesia systems) or are not in the network due to missing capacities (e.g. Ethernet sockets) or software add-ons (e.g. a proprietary application programming interface (API)).

\textbf{(2) Syntactic interoperability:} At this level, the structure of exchanged data is defined with basic semantic information. This level is arguably where most of today's efforts in medical data interoperability take place, and where a number of standards compete. 
A major player in the standardization is HL7 \citep{kalra_openehr_2005}, which has developed standards for the exchange of patient data since 1987. The eponymous HL7 standard has been continuously updated and most notably includes the Version 3 Messaging Standard, which specifies interoperability for health and medical transactions. HL7 has been criticized for the complexity of its implementation \citep{goldenberg_using_2017}, resulting in the proposal of HL7 Fast Healthcare Interoperability Resources (FHIR). HL7 FHIR simplifies implementation through the use of widely applied web technologies. Another important standard is provided by the openEHR foundation. In contrast to HL7, openEHR is not only a standard for medical data exchange, but an architecture for a data platform that provides tools for data storage and exchange. With this, however, come added complexity and challenges.

HL7 and openEHR provide the broadest scope of medical data exchange, but both build on standards that solve specific subtasks. While a complete listing is out of scope for this article, one notable exception is DICOM, which today is the undisputed standard for the management of medical imaging information. In 2019, DICOM was extended to include real-time video (DICOM Real-Time Video (DICOM-RTV)). This extension is an IP-based DICOM service for transmitting and broadcasting real-time video, with synchronized metadata, to subscribers (e.g. a monitor or SDS application server) with a quality comparable to standard OR video cables. 

The previously mentioned standards focus on enabling the exchange of patient-individual data between Hospital Information Systems (HIS). Inside the OR, requirements differ, since a host of devices create a real-time data stream that focuses on sensoric input instead of direct patient information (diagnosis, habits, morbidity). Accordingly, data exchange standards inside the OR are geared towards these data types. OpenIGTLink \citep{tokuda_openigtlink_2009}, for example, started as a communication protocol for Image Guided Therapy (IGT) applications. Today, OpenIGTLink has been expanded to exchange arbitrary types of data by providing a general framework for data communication. However, it does not define broad standards for the data format, instead relying on users to implement details according to their needs. Through this model, OpenIGTLink enabled data exchange inside the OR long before broad standards were feasible. Similarly, for the field of robotics, the Robot Operating System (ROS) \citep{koubaa_robot_2016} has been proposed.

More recent efforts by the OR.NET initiative (see Sec.~\ref{sec:technicalInfrastructure_keyInitiatives}) produced the IEEE 11073 SDC ISO standard which provides a means for general data and command exchange for devices and enables users to control devices in the OR. Standards less specific to the healthcare environment are also available. Similar to OpenIGTLink, The Internet of Things (IoT), for example, defines a standard for device communication without defining standards for the communicated data. While it has been used for data exchange between information systems \citep{xie_open_2018}, and between devices in the OR \citep{miladinovic_nfv_2018}, it has elicited mixed reactions.

\textbf{(3) Semantic interoperability:} This is the domain of clinical nomenclatures, terminologies and ontologies.
While modern standards like HL7 FHIR and openEHR already define basic semantics in data exchange, extending these annotations to more powerful nomenclatures like SNOMED CT (Systematized Nomenclature of Medicine - Clinical Terms) \citep{cornet_forty_2008} (see Sec.~\ref{sec:dataAnnotation}) enables systems to not only share data, but also their exact meaning and scope (i.e. what kind of data exactly falls under the given definition). To illustrate the difference between this level and the previous: HL7 FHIR defines less than 200 healthcare concepts (i.e. terms with a well-defined meaning) \citep{bender_hl7_2013}, while SNOMED CT defines more than 340,000 concepts \citep{minarro-gimenez_quantitative_2019}. Today, semantic interoperability is largely defined by terminologies (systematic lists of vocabulary), ontologies (definitions of concepts and categories along with their relationships) and taxonomies (classifications of entities, especially organisms) - the borders between which are often fluid. 
Standard languages such as the Resource Description Framework (RDF), Resource Description Framework Schema (RDFS) and the Web Ontology Language (OWL) \citep{bechhofer_owl_2009} have been defined by the World Wide Web Consortium (W3C), guaranteeing interoperability between ontology resources and data sets based on these ontologies. The aforementioned SNOMED CT is arguably the most complete terminology, spanning the whole field of clinical terms with a wide set of available translations. However, specialized alternatives may perform better on their respective field. Additionally, a host of medical ontologies are available. Most notable is the family of ontologies gathered under the OpenBiological and Biomedical Ontologies (OBO) Foundry \citep{smith_obo_2007}, which cover a wide array of topics from the biomedical domain and share the Basic Formal Ontology (BFO) \citep{grenon_snap_2004} as a common top-level ontology. Intraoperatively, the OntoSPM \citep{gibaud_toward_2018} provides terminology for the annotation of intraoperative processes, and has spawned efforts for the annotation of binary data \citep{katic_what_2017}. Common to all these efforts is that they serve best in combination with a standard addressing syntactic interoperability, where they can add semantic information to the data exchange. Semantic interoperability goes hand in hand with data annotation, and is expanded upon in Sec.~\ref{sec:dataAnnotation}.

It is important to note that semantic interoperability does not guarantee the availability of data. If two hospitals have agreed on a detailed semantic model but record different parameters for a specific procedures, then the two resulting data sets will contain well-defined but empty fields. Two avoid this, it is necessary to agree on lists of recorded parameters, e.g. in the form of CDE.

\textbf{(4) Pragmatic interoperability:} In order to define context, additional modeling is required to capture data context and involved processes. This can in part be achieved by extending modeling efforts from the semantic interoperability level to include these concepts. Furthermore, efforts to formalize the exchange processes themselves are required. In IEEE 11073 descriptions for architecture and protocol (IEEE 11073-20701) and in HL7 the IHE Patient Care Device (PCD) implementation guide and the conformance model are provided. 

For the remaining two levels, developments are more recent and less formalized. For Level \textbf{(5) Dynamic interoperability}, it is required to model how the meaning of data changes over time. This can range from simple state changes (planned operations becoming realized, proposed changes becoming effective) to new data types being introduced and old data types changing meaning or being deprecated. In IEEE 11073 the participant key purposes and in HL7 the workflow descriptions are created for supporting these aspects. Finally, Level \textbf{(6) Conceptual interoperability} allows for exchanging and understanding complex concepts. This requires a means to share the conceptual model of the system, its processes, state, architecture and use cases. This level can be achieved through defining use cases and profiles (e.g. IHE Services-oriented Device Point-of-care Interoperability (SDPi) Profiles) and/or provisioning reference architecture and frameworks.

\subsubsection{Data storage and distribution}

While current standards have focused on data exchange, they typically do not address data distribution and storage. Typically, data is exchanged between two defined endpoints (e.g. a tracking device and an IGT application, or a computed tomography (CT) scanner and a PACS system). To achieve a system that can be dynamically expanded with regard to its communication capabilities, it is necessary to implement messaging technology. Such tools allow arbitrary devices to take part in communication by registering via a message broker, where messages can typically be filtered by their origin, type, destination, for instance. Examples include Apache Kafka \citep{kim_reliable_2017, spangenberg_implementation_2018} or RabbitMQ\textsuperscript{\textregistered} \citep{ongenae_semantic_2016, trinkunas_research_2018}. Such systems enable developers to create flexible data exchange architectures using technologies that are mature and usually well documented thanks to their wide application outside the field of healthcare. However, they also create a level of indirection which introduces additional delay (which may be negligible with only a few milliseconds in local networks, or significant with several tens or even hundreds of milliseconds over the internet or wireless networks).

Finally, recording of the exchanged data requires distinct solutions as well. High-performance, high-reliability databases form an essential requirement for many modern businesses. Thanks to this demand, a large body of established  techniques exists, from which users can select the right tool for their specific needs. Binary medical data (images, videos, etc.) can be stored on premise in modern PACS systems, which provide extensive support for data annotation, storage and exchange. For clinical metadata, the selection of technology typically depends on the level of standardization of the recorded data. Highly standardized data can possibly be stored directly through interfaces of e.g. the IHE family of standards. If the target data are not standardized, but homogeneous, then a database model for classical database languages (e.g. Structured Query Language (SQL)) may be suitable. Use cases where a wide array of highly heterogeneous data is recorded may choose modern NoSQL databases. These databases do not (or not exclusively) rely on classical tabular data models, but instead allow the storage and querying of tree-like structures. The JavaScript Object Notation (JSON) Format is a popular choice for NoSQL databases for its wide support in toolkits and the immediate applicability with regard to Representational State Transfer (REST)-APIs. While initially applications of these databases were geared towards data lakes because of the relative ease of application, NoSQL databases have recently seen widespread application in big data and ML \citep{dasgupta_practical_2018}. A notable example is Elasticsearch (Elastic NV, Amsterdam, the Netherlands), which has achieved widespread distribution and is ranked among the most used search servers \citep{db-engines_db-engines_2020}.

Through the rising relevance of web technology, storing data in the cloud is increasingly becoming a viable option. A vast array of services are available and have been applied in the medical domain (e.g. Amazon Web Services (AWS) \citep{holmgren_health_2017}, Microsoft Azure \citep{hussain_cloud-based_2013}, and others). Storing data in the cloud has the potential to save money on HIT by eliminating the need to reduce the locally required storage capacity and maintenance personnel, but brings with it privacy concerns and slower local access to data than from local networks, which may be noticeable especially for large binary data like medical images and video streams. While data privacy options are available for all major services, the implementing personnel have to understand these options and align with them  the privacy needs of the institution and the respective data. Since answering these questions is complex, the privacy requirements  strict, and the consequences for failing to comply with the law severe, the created solutions are often conservative in nature with regard to privacy. Additionally, downloading large data sets may be costly, as in general, cloud storage providers incentivize performing computations in the cloud.

Finally, solutions to facilitate local storage have been proposed. Commercially available systems such as SCENARA\textsuperscript{\textregistered} .STORE (KARL STORZ) compress surgical images and video data over time to decrease storage needs.
Alternatively, SDS tools can be used to selectively store critical video sequences instead of entire procedural videos, as recently proposed \citep{MasacagniAlapatt2021_EndoDigest}

\begin{table}
\caption{Mission statement corresponding to technical infrastructure (Sec.~\ref{sec:technicalInfrastructure}) along with corresponding goals. The median priority for each goal is symbolized by a vertical priority bar representing the priorities "Not a priority" (one square filled), "Low priority" (two squares filled), "Medium priority" (three squares filled), "High priority" (four squares filled), "Essential priority" (five squares filled).}
\begin{tcolorbox}[title= Mission I: Technical infrastructure, colback=white, halign=left, subtitle style={boxrule=0.4pt,colback=gray}]
    %
    \textbf{Make the data needed for training, validating, evaluating and applying SDS algorithms accessible to and exchangeable between researchers, healthcare professionals and other stakeholders, both live and retrospectively}
    \tcbsubtitle{Goals}
    \hspace{-0.31cm}
    \begin{tabular}[c]{ m{0.9\textwidth} m{0.1\textwidth} }
        \begin{description}
            \item[Goal 1.1] Enhance health information technology including dedicated personnel resources in healthcare institutions such that relevant data can be prospectively, routinely, systematically acquired and stored
        \end{description} & \includegraphics[width=5pt]{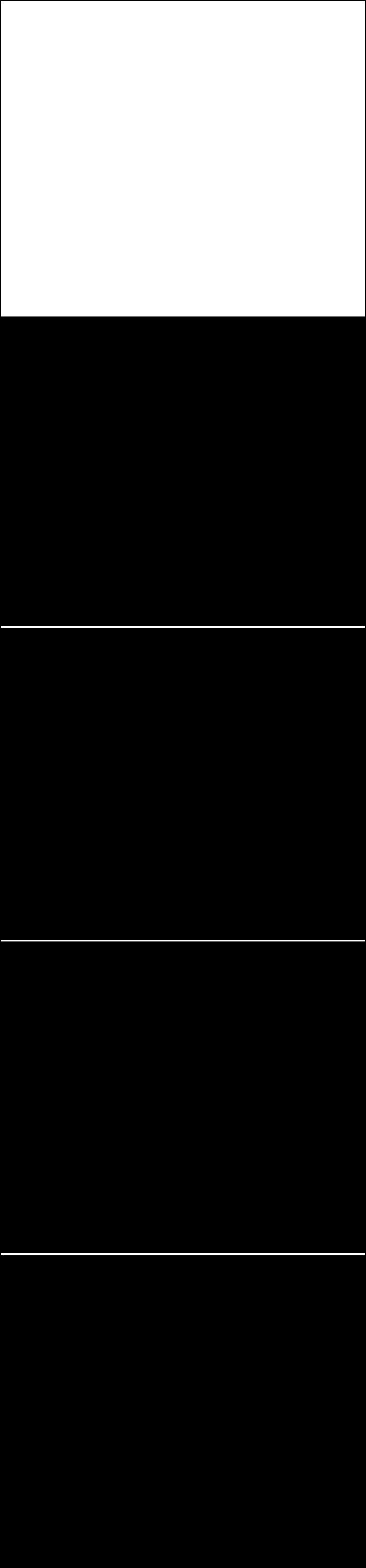} \newline \\ 
        \vspace{-0.4cm}
        \begin{description}
            \item[Goal 1.2] Develop standards for data storage with respect to key aspects including data structure, format and longevity
        \end{description} & \vspace{-0.4cm} \includegraphics[width=5pt]{figures/Priority4} \newline \\
        \vspace{-0.4cm}
        \begin{description}
            \item[Goal 1.3] Enhance health information technology in healthcare institutions such that data are relatable to their clinical context and can be transferred from/to acquisition/storage/display systems
        \end{description} & \vspace{-0.4cm} \includegraphics[width=5pt]{figures/Priority4} \newline \\
        \vspace{-0.4cm}
        \begin{description}
            \item[Goal 1.4] Develop new intraoperative imaging methods for obtaining relevant information on tissue function, morphology and pathology 
        \end{description} & \vspace{-0.4cm} \includegraphics[width=5pt]{figures/Priority4} \newline \\
        \vspace{-0.4cm}
        \begin{description}
            \item[Goal 1.5] Enhance health information technology to enable real-time inference in interventional settings
        \end{description} & \vspace{-0.4cm}
        \includegraphics[width=5pt]{figures/Priority4} \newline \\
        \vspace{-0.4cm}
        \begin{description}
            \item[Goal 1.6] Build local and international collaborations and partnerships involving clinical, research, public, regulatory and industrial stakeholders to implement the goals in accordance with internationally agreed standards
        \end{description} & \vspace{-0.4cm} \includegraphics[width=5pt]{figures/Priority4} \newline
    \end{tabular}
\end{tcolorbox}
\label{tab:goalsMission1}
\end{table}

\subsection{Current challenges and next steps}

The infrastructure-related mission as well as the corresponding goals generated by the consortium as part of the Delphi process 
are provided in Tab.~\ref{tab:goalsMission1}. This section elaborates on some of the most fundamental aspects:\\


\textbf{How to enable prospective capturing and storing of relevant perioperative data?} (goals 1.1/1.2): 
A major challenge we face is to capture all relevant perioperative data. While several initiatives and standards are already dedicated to this problem, a particular focus should be put on the recording and integration of patient outcome measures, including measures that need to be captured long after the patient has left the hospital (e.g. 5-year-survival).
The field of SDS stands in contrast to the field of radiology, where the DICOM standard now covers the exchange of medical images and related data. This standard can be seen as a direct result of market pressure: Early medical imaging devices did not prioritize communication standards, instead relying on manufacturer-supplied software specific to the hardware purchased. This behaviour did not change until PACS systems became widespread, providing specialized software that offered a benefit to clinical workflows, and the ability to transmit images to them became a driving requirement for the purchase of new imaging hardware. However, the previously mentioned domain complexity also affects standard development. For example, the DICOM specification document alone consists of 6,864 pages\footnote{\url{http://dicom.nema.org/medical/dicom/current/} (accessed 2020-07-30)}, indicating the effort to develop and maintain such a standard.
Evolving standards for the exchange of medical data like IEEE 11073 SDC and HL7 FHIR are a step in the right direction, but in order to create a driving force, incentivizing  the industry to enable widespread interconnection appears useful. 

Storing acquired data is, in theory, largely possible with modern technologies. 
 Missing, however, are standards for storage format, duration and data quality. These should be developed with the involvement of industrial stakeholders and the respective clinical/technical societies and should specifically include recommendations with respect to minimum standards for storage and annotation. The international Society of American Gastrointestinal and Endoscopic Surgeons (SAGES), for example, created an AI task force with the mission to propose and establish best practices for structured video data acquisition and storage, including recommendations for resolution and compression. Generally speaking, a clear distribution of roles between different stakeholders, particularly regarding who takes the initiative, as well as a clear definition of the subject matter to be standardized are now needed.\\

\textbf{How to link data from different sources and sites?} (goal 1.3) The need for exchanging data between different sources and sites calls for semantic operability (Sec.~\ref{sec:technicalInfrastructure_standardsAndTools}): Simply storing all data in a data lake without sufficient metadata management poses the risk of creating a data swamp that makes data extraction hard to impossible \citep{hai_constance_2016}. 
Data distribution among several systems is a healthy approach since it reduces load on a single system and enables engineers to choose the system best suited for the specific types of data stored within. As long as metadata models \citep{gibaud_toward_2018, marz_toward_2015, soualmia_efficient_2016} exist that are able to sufficiently describe the data and where to find them, retrieval will be possible through querying the model. Accordingly, efforts should focus on enhancing current clinical information infrastructures from the level of syntactic operability to semantic interoperability. Metadata also becomes essential for data sharing. 
An increasingly popular approach to data sharing is federated learning \citep{konecny_federated_2016, rieke_future_2020}. Instead of sharing data between institutions, the training of algorithms is distributed among participants. While this presumably reduces the ethical and legal complications associated with large-scale data sharing, it is still necessary to achieve semantic interoperability, and the regulatory issues regarding the exchange of models that contain encoded patient data are not fully understood yet.\\

\textbf{How to perceive relevant tissue properties dynamically?} (goal 1.4) 
Surgical imaging modalities should provide discrimination of local tissue with a high contrast-to-noise-ratio, should be quantitative and digital, ideally be radiation- and contrast agent-free, enable fast image acquisition and be easy to integrate into the clinical workflow. The approach of registering 3D medical image data sets to the current patient anatomy for augmented reality visualization of subsurface anatomical details has proven ill-suited for handling tissue dynamics such as perfusion or oxygenation (e.g. for ischemia detection).
The emerging field of biophotonics refers to techniques that take advantage of the fact that different tissue components feature unique optical properties for each wavelength. 
Specifically, spectral imaging uses multiple bands across the electromagnetic spectrum~\citep{clancy_surgical_2020} to extract relevant information on tissue morphology, function and pathology (see e.g. \citet{wirkert_robust_2016, moccia_uncertainty-aware_2018, ayala_video-rate_2021}). Benefiting from a lack of ionizing radiation, low hardware complexity and easy integrability into the surgical workflow, spectral imaging could be leveraged to inform surgical operators directly or be used for the generation of relevant input for SDS algorithms \citep{Mascagni2018_ORimaging}. Open research questions are, among others, related to reproducibility of measurements, possible confounders in the data~\citep{dietrich_machine_2021}, inter-patient variability and the robust quantification of tissue parameters in clinical settings.\\

\textbf{How to enable real-time inference in interventional settings?} (goal 1.5)
While processing times of several seconds or even minutes may be acceptable in some scenarios, other SDS applications, such as autonomous robotics, require real-time inference. Real-time inference requires a number of complex prerequisites to be fulfilled. Relevant data needs to be streamed to a common endpoint where it can be processed; data streams need to be sufficiently formalized to enable fully automatic decoding; the hardware and networks receiving these streams must be sufficiently fast to decode the streams with minimal latency and high resilience, and the algorithms that provide inference need to be implemented efficiently and run on sufficiently fast hardware to enable real-time execution. If additional data (e.g. preoperative imaging, patient-specific data, etc.) is required, the algorithms need to be able to access this data, and inferred information needs to be relayed to the OR team in an adequate manner. These problems can potentially be addressed in a variety of ways, however, it seems prudent to integrate the necessary infrastructure (acquisition, computation, display) directly on site in or near the OR. In a first step, test environments such as experimental operating rooms can serve as platforms where technical concepts for real-time interference can be developed, validated and evaluated in a realistic setting.\\

\textbf{How to overcome regulatory and political hurdles?} (goal 1.6) Timelines and associated costs of data privacy management (discussed further in Sec.~\ref{sec:dataAnnotation_currentChallengesAndNextSteps}) and regulatory processes need to be supported in both academic and commercial projects: Academic work requires funding and appropriate provision for delays in the project timeline. 
Notably, the COVID-19 pandemic may have stimulated rapid response from both academic and regulatory bodies in response to urgent needs, and perhaps some of this expedience will remain (examples in Continuous Positive Airway Pressure (CPAP) devices such as \citeauthor{ucl-ventura_ucl-ventura_2020}). Industry also needs to allocate costs, adhere and maintain standards, cover liability and have clear expectations on the required resources. While these processes are well developed and supported in large organizations, smaller companies, in particular startups,  have less capacities for them at their disposal. A variety of additional standards would also need to be met since a prospective SDS system approaches a medical device as defined by The U.S. Food and Drug Administration (FDA) (USA) or the Medical Device Regulation (MDR) (EU). These may be ISO-certified or require audits and approval from regulatory agencies and notified bodies, compliance with data protection regulations (e.g. GDPR), more stringent (cyber-)security features and testing adherence.
As the field of AI and its regulation is increasingly discussed in public venues, political visibility is rising. By clearly identifying the limiting effects of insufficient infrastructure on the one hand, and potential benefits of improving it on the other, it should become possible to convince political and clinical stakeholders that an investment in HIT as well as dedicated data management and processing personnel is key to exploiting the potential of AI for interventional healthcare. Furthermore, industrial engagement in creating the necessary infrastructure needs to be fostered within the boundaries of global standardization while considering the specific market needs. Healthcare institutions thus need to engage globally with industry to put forth common standards and processes enabling SDS applications compatible with strategic business needs. Of note, existing infrastructures can be leveraged and enhanced in this process. The SDS community should be aware of the complexity of the topic and the messages that are publicized (i.e. premature success stories) and create constructive proposals with realistic outlooks on potential benefits, focusing on long-term investments with the potential to drive change. Specifically, market studies could identify for each individual stakeholder the benefits of SDS solutions compared to their expected costs. Consider for instance a "number needed to treat" type of example, where for every X number of patients for which data insights are applied, one complication costing USD Y may be avoided. By providing estimated returns on investment for improvements to clinical delivery based on reducing person-hours, complications, or duplicative work, such studies would in turn provide key arguments for future investments. 

Overall, local and international collaborations and partnerships involving clinical, patient, academic, industry and political stakeholders are needed (see Tab.~\ref{tab:stakeholders}). Policies and procedures regarding data governance within an institution have to be defined that involve all stakeholders within  the SDS data lifecycle. Already existing multinational political entities or governing bodies, as exemplified by the EU, can be leveraged in a first step towards international collaboration and standardization. When implementing the goals put forth in Tab.~\ref{tab:goalsMission1}, internationally agreed standards should be respected. These include, but are not limited to, ethical guidelines. In fact, the world health organization (WHO) recently put forth  a guidance document on \textit{Ethics \& Governance of Artificial Intelligence for Health} \citep{health_ethics__governance_ethics_2021}, which was compiled by a multidisciplinary team of experts from the fields of ethics, digital technology, law and human rights, as well as experts from Ministries of Health. The report identifies the ethical challenges and risks associated with the use of AI in healthcare and puts forth several internationally agreed on best practices for both the public and the private sector. 


\section{Data annotation and sharing}
\label{sec:dataAnnotation}

The access to annotated data is one of the most important prerequisites for SDS.
There are different requirements that impact the quality of the annotated data sets. Ideally, they should include multiple centers to capture possible variations using defined protocols regarding acquisition and annotation, preferably linked to patient outcome. In addition, the data set has to be representative for the task to be solved and combined with well-defined criteria for validation and replication of results. Broadly, the key considerations when generating an annotated data set include reliability, accuracy, efficiency, scalability, cost, representativeness and correct specification.

\subsection{Current practice}
A comprehensive list of available curated data sets that are relevant to the field of SDS is provided in appendix \ref{app:publicDataRepositories}. In general, they serve as a good starting point, but are still relatively small, often tied to a single institution, and extremely diverse in structure, nomenclature, and target procedure.

Surgical data such as video involves diverse annotations with different granularity depending on the clinical use case to be solved. It can be distinguished between spatial, temporal or spatio-temporal annotations. Examples for spatial annotations include image-level classification (e.g. what tissue/tools/events are visible in an image), semantic segmentation (e.g. which pixels belong to which tissue/tools/events in an image) and numerical regression (e.g. what is the tissue oxygenation at a certain location). Temporal annotations involve the surgical workflow and can have different levels of granularity, e.g. surgical phases at the highest level, which consist of several steps, which are in turn composed of activities such as suturing or knot-tying \citep{lalys_surgical_2014}. In addition, specific events such as complications, performance or quality assessment of specific tasks complement temporal annotations. Spatio-temporal annotations involve both spatial and temporal information. While simple annotation tasks such as labeling surgical instruments may be accomplished by non-experts~\citep{maier-hein_can_2014}, more complex tasks such as tissue labeling or quality assessment of anastomoses most likely require domain experts.

The major bottleneck for data annotation in surgical applications is access to expert knowledge.
Reducing the annotation effort is therefore of utmost importance, and various methods have been proposed. Crowdsourcing \citep{maier-hein_can_2014} has proven to be a successful method, but designing the task such that non-experts are able to provide meaningful annotations is still one of the biggest challenges. Recently, active learning approaches that determine which unlabeled data points would provide the most information and  thus reduce the annotation effort to these samples have been proposed \citep{bodenstedt_active_2019}. Similarly, error detection methods reduce the annotation effort to  erroneous samples only \citep{lecuyer_assisted_2020}. Data can also be annotated directly during acquisition
\citep{padoy_statistical_2012, sigma_surgical_corporation_sigma_nodate}.

\subsection{Key initiatives and achievements}


One of the most successful initiatives fostering access to open data sets is \citeauthor{grand_challenge_grand_nodate} which provides infrastructure and tools for organizing challenges in the context of biomedical image analysis. The platform hosts several challenges including data sets and also serves as a framework for end-to-end development of ML solutions. Notably, the Endoscopic Vision Challenge \citeauthor{endovis_endovis_2015}, an initiative that takes place at the international conference hosted by the Medical Image Computing and Computer Assisted Intervention (MICCAI) Society, is the largest source of SDS data collections~\citep{bernal_comparative_2017, endovis15_instrument_subchallenge_dataset_endovis15_nodate, endovis-giana_gastrointestinal_nodate, allan_2017_2019, hattab2020kidney, ac97-8m18-21, allan_2018_2020, maier-hein_heidelberg_2021, allan_stereo_2021, endovis-workflowandskill_endovissub-workflowandskill_nodate, ros_comparative_2021, zia_surgical_2021, huaulme_micro-surgical_2021, EndoVis-HeiSurf_nodate, EndoVis-GIANA21_nodate, EndoVis-CholecTriplet2021_nodate, EndoVis-FetReg_nodate, EndoVis-PETRAW_nodate, EndoVis-SimSurgSkill_nodate}. It consists of several sub-challenges every year which support the availability of new public data sets for developing and benchmarking methods. Generally speaking, however, quality control in biomedical challenges and data sharing is still an issue \citep{maier-hein_why_2018, maier-hein_bias_2020}. 

The importance of public data sets in general is illustrated through new journals dedicated to only publishing high quality data sets, such as
\textit{Nature Scientific Data}.
An important contribution in this context are the FAIR data principles\citep{wilkinson_fair_2016}, already introduced in the \textit{context statement} above.  
Recently, the Journal of the American Medical Association (JAMA) Surgery partnered with the \citeauthor{surgical_outcomes_club_surgical_nodate} and launched a series consisting of statistical methodology articles and a checklist that aims to elevate the science of surgical database research \citep{haider_checklist_2018}. It also includes an overview of the most prominent surgical registries and databases, e.g. the National Cancer Database \citep{merkow_practical_2018}, the National Trauma Data Bank \citep{hashmi_practical_2018} or the National Surgical Quality Improvement Program \citep{raval_practical_2018}.

Annotation of data sets requires consistent ontologies for SDS. The OntoSPM project \citep{gibaud_ontospm_2014} is the first initiative whose goal is to focus on the modeling of the entities of surgical process models, as well as the derivation LapOntoSPM \citep{katic_erratum_2016} for laparoscopic surgery. OntoSPM is now organized as a collaborative action associating a dozen research institutions in Europe, with the primary goal of specifying a core ontology of surgical processes, thus gathering the basic vocabulary to describe surgical actions, instruments, actors, and their roles. An important endeavor that builds upon current initiatives was recently initiated by SAGES, which hosted an international consensus conference on video annotation for surgical AI. The goal was to define standards for surgical video annotation based on different working groups regarding temporal models, actions and tasks, tissue characteristics and general anatomy as well as software and data structure \citep{meireles2021sages}.

\subsection{Standards, platforms and tools}

In SDS, images or video are typically the main data sources since they are ubiquitous and can be used to capture information at different granularities ranging from cameras observing the whole interventional room or suite to cameras inserted into the body endoscopically or observing specific sites through a microscope \citep{chadebecq_computer_2020}. Different image/video annotation tools regarding spatial, temporal and spatio-temporal annotations already exist (Table~\ref{tab:annotationTools}), but to date no gold standard framework enabling different annotation types combined with AI-assisted annotation methods exists in the field of SDS.

Consistent annotation requires well-defined standards and protocols taking different clinical applications into account. Current initiatives are working on the topic of standardized annotation, but no widely accepted standards have resulted from the efforts yet. Notable exceptions can be seen in the fields of skill assessment, where annotations have been required for a long time to rate students and can serve as an example for different kinds of SDS annotation protocols \citep{vedula_objective_2017}, and in  cholecystectomy, where methods for consistent assessment of photos \citep{Sanford2014_Doublet} and videos \citep{mascagni_formalizing_2020} of the Critical View of Safety (CVS) were developed to favour documentation of this important safety step.

Data annotation also requires a consistent vocabulary, preferable modeled as ontology (Sec.~\ref{sec:technicalInfrastructure}). Several relevant ontologies with potential use in surgery such as the Foundational Model of Anatomy (FMA), SNOMED CT or RadLex \citep{langlotz_radlex_2006} are already available. Existing initiatives like the OBO Foundry project that focuses on biology and biomedicine provide further evidence that building and sharing interoperable ontologies stimulate data sharing within a domain. In biomedical imaging, ontologies have been successfully used to promote interoperability and sharing of heterogeneous data through consistent tagging \citep{gibaud_neurolog_2011, smith_biomedical_2015}. 

The challenges and needs for gathering large-scale, representative and high-quality annotated data sets are certainly not limited to SDS. In response, a new industry branch has emerged offering online data set annotation services through large organized human workforces. A listing of the major companies is provided in Table~\ref{tab:annotationServices}. Interestingly, the market was estimated to grow to more than USD~1~billion by 2023 in 2019 \citep{cognilytica_data_2019}, but the consecutive annual report this year estimates the market to grow to more than USD~4.1~billion by 2024 \citep{cognilytica_data_2020}. Most companies recruit non-specialists who can perform conceptually simple tasks on image and video data, such as urban scene segmentation and pedestrian detection for autonomous driving. Recently, several companies such as Telus International (Vancouver, BC, CA) and Edgecase AI LLC (Hingham, MA, US) have started offering medical annotation services performed by networks of medical professionals. However, it is unclear to what extent medical image data annotation can be effectively outsourced to such companies, particularly in the case of surgical data, where important context information may be lost. Furthermore, the associated costs of medical professionals as annotators and annotation reviewers for quality assurance may render these services out of reach for many academic institutes and small companies.

\subsection{Current challenges and next steps}
\label{sec:dataAnnotation_currentChallengesAndNextSteps}


The data annotation-related mission as well as corresponding goals generated by the consortium  
are provided in Tab.~\ref{tab:goalsMission2}. This section elaborates on some of the most fundamental aspects:

\begin{table}
\caption{Mission statement corresponding to data annotation and sharing (Sec.~\ref{sec:dataAnnotation}) along with corresponding goals. The median priority for each goal is symbolized by a vertical priority bar representing the priorities "Not a priority" (one square filled), "Low priority" (two squares filled), "Medium priority" (three squares filled), "High priority" (four squares filled), "Essential priority" (five squares filled).}
\begin{tcolorbox}[title= Mission II: Data annotation and sharing, colback=white, halign=left, subtitle style={boxrule=0.4pt,colback=gray}]
    %
    \textbf{Facilitate data sharing across institutions and establish large-scale, representative and quality-checked annotated databases}
    \tcbsubtitle{Goals}
    \hspace{-0.31cm}
    \begin{tabular}[c]{ m{0.9\textwidth} m{0.1\textwidth} }
        \begin{description}
            \item[Goal 2.1] Establish standardized ontologies for surgical data science
        \end{description} & \includegraphics[width=5pt]{figures/Priority4} \newline \\
        \vspace{-0.2cm}
        \begin{description}
            \item[Goal 2.2] Establish standards for addressing data biases via metadata annotations
        \end{description} &  \includegraphics[width=5pt]{figures/Priority4} \newline \\
        \vspace{-0.2cm}
        \begin{description}
            \item[Goal 2.3] Establish new methods for efficient data annotation
        \end{description} &  \includegraphics[width=5pt]{figures/Priority4} \newline \\
        \vspace{-0.2cm}
        \begin{description}
            \item[Goal 2.4] Establish annotation standards and protocols
        \end{description} &  \includegraphics[width=5pt]{figures/Priority4} \newline \\
        \vspace{-0.2cm}
        \begin{description}
            \item[Goal 2.5] Establish best practices for assessing and assuring annotation quality
        \end{description} & \includegraphics[width=5pt]{figures/Priority4} \newline \\
        \vspace{-0.2cm}
        \begin{description}
            \item[Goal 2.6] Develop and disseminate openly accessible platforms to enable adherence to annotation standards and protocols
        \end{description} & \includegraphics[width=5pt]{figures/Priority4} \newline \\
        \vspace{-0.2cm}
        \begin{description}
            \item[Goal 2.7] Establish platforms and workflows for sharing data in a controlled manner
        \end{description} & \includegraphics[width=5pt]{figures/Priority4} \newline \\
        \vspace{-0.2cm}
        \begin{description}
            \item[Goal 2.8] Create more incentives for SDS stakeholders to share and annotate data
        \end{description} & \includegraphics[width=5pt]{figures/Priority4} \newline \\
        \vspace{-0.2cm}
        \begin{description}
            \item[Goal 2.9] Incentivize adherence to best practices, standards and protocols relevant to data annotation and sharing
        \end{description} & \includegraphics[width=5pt]{figures/Priority4} 
    \end{tabular}
\end{tcolorbox}
\label{tab:goalsMission2}
\end{table}

\textbf{How to develop standardized ontologies for surgical data science?} (goal 2.1)
As current  practices and standards differ greatly between different countries, clinical sites, and healthcare professionals, publicly available surgical data sets generally display vast variation in terms of their annotations. The field, however, is in need of standardized annotations based on a common vocabulary which can be achieved by shared ontologies.  For example, evaluating the efficacy of a particular procedure requires a standardized definition and nomenclature for the different hierarchy levels, e.g. the phases, steps/tasks and activities/actions. A standardized nomenclature along with specifics such as beginning and end of temporal events does not exist yet. Studies can help standardize these definitions and reach a consensus. This is for instance demonstrated by \citet{kaijser_delphi_2018} who conducted a Delphi consensus study to standardize the definitions of crucial steps in the common procedures of gastric bypass and sleeve gastrectomy. Such processes could be adopted for other domains, with the Delphi methods being a particularly useful tool to agree on terminology. Once available and broadly adopted, a shared ontology would stimulate the community as well as boost data and knowledge exchange in the entire domain of SDS. Less formal options such as terminologies are also an alternative but may risk to reach some limits in the long term.\\

\textbf{How to account for biases?} (goal 2.2)
Various sources and types of bias with potential relevance to SDS have been identified in the past \citep{ho_biases_2020}. Among the most critical are \textit{selection bias} and \textit{confounding bias}. Selection bias, also called \textit{sample bias}, refers to a selection of contributing data in a way that does not allow for proper randomization or representativeness to be achieved. Crucially, in the context of SDS, representativeness refers to numerous factors including variances related to patients (e.g. age, gender, origin), the surgical procedure (e.g. adverse events), input data (e.g. device type, protocol; preprocessing methods), and surgeons (e.g. level of expertise). Creating a fully representative data set is thus highly challenging and only possible in a multi-center setting. Unrepresentative data, on the other hand, leads to biased algorithms. A recent study published in the context of radiological data science \citep{larrazabal_gender_2020}, for example, showed that the performance of AI algorithms for a specific sex (e.g. female) crucially depends on the ratio of samples from the respective sex in the training data set. Another source of overestimation regarding algorithm performance is confounding bias. Confounding ``arises when variables that are not mediators of the effect under study, and that can explain part or all of the observed association between the study exposure and the outcome, are not measured and controlled for during study design or analysis''~\citep{arah_bias_2017}. Recent work in biomedical image analysis \citep{badgeley_deep_2019, roberts_common_2021, dietrich_machine_2021} showed that knowledge of confounding variables is crucial to the development of successful predictive models. Conversely, a striking recent example of a confounder rendering results meaningless can be seen in the many papers using a particular pneumonia dataset as a control group in the development of COVID-19 detection and prognostication models. Since this dataset solely consists of young paediatric patients, any model using adult COVID-19 patients and these patients as a control group would likely overperform merely by detecting children \citep{roberts_common_2021}. Other examples of confounders (also called \textit{hidden variables}) are chest drains and skin markings in the context of melanoma \citep{Winkler2019_markedmelanoma} and pneumothorax diagnosis \citep{Oakden-Rayner2020_hidden}. Recognizing and minimizing potential biases in SDS by enhancing data sets with, for example, relevant metadata is thus of eminent importance.  \\

\textbf{How to make data annotation more efficient?} (goal~2.3)
Overcoming the lack of experienced observers might be possible through embedding clinical data annotation in the education and curricula of medical students. In fact, early evidence suggests that annotating surgical skills during video-based training improves the learning experience \citep{delagarza2019_eduannot}. The annotation process could also involve several stages, starting with annotations by non-experts that are reviewed by experts. In a similar fashion, active learning methods reduce the annotation effort to the most uncertain samples \citep{bodenstedt_active_2019, maier-hein_crowd-algorithm_2016}.
 An alternative approach to overcome the lack of annotated data sets is to generate realistic synthetic data based on simulations. A challenge in this context is to bridge the domain gap, so that models trained on synthetic data generalize well to real data. Promising approaches already studied in the context of SDS are for example generative adversarial networks (GANs) for image-to-image translation of laparoscopic images \citep{pfeiffer_generating_2019, rivoir2021long} or transfer learning-based methods for physiological parameter estimation \citep{wirkert_physiological_2017}. In the context of photoacoustic imaging, recent work has further explored the GAN-based generation of plausible tissue geometries from available imaging data~\citep{schellenberg_data-driven_2021}.\\


\textbf{How to establish common standards, protocols and best practices for quality-assured data annotation?} (goals 2.3-2.6/2.9)
Standardized open-source protocols that include well-defined guidelines for data annotation are needed to provide accurate labels. Ideally, the annotations should be generated by multiple observers and the protocol should be defined to reduce inter-observer variability and bias. A recent study in the context of CT image annotation concluded that more than three annotators might be necessary to establish a reference standard \citep{joskowicz_inter-observer_2019}. Comprehensive labeling guides and extensive training are necessary to ensure consistent annotation. \citet{shankar_evaluating_2020}, for example, proposed a 400-page labeling guide in the context of ImageNet annotations to reduce common human failure modes such as fine-grained distinction of classes.  In SDS, a protocol with checklists and examples on how to consistently segment hepatocystic anatomy and assess the CVS in laparoscopic cholecystectomy was recently published to favour reproducibility and trust in the clinical relevance of annotations \citep{MascagniAlapatt2021_AnnotationProtocol}. Such detailed annotation protocols and extensive user training supported by adequate training material are now required.
However, establishing annotation guides for surgical video data is a particularly challenging task since it involves complex actions that require understanding of the surgical intent based on visual cues. In particular, temporal annotations such as phase transitions are often challenging as the start and end of a specific phase is hard to define.  \cite{ward_2021} provide a comprehensive list regarding challenges associated with surgical video annotation. Taking into account the variety of surgical techniques this may lead to annotation inconsistencies even amongst experts, but these could also be used as a hint to estimate the difficulty associated with a surgical situation \citep{ward_2021}. In this context, research on the needs with respect to data and annotation quality in the context of the clinical goals is also required. 
As data sets and annotations evolve over time, another aspect to be taken into account involves versioning of data sets and annotations, similar to code, which is a non-trivial task \citep{marzahl2021exact}. For all tasks related to data annotation, it will be prudent to establish and enforce best practices that can easily be integrated into the surgical workflow. This could be achieved by journal editors explicitly requesting annotation protocols to be submitted along with a paper that is based on annotated data. Journals could also allow for the explicit publication of annotation protocols in analogy to study protocols. 
Finally, platforms that enable spatial as well as temporal annotation in a collaborative manner and  share common annotation standards and protocols as well as ML based methods to facilitate automatic annotations are crucial.One means is to adapt already existing annotation platforms (see Table~\ref{tab:annotationTools}) to fit the specific needs of SDS. Funding agencies should explicitly support efforts to make progress in this regard.
Overall, a particularly promising approach to generating progress with respect to annotation standards is to start from the respective societies, such as SAGES. Alternatively or additionally, international working groups, similar to the one developing the DICOM standard, should be established. Such working groups should collaborate with existing initiatives, such as DICOM or HL7.
In the end, standards will only be successful if enough resources are invested into the actual data annotation. In this case various non-monetary incentives should be considered, including gamification and the issuing of certificates (e.g. for \textit{Certified Professional for Medical Data Annotation} in analogy to \textit{Certified Professional for Medical Software})\\

\textbf{How to incentivize and facilitate data sharing across institutions?} (goals 2.7-2.9)
Data anonymization is a key enabler for sharing medical data and advancing the SDS field. By definition, anonymized data cannot be traced back to the individual and in both the USA and EU, anonymized data are not considered personal data, rendering them out of the scope of privacy regulation such as the GDPR. However, achieving truly anonymized data is usually difficult, especially when multiple data sources from an individual are linked in one data set. Removing identifiable metadata such as sensitive DICOM fields linking the patient to the medical image is necessary but not always sufficient for anonymization. For example, removing DICOM fields in a Magnetic Resonance Imaging (MRI) scan of a patient’s head is not sufficient because the individual may be identified from the image data through facial recognition \citep{schwarz_identification_2019}. Pseudonymization is a weaker form of anonymization where data cannot be attributed to an individual unless it is linked with other data held separately \citep[Article 4, Definitions]{general_data_protection_regulation_gdpr_regulation_2016}. This is often easier to achieve compared to true anonymization, however, pseudonymized data are still defined as personal data, and as such remain within the scope of the GDPR. The public data sets used in SDS research such as endoscopic videos recorded within the patient’s body are generally assumed to be anonymized but clear definitions and regulatory guidance are needed.
Recent advances in federated learning could reduce security and privacy concerns since they rely on sharing machine learning models rather than the data itself~\citep{kaissis2020secure} (see Sec.~\ref{sec:technicalInfrastructure}).
A complementary strategy for bypassing current hurdles related to data sharing is \textit{data donation}. \textit{Medical Data Donors e.V.}, for example, is a registered German non-profit organization, designed to build a large annotated image database which will serve as a basis for medical research. It can be supported by the public via donation of medical imaging data or by shopping at Amazon Smile. In the broader context of data donation, the SDS initiative discussed the concept of a \textit{data donation  card} in analogy to the existing \textit{organ donation card}. With such a card, patients could explicitly state which kind of data they are willing to share with whom and under which circumstances. 
 Overall, making progress on large public databases will require establishing an interlocking set of standards, technical methods, and data analysis tools tied to metrics to support reproducible SDS \citep{nichols_best_2017} and provide value for the community. Clinical registries provide a good example of such a mechanism. In a registry, a specific area of practice agrees on data to be shared, outcome measures to be assessed, and standardized formats as well as quality measures for the data \citep{arts_defining_2002}. Identifying areas of SDS where the value proposition exists to drive the use of registries would provide much-needed impetus to create data archives. So would creating more monetary and non-monetary incentives for institutions, clinical staff and patients to share and annotate data, although particularly the issue of incentivizing patients to share data presents an ethical gray area.\\

\section{Data analytics}
\label{sec:dataAnalytics}

Data analytics (addressing the \textit{interpretation} task in Fig.~\ref{fig:sdsComponents}) is often regarded as the core of any SDS system. The perioperative data is processed to derive information addressing  a specific clinical need, where applications may range from prevention and training to interventional diagnosis, treatment assistance and follow-up \citep{maier-hein_surgical_2017}. 


\subsection{Current practice}

Surgical practice has traditionally been based on observational learning, and decision making before, during and after surgical procedures highly depends on the domain knowledge and past experiences of the surgical team \citep{maier-hein_surgical_2017}. SDS has the potential to initiate a paradigm shift with a data-driven approach \citep{hager_chapter_2020, vercauteren_cai4cai_2020}. Bishop and others classify data analytics tools as descriptive, diagnostic, predictive, and prescriptive  \citep{bishop_pattern_2006, tukey_exploratory_1977}:\\

\textbf{Descriptive analytics tools - what happened?} Descriptive analytics primarily provide a global, comprehensive summary of data made available through data communication such as simple reporting features. 
\citeauthor{syus_syus_nodate}’ Periop Insight (Syus, Inc., Nashville, TN, USA) is an example of how descriptive analytics are used to access data, view key performance metrics, and support operational decisions through documentation and easy interpretation of historical data on supply costs, delays, idle time etc., relating overall operating room efficiency and utilization. Business Intelligence (BI) \citep{chen_business_2012} tools are a typical form of descriptive analysis tools which comprise an integrated set of IT tools to transform data into information and then into knowledge, and have been used in healthcare settings \citep{ward_applications_2014} (e.g. Sisense\texttrademark{} (Sisense Ltd., New York City, NY, USA), Domo\texttrademark{} (Domo, Inc., American Fork, UT, USA), MicroStrategy\texttrademark{} (MicroStrategy Inc., Tysons Corner, VA, USA), Looker\texttrademark{} (Looker Data Sciences Inc., Santa Cruz, CA, USA), Microsoft Power BI\texttrademark{} (Microsoft Corporation, Redmond, WA, USA) and Tableau\texttrademark{} (Tableau Software Inc., Seattle, WA, USA)). These tools often incorporate features such as interactive dashboards \citep{upton_heart_2019} that provide customized graphical displays of key metrics, historical trends, and reference benchmarks and can be used to assist in tasks such as surgical planning, personalized treatment, and postoperative data analysis. \\


\textbf{Diagnostic analytics tools - why did it happen?} Diagnostic analytics tools, on the other hand, explore the data, address the correlations and dependencies between variables, and focus on interpreting the factors that contributed to a certain outcome through data discovery and data mining. These tools can facilitate the understanding of complex processes and reveal relationships between variables, or find root causes. For example, clinicians can use data on postoperative care to assess the effectiveness of a treatment \citep{bowyer_importance_2016, kehlet_evidence-based_2008}. \\


\textbf{Predictive and prescriptive analytics tools - What will happen? How can we make it happen?} 
Predictive analytics uses historical data, performs an in-depth analysis of historical key trends underlying patterns and correlations, and uses the insights gained to make predictions about what will likely happen next (\textit{What will happen?}).
Prescriptive analytics complement predictive analytics by offering insights into what actions can be taken to achieve target outcomes (\textit{How can we make it happen?}).
ML can meet these needs, but the challenges specific to surgery are manifold, as detailed in \cite{maier-hein_surgical_2017}. Importantly, the preoperative, intraoperative and postoperative data processed are potentially highly heterogeneous, consisting of 2D/3D/4D imaging data (e.g. diagnostic imaging data), video data (e.g. from medical devices or room cameras), time series data (e.g. from medical devices or microphones), and more (e.g. laboratory results, patient history, genome information). Furthermore, while the diagnostic process follows a rather regular flow of data acquisition, the surgical process varies significantly and is highly specific to patient and procedure. Finally, team dynamics play a crucial role. In fact, several studies have demonstrated a correlation between nontechnical skills, such as team communication, and technical errors during surgery \citep{hull_impact_2012}.
While first steps have been taken to apply ML in open research problems with applications ranging from decision support (e.g. determining surgical resectability \citep{marcus_improved_2020}) to data fusion for enhanced surgical vision (e.g. \citet{akladios_augmented_2020}), and OR logistics (e.g. \citet{twinanda_rsdnet_2019, bodenstedt_prediction_2019, hager_chapter_2020}), the vast majority of research has not yet made it to clinical trial stages. Sec.~\ref{sec:dataAnalytics_currentChallengesAndNextSteps} highlights several challenges that need to be addressed in order to effectively adopt ML as an integral part of surgical routine.

\subsection{Key initiatives and achievements}

This section reviews some key initiatives and achievements from both an industrial and an academic perspective,

\textbf{Industry initiatives:} Commercial platforms and projects have conventionally focused on analysing multidimensional patient data for clinical decision-making - primarily outside the field of surgery. The most widely discussed initiative so far is probably IBM\textsuperscript{\textregistered} Watson\texttrademark{} Health\textsuperscript{\textregistered} (International Business Machines Corporation (IBM), Armonk, NY, USA), which initiated several projects such as \textit{Watson Medical Sieve}, \textit{Watson For Oncology} or \textit{Watson Clinical Matching} that apply the Watson cognitive computing technology to different challenges in healthcare \citep{chen_ibm_2016}. The goal of \textit{Watson Medical Sieve}, for example, is to filter relevant information from patient records consisting of multimodal data to assist clinical decision making in radiology and cardiology. \textit{Watson Clinical Matching} finds clinical studies that match the conditions of individual patients. With its vast capability to reach patient records and medical literature, Watson was believed to be the future of medicine. However, after it was put to use in the real world, it quickly became clear that the powerful technology has its limitations, as reported by Strickland: It performed poorly in India for breast cancer, where only 73\% of the treatment recommendations were in concordance with the experts. 
Another critical example is the \textit{Watson-powered Oncology Expert Advisor} which had only around 65\% accuracy in extracting time-dependent information like therapy timelines from text documents in medical records \citep{strickland_ibm_2019}. Despite its limitations, Watson Health has shown to be efficient in certain, narrow and controlled applications. For example, \textit{Watson for Genomics} is used by genetics labs that generate reports for practicing oncologists. Given the information on a patient's genetic mutations, it can generate a report that describes all relevant drugs and clinical trials \citep{strickland_ibm_2019}. 
Other companies, societies and initiatives, such as Google (Mountain View, CA, USA) DeepMind Health~\citep{graves_hybrid_2016, tomasev_clinically_2019}, Intel (Santa Clara, CA, USA) \citep{healthcare_it_news_leveraging_2012} and the American Society of Clinical Oncology (ASCO) CancerLinQ\textsuperscript{\textregistered} \citep{sledge_cancerlinq_2013} have also been focusing on clinical data, and industrial success stories in surgery at scale are still lacking, as detailed in Sec.~\ref{sec:clinicalTranslation}.

\textbf{Academic initiatives:} In academia, interdisciplinary collaborative large-scale research projects have developed data analytics tools to address different aspects of SDS. The Transregional Collaborative Research Center “Cognition Guided Surgery” focused on the development of a technical-cognitive assistance system for surgeons that explores new methods for knowledge-based decision support for surgery \citep{marz_toward_2015} as well as intraoperative assistance \citep{katic_bridging_2016}. First steps towards the operating room of the future have recently been taken, focusing on different aspects like advanced imaging and robotics, multidimensional data modelling, acquisition and interpretation, as well as novel human-machine interfaces for a wide range of surgical and interventional applications (e.g. \citeauthor{advanced_multimodality_image-guided_operating_amigo_advanced_nodate}, \citeauthor{computer-integrated_surgical_systems_and_technology_cisst_computer-integrated_nodate} Engineering Research Center, \citeauthor{hamlyn_centre_hamlyn_nodate}, \citeauthor{university_college_london_ucl_ucl_nodate}, \citeauthor{innovation_center_computer_assisted_surgery_iccas_iccas_nodate}, \citeauthor{ihu_strasbourg_ihu_nodate}, \citeauthor{national_center_for_tumor_diseases_dresden_nctucc_national_nodate} and \citeauthor{national_center_for_tumor_diseases_heidelberg_surgical_nodate}).

Broadly speaking, much of the academic work in SDS is currently focusing on the application of ML methods in various contexts \citep{navarrete-welton_current_2020, zhou_artificial_2019, alapatt_neural_2020}, but clinical impact remains to be demonstrated (see Sec.~\ref{sec:clinicalTranslation}).

\subsection{Standards, platforms and tools}
\label{sec:dataAnalaytics_standardsAndTools}

 A broad range of software tools are used by the SDS community each day, reflecting the interdisciplinary nature of the field. Depending on the SDS application, tools may be required from the following technical disciplines that intersect with SDS: classical statistics, general ML, deep learning, data visualization, medical image processing, registration and visualization, computer vision, Natural Language Processing (NLP), signal processing, surgery simulation, surgery navigation and Augmented Reality (AR), robotics, BI and software engineering. Many established and emerging software tools exist within each discipline and a comprehensive list would be vast and continually growing. In Table~\ref{tab:commonTools}, we have listed software tools that are commonly used by SDS practitioners today, organized by the technical disciplines mentioned above. In this section, we focus on ML frameworks and the regulatory aspects of software development for SDS.

\textbf{ML frameworks and model standards:} ML is today one of the central themes of SDS analytics, and many frameworks are used by the SDS community. The \citeauthor{scikit-learn_scikit-learn_nodate} library in Python is the most widely used framework for ML-based classification, regression and clustering using non-DL models such as Support Vector Machines (SVMs), decision trees and multi-layer perceptron (MLPs). DL, the sub-field of ML that uses Artificial Neural Networks (ANNs) with many hidden layers, has exploded over the past 5 years, 
also due to the mature DL frameworks. The dominating open-source frameworks today are \citeauthor{tensorflow_tensorflow_nodate} by Google and \citeauthor{pytorch_pytorch_nodate} by Facebook (Menlo Park, CA, USA). These provide mechanisms to construct, train and test ANNs with comprehensive and ever-growing APIs and they are backed up by large industrial investment and community involvement. Other important, but less widely used frameworks include \citeauthor{caffe_caffe_nodate}, \citeauthor{caffe2_caffe2_nodate} (now a part of PyTorch), \citeauthor{apache_mxnet_apache_nodate}, \citeauthor{flux_fluxmlfluxjl_2020}, \citeauthor{chainer_chainer_nodate}, MATLAB's \citeauthor{deep_learning_toolbox_deep_nodate} and Microsoft's \citeauthor{microsoft_cognitive_toolkit_cntk_microsoft_nodate}. Wrapper libraries have been constructed on top of several frameworks with higher level APIs that simplify DL model design and promote reusable components. These include TensorFlow’s \citeauthor{keras_keras_nodate} (now native to TensorFlow), \citeauthor{tensorlayer_tensorlayer_nodate}, \citeauthor{tflearn_tflearn_nodate} and \citeauthor{niftynet_niftynet_nodate} (specifically for medical image data), and PyTorch’s TorchVision \citep{nguyen_machine_2019}. Other useful tools include training progress visualization with Tensorboard, and AutoML systems for efficient automatic hyperparameter and model architecture search, such as \citeauthor{h2o_automl_nodate}, \citeauthor{auto-sklearn_auto-sklearn_nodate}, \citeauthor{autokeras_autokeras_nodate} and Google \citeauthor{cloud_automl_cloud_nodate}. \citeauthor{nvidia_digits_nvidia_nodate} takes framework abstraction a step further with a web application to train DL models for image classification, segmentation and object detection, and a graphical user interface (GUI) suitable for non-programmers. 
Such tools are relevant in SDS where clinical researchers can increasingly train standard DL models without any programming or ML experience \citep{faes_automated_2019}. On the one hand this is beneficial for technology democratization, but on the other hand it elevates known risks of treating ML and DL systems as \enquote{black boxes} \citep{phg_foundation_black_2020}. 
Recently NVIDIA has released \citeauthor{nvidia_clara_nvidia_nodate}, a software infrastructure to develop DL models specifically for healthcare applications with large-scale collaboration and federated learning.

Each major framework has its own format for representing and storing ML models and associated computation graphs. There are now efforts to standardize formats to improve interoperability, model sharing, and to reduce framework lock-in. Examples include the Neural Network Exchange Format (NNEF), developed by the \citeauthor{khronos_group_khronos_2020} with participation from over 30 industrial partners, \citeauthor{onnx_onnx_nodate} and Apple’s (Cupertino, CA, USA) \citeauthor{core_ml_core_nodate} for sharing models, and for sharing source code to train and test these models. \citeauthor{github_github_nodate} is undeniably the most important sharing platform, used extensively by SDS practitioners, which greatly helps to promote research code reusability and reproducibility. \enquote{Model Zoos} (e.g. \citeauthor{model_zoo_model_nodate}, \citeauthor{onnxmodels_onnxmodels_2020}) are also essential online tools to allow easy discovery and curation of many of the landmark models from research literature.

\textbf{Regulatory software standards:} The usual research and development pipeline for an SDS software involves software developed at various stages including data collection and curation, model training, model testing, application deployment, distribution, monitoring, model improvement, and finally a medically approved product. For the classification as a medical product, the intended purpose by the manufacturer is more decisive than the functions of the software. Software is a \enquote{medical device software} (or \enquote{software as a medical device} (SaMD)) if \enquote{intended to be used, alone or in combination, for a purpose as specified in the definition of a medical device in the medical devices regulation or in vitro diagnostic medical devices regulation} (\citeauthor{mdcg_2019-11_guidance_2019}), i.e. if intended to diagnose, treat or monitor diseases and injuries. The manufacturer of an SDS software application as SaMD needs to ensure that the safety of the product is systematically guaranteed, prove that they have sufficient competencies to ensure the relevant safety and performance of the product according to the state of the art (and keep evidence for development, risk management, data management, verification and validation, post-market surveillance and vigilance, service, installation, decommissioning, customer communication, monitoring applicable new or revised regulatory requirements).

Yet, ML-based software requires particular considerations \citep{gerke_need_2020}. For example, the fact that models can be improved over time with more training data (often called the \enquote{virtuous cycle}) is not well handled by these established standards. In 2019, the FDA published a \enquote{Proposed Regulatory Framework for Modifications to Artificial Intelligence/Machine Learning (AI/ML)-Based Software as a Medical Device (SaMD)}, specifically aimed to clarify this subject \citep{fda_proposed_2019}. In contrast to the previously \enquote{locked} algorithms and models, this framework formulates requirements on using Continuous Learning Systems (CLS) and defines a premarket submission to the FDA when the AI/ML software modification significantly affects device performance, or safety and effectiveness; the modification is to the device's intended use; or the modification introduces a major change to the SaMD algorithm. The implementation of these requirements, especially with regard to the actual product development, is an unsolved problem.

\begin{table}
\caption{Mission statement corresponding to data analytics (Sec.~\ref{sec:dataAnalytics}) along with corresponding goals. The median priority for each goal is symbolized by a vertical priority bar representing the priorities "Not a priority" (one square filled), "Low priority" (two squares filled), "Medium priority" (three squares filled), "High priority" (four squares filled), "Essential priority" (five squares filled).}
\begin{tcolorbox}[title= Mission III: Data analytics, colback=white, halign=left, subtitle style={boxrule=0.4pt,colback=gray}]
    %
    \textbf{Align SDS methods research with clinical objectives and priorities}
    \tcbsubtitle{Goals}
    \hspace{-0.31cm}
    \begin{tabular}[c]{ m{0.9\textwidth} m{0.1\textwidth} }    
        \begin{description}
            \item[Goal 3.1] Develop methods and validation concepts focused on robustness including generalization
        \end{description} & \includegraphics[width=5pt]{figures/Priority4} \newline \\ 
        \vspace{-0.4cm}
        \begin{description}
            \item[Goal 3.2] Develop methods and validation concepts focused on transparency and explainability
        \end{description} & \includegraphics[width=5pt]{figures/Priority4} \newline \\ 
        \vspace{-0.4cm}
        \begin{description}
            \item[Goal 3.3] Develop methods and validation concepts focused on learning efficiently from limited data
        \end{description} & \includegraphics[width=5pt]{figures/Priority4} \newline \\ 
        \vspace{-0.4cm}
        \begin{description}
             \item[Goal 3.4] Develop methods and validation concepts focused on understanding, avoiding and dealing with biases
        \end{description} & \includegraphics[width=5pt]{figures/Priority4} \newline \\ 
        \vspace{-0.4cm}
        \begin{description}
             \item[Goal 3.5] Develop methods and validation concepts focused on the understanding of surgical processes, including variability resulting  from technical-, personnel-, patient- and environment-specific factors
        \end{description} & \includegraphics[width=5pt]{figures/Priority4} \newline \\ 
        \vspace{-0.4cm}
        \begin{description}
             \item[Goal 3.6] Develop methods and validation concepts focused on fast inference
        \end{description} & \includegraphics[width=5pt]{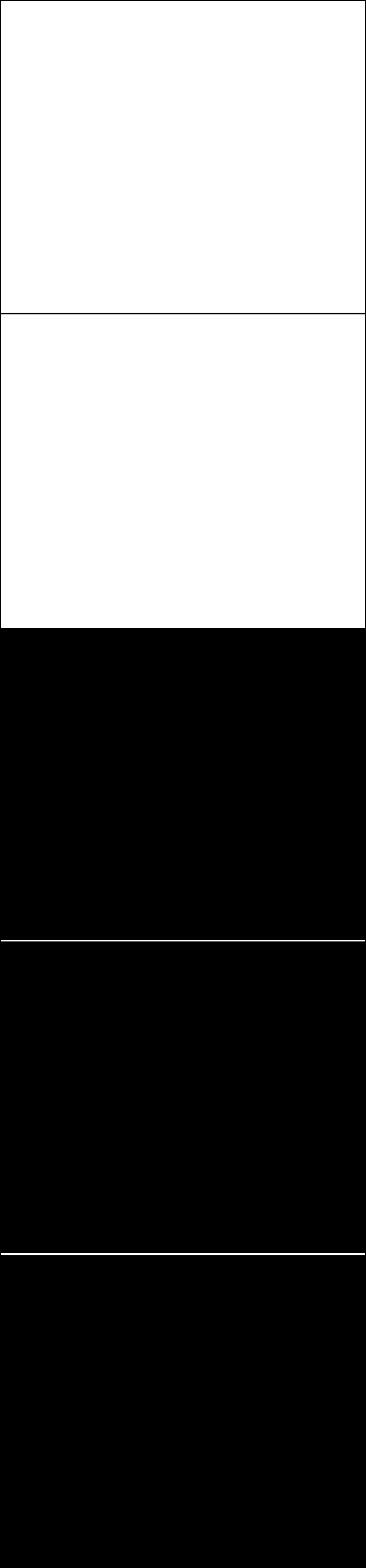} \newline \\ 
        \vspace{-0.4cm}
        \begin{description}
             \item[Goal 3.7] Develop concepts and workflows for facilitating regulatory approval of algorithms
        \end{description} & \includegraphics[width=5pt]{figures/Priority4}
    \end{tabular}
\end{tcolorbox}
\label{tab:goalsMission3}
\end{table}

\subsection{Current challenges and next steps}
\label{sec:dataAnalytics_currentChallengesAndNextSteps}


The data analytics-related mission as well as corresponding goals generated by the consortium 
are provided in Tab.~\ref{tab:goalsMission3}. This section elaborates on the most important research questions from a ML methodological perspective:\\

\textbf{How to ensure robustness and generalization?} (goal 3.1) Models trained on the data from one clinical site may not necessarily generalize well to others due to variability in devices, individual practices of the surgical team or the patient demographic. 
While data augmentation~ \citep{itzkovich_using_2019} can address this issue to some extent, an alternative promising approach is to develop
architectures designed to generalize across domains. 
Early approaches focused on \textit{domain adaptation} \citep{heimann_learning_2013, wirkert_physiological_2017} or more generically \textit{transfer learning} \citep{pan_survey_2010} to compensate for domain shifts in the data. Other attempts have focused on converting data into a domain-invariant representation and on decoupling generic task-relevant features from domain-specific ones \citep{dai_deformable_2017, mitchell_artificial_2019, sabour_dynamic_2017, sarikaya_towards_2020}. Generally speaking, however, ML methods trained in a specific setting (e.g. hospital) still tend to fail to generalize to new settings.\\

\textbf{How to improve transparency and explainability?}\\ (goal 3.2) 
The WHO document on \textit{Ethics \& Governance of Artificial Intelligence for Health} \citep{health_ethics__governance_ethics_2021} (see Sec.~\ref{sec:technicalInfrastructure}) states that ``AI technologies should be intelligible [...] to developers, medical professionals, patients, users and regulators'' and that ``two broad approaches to intelligibility are to improve the transparency of AI technology and to make AI technology explainable.'' In this context, transparency also relates to the requirement that ``sufficient information be published or documented before the design or deployment of an AI technology and that such information facilitate meaningful public consultation and debate on how the technology is designed and how it should or should not be used''. Explainability stems from the urge to understand why an algorithm produced a certain output.
 In fact, the complexity of neural network architectures with typically millions of parameters poses a difficulty for humans to understand how these models reach their conclusions \citep{reyes_interpretability_2020}. As a result, the EU’s GDPR, implemented in 2018, also discourages the use of black-box approaches, thus providing explicit motivation for the development of models that provide human-interpretable information on how conclusions were reached.
Interpretable models are still in their infancies and are primarily studied by the ML community~\citep{adebayo_sanity_2018, bach_pixel-wise_2015, koh_understanding_2017, shrikumar_learning_2017}.
These advances are being adopted within medical imaging communities in applications that are used to make a diagnosis (e.g. detecting/segmenting cancerous tissue, lesions on MRI data) \citep{gallego-ortiz_interpreting_2016}, and to generate reports that are on par with human radiologists \citep{gale_producing_2018}, for example. Open research questions are related to how to validate the explanation of the models (lack of ground truth) and how to best communicate the results to non-experts. 
A concept related to explainability is causality.
To date, it is generally unknown how a given intervention or change is likely to affect outcome, which is influenced by many factors even beyond the surgeon and the patient. Furthermore, randomized controlled trials (RCTs) to evaluate surgical interventions are difficult to perform \citep{mcculloch_randomised_2002}. Thus, it is hard to provide the same quality of evidence and understanding of surgery as, for example, for a drug treating a common non-life-threatening condition \citep{hager_chapter_2020}. While large-scale data may help reveal relationships among many factors in surgery, correlation does not equal causation. Recent work on causal analysis \citep{peters_elements_2017,scholkopf_causality_2019, castro_causality_2020}, however, may help in this regard.\\

\textbf{How to address data sparsity?} (goal 3.3) One of the most crucial problems in SDS is the data sparsity (see Sec.~\ref{sec:lackOfSuccessStories}). Several complementary approaches have been proposed to address this bottleneck. 
These include \textit{crowdsourcing}~\citep{maier-hein_can_2014, maier-hein_crowdtruth_2015, malpani_study_2015,heim_clickstream_2018,albarqouni_aggnet_2016, maier-hein_crowd-algorithm_2016} and \textit{synthetic data generation} \citep{pfeiffer_generating_2019, ravasio_learned_2020, wirkert_physiological_2017, rivoir2021long} briefly mentioned above. Unlabeled data can also be exploited by using \textit{self-supervised} (see e.g. \citep{ross_exploiting_2018}) and \textit{semi-supervised} learning (see e.g. \citep{yu_learning_2019, srivastav_self-supervision_2020}). Self-supervised methods solve an alternate, pretext or auxiliary task, the result of which is a model or representation that can be used in the solution of the original problem. Semi-supervised methods can exploit the unlabelled data in many different ways. In \citep{yu_learning_2019, srivastav_self-supervision_2020}, for example, pseudo-annotations are generated on the unlabelled data using a teacher model, and the resulting pseudo-annotated dataset is then used to train another (student) model. 
Recent studies have further shown that exploiting the relationship across different tasks with the concept of  \textit{multi-task learning} \citep{twinanda_endonet_2017} may be used to address data sparsity as well. It has been demonstrated to be beneficial to jointly reason across multi-tasks \citep{kokkinos_ubernet_2017, long_learning_2017, yao_describing_2012, sarikaya_joint_2018} and take advantage of a combination of shared and task-specific representations \citep{misra_cross-stitch_2016}. However, the performance of some tasks may also worsen through such a paradigm \citep{kokkinos_ubernet_2017}. A possible solution to this problem might lie in the approach of attentive single-tasking \citep{maninis_attentive_2019}. Finally, \textit{meta-learning} \citep{vanschoren_meta-learning_2018, godau2021} and more generally \textit{lifelong learning} \citep{parisi_continual_2019} are further potential paradigms for addressing the problem of data sparsity in the future. Progress in this field will, at any rate, crucially depend on the availability of more public multi-task data sets, such as~\cite{maier-hein_heidelberg_2021}\\

\textbf{How to detect, represent and compensate for uncertainties and biases?} (goal 3.4) A common criticism of ML-based solutions is the way that they handle \enquote{anomalies}. If a measurement is out-of-distribution (ood; i.e. it does not resemble the training data), the algorithm cannot make a meaningful inference, and the probability of failure (error) is high. This type of \textit{epistemic uncertainty} \citep{kendall_what_2017} is particularly crucial in medicine as not all anomalies/pathologies can be known beforehand. As a result, current work is dedicated to this challenge of anomaly/novelty/ood detection \citep{adler_uncertainty-aware_2019}. Even if a sample is in the support of the training distribution, a problem may not be uniquely solvable \citep{ardizzone_analyzing_2018} or the solution may be associated with high uncertainty. Further research has therefore been directed at estimating and representing the certainty of AI algorithms \citep{adler_uncertainty-aware_2019, nolke_invertible_2021}. Future work should focus on making use of the uncertainty estimates in clinical applications and increasing the reliability of ood methods, as well as systematically understanding and addressing the issue of biases and confounders (see Sec.~\ref{sec:dataAnnotation_currentChallengesAndNextSteps}). In this context the increased involvement of statisticians and experts from clinical epidemiology, such as in the biomedical image analysis initiative~\citep{maier-hein_bias_2020,ros_how_2021}, would be desirable. Adopting the necessity of reporting data biases and confounders in publications should be a natural progression for the field of SDS.
\\

\textbf{How to address data heterogeneity and complexity?} (goal 3.5) The surgeons and surgical team dynamics play a significant role in intraoperative care. While the main surgeon has the lead and makes decisions based on domain knowledge, experience and skills, anesthesiologists, assistant surgeons, nurses and further staff play crucial roles at different steps of the workflow. Their smooth, dynamic collaboration and coordination is a crucial factor for the success of the overall process. Data analytics can play a key role in quantifying these intangibles by modeling workflows and processes. Surgeon skill evaluation, personalized and timely feedback during surgical training, optimal surgeon and patient/case or surgeon and surgical team matches are among the issues that can benefit from data analytics tools.
 Furthermore, data collected from multiple sources such as vital signs from live monitoring devices, electronic health records, patient demographics, or preoperative imaging modalities require analysis approaches that can accommodate their heterogeneity. Recent approaches in fusion of heterogeneous information include the use of specialized frameworks such as iFusion \citep{guo_ifusion_2019}. Other work has specifically focused on handling incomplete heterogeneous data with Variational Autoencoders (VAEs) \citep{nazabal_handling_2020}. Graph neural networks \citep{zhou_graph_2019} appear to be another particularly promising research direction in this regard. Here as well, however, the lack of large amounts of annotated data is a limiting factor. \citep{raghu_transfusion_2019}. Heterogeneity may also occur in labels~\citep{joskowicz_inter-observer_2019}. This could potentially be addressed with fuzzy output/references as well as with probabilistic methods capable of representing multiple plausible solutions in the output, as suggested by some early work on the topic~\citep{kohl_probabilistic_2018, adler_uncertainty-aware_2019, trofimova_representing_2020}.\\

\textbf{How to enable real-time assistance?} (goal 3.6) Fast inference in an interventional setting relies on (1) an adequate hardware and communication infrastructure (covered in Sec.~\ref{sec:technicalInfrastructure}) and on (2) fast algorithms. The trade-off between algorithm and software optimization should be finely balanced between the available edge compute power and the latency requirements of the specific application. Moving high resolution video between devices or displays inherently adds delays and should be minimized for dynamic assistance applications or whether data inference links to control systems. This means that edge compute solutions should carefully consider the input to the display pipeline and the size of the inference models that can be loaded into an edge processor. Where latency is less critical, cloud execution of AI models has already been shown to be viable in assistive systems (e.g. Cydar EV from Cydar Medical (Cambridge, UK) for endovascular navigation, or CADDIE / CADDU from Odin Vision Ltd (London, UK) for AI assisted endoscopy). Cloud computing for real-time assistance relies on good connectivity to move data but offers the possibility of running potentially large inference models and returning results for assistance to the OR. Recent advances in the emerging research field of Tactile Internet with Human-in-the-Loop (TaHiL) \citep{fitzek2021tactile}, which involves intelligent telecommunication networks and secure computing infrastructure is an enabling technology for real-time remote SDS application. To trigger progress in the field, specific clinical applications requiring real-time support should be identified and focused on. Dedicated benchmarking competitions in the context of these applications could further guide methodological development.\\

\textbf{How to train and apply algorithms under regulatory constraints?} (goal 3.7) When an SDS data set contains personal medical data, an open challenge lies in how to perform data analytics and train ML models without sensitive information being exposed in the results or models. A general solution that is gaining increasing traction in ML is differential privacy \citep{dwork_calibrating_2006}. This offers a strong protection mechanism against linkage, de-anonymization and data reconstruction attacks, with rigorous privacy guarantees from cryptography theory. 
A limitation of differential privacy can be seen in the resulting compromise in terms of model accuracy, which may conflict with accuracy targets. 
Differential privacy may ultimately be mandatory for federated learning \citep{li_privacy-preserving_2019} and publicly releasing SDS models built from personal medical data. Since patients have the right to delete their data, privacy questions also arise regarding models that were trained on their data. In addition, it might be an attractive business model for companies to sell their annotated data or make them publicly available for research purposes. This requires methods to detect whether specific data has been used to train models, e.g. using concepts of \enquote{radioactive data} \citep{sablayrolles_radioactive_2020}, or methods that detect whether a model has forgotten specific data \citep{liu_have_2020}. A complementary approach to preserving privacy is to work with a different representation of the data. For example, \citep{twinanda_data-driven_2015, SharghiHOM20} evaluate the use of depth images rather than RGB images to recognize human activity in the hospital, while \cite{haque2018activity,srivastav_human_2019} performs the analysis on low-resolution images. \\\\

\textbf{How to ensure meaningful validation and evaluation?} (all goals) Validation - defined as the demonstration that a system does what it has been designed to do - as well as evaluation - defined as the demonstration of the short-, mid- and long-term added values of the system - are crucial for the development of SDS solutions. The problem with the assessment of ML methods today is that models trained on a particular data set are evaluated on new data 
taken from the same distribution as the training data. Although recent efforts have been made in healthcare \citep{mckinney_international_2020} to include test data from different clinical sites, these still remain limited. This situation poses a challenge particularly for healthcare applications, as real-world test data, after the model is deployed for clinical use, will typically  not have ground-truth annotation, making its assessment difficult \citep{castro_causality_2020}. A recent example of this is Google Health’s deep learning system that predicts whether a person might be at risk for diabetic retinopathy. In this case, after its deployment at clinics in rural Thailand, despite having high theoretical accuracy, the tool was reported to be impractical in real-world testing \citep{techcrunch_google_2020}. In the future, evaluation of methods should be performed increasingly in multi-center settings and incorporate the important aspects of robustness to domain shifts, data imbalance and bias. Global initiatives such as \citeauthor{mlcommons_2018} and its Medical Working Group will play a central role in designing benchmarks and propose best practices in this regard. Furthermore, matching performance metrics to the clinical goals should be more carefully considered, as illustrated in recent work~\citep{reinke_common_2021}. Finally, specific technical aspects (e.g. explainability, generalization) should be comparatively benchmarked with international challenges and covered at dedicated workshops. In this context, acquiring dedicated sponsor money for annotations could help generate more high-quality public data sets.

\section{Clinical translation}
\label{sec:clinicalTranslation}

The process of clinical translation from bench to bedside has been described as a valley of death, not only for surgical (software) products, but biomedical research in general \citep{butler_translational_2008}. 
In this section, we will begin by describing current practice and key initiatives in clinical translation of SDS. We elaborate on the concept of \enquote{low-hanging fruit} that may be reached in a comparatively straightforward manner through collaboration of surgeon scientists, computer scientists and industry leaders. Finally, we will outline current challenges and next steps for those low-hanging fruit to cross the valley of death, rendering SDS applications from optional translational research projects to key elements of the product portfolio for modern operating room vendors, which in turn will increase engagement on the part of researchers, industry, funding agencies and regulatory bodies alike.

\subsection{Current practice}

Clinical translation of products developed through SDS is regulated under existing rules and guidelines. Ultimately, systems or products using SDS components must be able to provide value before, during or after surgery or interventions. Validating such capabilities requires prospective clinical trials in real treatment practices, which require ethics and safety approval by relevant bodies as well as adherence to software standards described in Sec.~\ref{sec:dataAnalytics_currentChallengesAndNextSteps}. System documentation and reliability is critical to pass through such approval procedures, which can however also exceptionally be obtained for research purposes without proof of code stability.

From a clinical research perspective, meta-analyses of RCTs are considered the gold standard. However, the field of surgery exhibits a notable lack of high-quality clinical studies as compared to other medical disciplines \citep{mcculloch_randomised_2002}. While long-term clinical studies are a common prerequisite for clinical translation, despite intense research, the number of existing clinical studies in AI-based medicine is extremely low \citep{nagendran_artificial_2020}. As a result, most current clinical studies in the field are based on selected data that are retrospectively analyzed, leading to a lack of high quality evidence that in turn hampers clinical progress. A recent scoping review on AI-based intraoperative decision support in particular named the small size, single-center provenance and questionable representability of the data sets, the lack of accounting for variability among human comparators, the lack of quantitative error analysis, and a failure to segregate training and test data sets as the prevalent methodological shortcomings \citep{navarrete-welton_current_2020}. 

Despite these shortcomings, it should be noted that not all questions that arise in the process of clinical translation of an algorithm necessarily need to be addressed by RCTs. For example, a recent deep learning algorithm to diagnose diabetic retinopathy was approved by the FDA based on a pivotal cross-sectional study \citep{abramoff_pivotal_2018}. Translational research on SDS products for prognosis also leverages existing methodology on prospective and retrospective cohort studies for the purposes of internal and external validation.

Generally speaking, the field of SDS still faces several domain-specific impediments. For instance, digitalization has not percolated the OR and the surgical community in the same way as other areas of medicine. A lack of standardization of surgical procedures hampers the creation of standardized annotation protocols, an important prerequisite for large-scale multi-center studies. Pioneering clinical success stories are important motivators to help set in motion a virtuous circle of advancement in the OR and beyond.

\subsection{Key initiatives and achievements}

The following section will provide an overview of existing SDS products and clinical studies in SDS.

\textbf{SDS products:} Over the past few years, modest success in clinical translation and approval of SDS products has been achieved, as summarized in Table~\ref{tab:products}. This predominantly includes decision support in endoscopic imaging. Endoscopic AI (AI Medical Service, Tokyo, Japan) and GI Genius\texttrademark{} (Medtronic, Dublin, Ireland) support gastroenterologists in the detection of cancerous lesions, the former albeit struggling with a low positive predictive value \citep{hirasawa_application_2018}. Other successful applications include OR safety algorithms or computer vision-based data extraction.

\begin{table*}[width=2.025\linewidth]
\caption{\label{tab:products}Selection of SDS products with machine learning (ML)-based components as of October 2020.}
\centering
\begin{tabular}{p{4cm} p{2cm} p{2.5cm} p{4.2cm} p{2.8cm}}
\toprule
Manufacturer &
Product &
Purpose &
SDS functionality &
Approval \\
\midrule
\multicolumn{5}{l}{Decision Support} \\
\midrule
\citeauthor{ai_medical_service_inc_tokyo_japan_aim_nodate} &
  Endoscopic AI &
  Early detection of gastrointestinal cancers &
  Data-driven detection of cancer lesions in endoscopic videos &
  FDA: Breakthrough Device Designation \newline Europe: none \\
\citeauthor{medtronic_plc_dublin_ireland_intelligentes_nodate} &
  GI Genius\texttrademark &
  Early detection of colorectal cancer  &
  Data-driven anomaly detection in colonoscopy videos &
  FDA: none \newline Europe: none \\
\citeauthor{gauss_surgical_inc_menlo_park_caus_gauss_nodate} &
  Triton\texttrademark &
  Improvement of safety in the operating room  &
  Data-driven obstetric hemorrhage quantification through scans of sponges and canisters and sponge counting through scans of surgical field or counter bags &
  FDA: De Novo and 510(k) \newline Europe: CE mark \\
\midrule
\multicolumn{5}{l}{Surgical Education} \\
\midrule
\citeauthor{theator_inc_san_mateo_ca_us_theator_nodate} &
  Surgical Intelligence Platform &
  Surgical training &
  Computer vision-based key moment extraction and annotation on surgical videos and video-based training &
  FDA: none \newline Europe: none \\
\bottomrule
\end{tabular}
\end{table*}

\textbf{Translational progress in academia:}  While most of the work has focused on preoperative decision support, here, we place a particular focus on intraoperative assistance. Table~\ref{tab:clinicalStudies} shows several exemplary studies in academia that illustrate how far SDS products have been translated to clinical practice in this regard. 

\textit{Intraoperative assistance:} A recent review on AI for surgery mainly found studies that use ML to improve intraoperative imaging such as hyperspectral imaging or optical coherence tomography \citep{navarrete-welton_current_2020}. Further notable intraoperative decision support efforts have focused on hypoxemia prevention \citep{lundberg_explainable_2018}, sensor monitoring to support anesthesiologists with proper blood pressure management \citep{wijnberge_effect_2020} and intelligent spinal cord monitoring during spinal surgery \citep{fan_intelligent_2016}. A number of models have been developed to promote safety in laparoscopic cholecystectomy, a very common and standardized minimally invasive abdominal procedure. For instance, a model for bounding box detection of hepatocystic anatomy was recently tested in the operating room \citep{Tokuyasu2021_LCanatomydet}. Another example of SDS for safe cholecystectomy is DeepCVS, a neural network trained to semantically segment hepatocystic anatomy and assess the criteria defining the CVS \citep{MascagniVardazaryanAlapatt2021_DeepCVS}. A recent study based on 290 laparoscopic cholecystectomy videos from 37 countries showed that DL-based image analysis may be able to identify safe and dangerous zones of dissection~\citep{madani_artificial_2021}. Finally, a cross-sectional study using deep learning algorithms developed on videos of the surgical field from more than 1000 cholecystectomy procedures from two institutions showed an association between disease severity and surgeons’ ability to verify the CVS \citep{korndorffer_situating_2020}. 
Another example of intraoperative decision support is a study by \citet{harangi_recognizing_2017}, who developed a neural network-based method to classify a structure specified by a surgeon (by drawing a line in the image) into the uterine artery or ureter. The authors reported a high accuracy, but the study was a cross-sectional design with a convenience sample. In fact, convenience samples are the norm in most existing studies in SDS addressing recognition of objects or anatomical structures in the surgical field. This sampling mechanism makes the findings susceptible to selection bias, which affects generalizability or external validation of the methods.

\textit{Perioperative decision support and prediction:} A selection of studies on perioperative assistance can be found in Appendix~\ref{app:clinicalStudies-peri-op}. One important application of academic SDS is clinical decision support systems (CDSS) that integrate various information sources and compute a recommendation for surgeons about the optimal treatment option for a certain patient. Many of these CDSS are prediction systems that integrate into a mathematical model clinical, radiological and pathological attributes collected in a routine setting and weigh these parameters automatically to achieve a novel risk stratification \citep{shur_clinical-radiomic_2020}. Trained with a specifically selected subpopulation of patients, these prediction systems may help improve current classification systems in guiding surgical decisions \citep{tsilimigras_utilizing_2020}. Relevant information like overall- and recurrence-free survival \citep{schoenberg_novel_2020} or the likelihood of intra- and postoperative adverse events to occur \citep{bhandari_predicting_2020} can be assessed and obtained quickly via online applications such as the \textit{pancreascalculator.com} \citep{van_roessel_international_2020}. In contrast to these score-based prediction systems, ML-based systems are more flexible. The most prominent ML-based system, IBM's Watson for Oncology, is based on natural language processing and iterative features and demonstrated good accordance with treatments selected by a multidisciplinary tumor board in hospitals in India \citep{somashekhar_watson_2018} and South Korea \citep{lee_assessing_2018}. Weaknesses of this system include the necessity of skilled oncologists to operate the program, low generalizability to different regions, and the fact that not all subtypes of a specific cancer can be processed \citep{yao_real_2020, strickland_ibm_2019}.

Another important application besides decision support is prediction of adverse events. A widely discussed work showed that DL may predict kidney failure up to 48 hours in advance \citep{tomasev_clinically_2019}. In the intensive care unit (ICU), where surgeons face enormous quantities of clinical measurements from multiple sources, such as monitoring systems, laboratory values, diagnostic imaging and microbiology results, data-driven algorithms have demonstrated the ability to predict circulatory failure \citep{hyland_early_2020}. 

Table~\ref{tab:registeredClinicalStudies} provides an overview of currently registered SDS clinical studies. While most aim for evaluation of specific applications, a number of ongoing clinical trials focus on data collection for the original development of future CDSS or other SDS applications.

\begin{table*}[width=2.025\linewidth]
\caption{\label{tab:clinicalStudies}Selection of SDS clinical studies. Searches were performed in June 2021  using [machine learning] AND [surgery] or [deep learning] AND [surgery] or [artificial intelligence] AND [surgery] or [decision support] AND [surgery] or [surgical data science] AND [clinical] in PubMed and Google. Search results were manually evaluated and all studies that analyzed an intraoperative SDS system with a machine learning (ML)-based component were included.}
\centering
\begin{tabular}{p{3.3cm} p{8cm} p{2.2cm} p{1.9cm}}
\toprule
Publication &
Subject &
Type of study &
Study size \newline (\# patients)\\
\citet{fan_intelligent_2016} &
ML-based intraoperative somatosensory evoked potential monitoring based on somatosensory evoked potential measurements & 
Cross-sectional & 10\\
\citet{harangi_recognizing_2017} &
ML-based classification of uterine artery and the ureter based on video images from gynecologic surgery & 
Cross-sectional & 35\\
\citet{korndorffer_situating_2020} &
ML-based detection of intraoperative events of interest and case severity based on laparoscopic cholecystectomy videos &
Cross-sectional & n/a \newline (1,051 videos)\\
\citet{lundberg_explainable_2018} &
Explainable ML-based predictions for the prevention of hypoxemia during surgery based on minute-by-minute data from electronic health records &
Prospective \newline cohort & n/a \newline (53,126 procedures)\\
\citet{madani_artificial_2021} &
ML-based segmentation of safe and dangerous zones of dissection based on laparoscopic cholecystectomy videos &
Cross-sectional & n/a \newline (290 videos)\\
\citet{MascagniVardazaryanAlapatt2021_DeepCVS} &
ML-based segmentation of anatomy and assessment of CVS criteria based on laparoscopic cholecystectomy videos &
Cross-sectional & n/a \newline (201 videos)\\
\citet{Tokuyasu2021_LCanatomydet} &
ML-based bounding box detection of hepatocystic anatomy on laparoscopic cholecystectomy videos &
Cross-sectional & 1 \newline (99 videos)\\
\citet{wijnberge_effect_2020} &
ML-based early warning system for intraoperative hypotension based on continuous invasive blood pressure monitoring & 
Randomized \newline controlled trial & 68\\
\bottomrule
\end{tabular}
\end{table*}

\subsection{Low-hanging fruit}

In light of the lack of a critical number of clinical success stories, a viable approach to clinical translation initially should focus on \enquote{low-hanging fruit}. We believe the following criteria influence the likelihood of successful translation of an SDS application: high patient safety, technical feasibility - especially regarding data needs and performance requirements -  easy workflow integration, high clinical value and high business value to encourage industry adoption.
Low-hanging fruit typically also avoid being classified as a high-risk medical product, thereby reducing regulatory demands and development barriers. However, it is difficult to satisfy all of these often conflicting criteria simultaneously. For example, applications of significant clinical value such as real-time decision support are highly technically challenging. By contrast, low-level video processing applications such as uninformative frame detection are technically simple but of limited clinical value. SDS applications that are low-hanging fruit are ones that offer a good balance between most or all of these criteria.

An example for a low-risk medical device in the broader scope of SDS is the aforementioned GI Genius that uses AI for real-time detection and localization of polyp detection during  colonoscopy, supporting the examination but not replacing the clinical decision making and diagnostics by clinicians. Considering the low risk to patients, GI Genius is classified as a Class II medical device (with special controls) by the FDA~\citep{fda_news_release_2021}.

In surgery, a framework that may help determine the next steps for low-hanging fruit is the digital technology framework that categorizes data-centric product innovations in \textit{descriptive}, \textit{diagnostic}, \textit{predictive} and \textit{prescriptive}, as detailed in section~\ref{sec:dataAnalytics}. Currently, the overwhelming focus for SDS researchers is in the prescriptive technology area – for example on tools that provide surgical decision support or predict adverse events. Changing the development lens from prescriptive to descriptive SDS applications, however, may open up entirely new avenues. For instance, a low-hanging fruit may lie in a descriptive decision support tool that informs surgeons on how many surgeons performed certain steps within an intervention and the consequences. Such a data-centric SDS product would not require embedded surgical expertise in order to provide value to the surgeon, but only a database of surgical videos and automated recognition of anatomical structures and surgical instruments, which is technically feasible. In essence, instead of the very difficult automation of surgical decisions, value can be found in providing surgeons and surgical teams with moment-to-moment risk stratification data to facilitate their decisions. An additional benefit of this approach is that it can be combined with real-time data acquisition regarding \textit{how} surgeons interact with the risk stratification data, which would greatly facilitate the development of both predictive and prescriptive decision support tools.

Importantly, presenting statistical data and evidence-based risk stratification information to the surgeon would also have a different regulatory path than a prescriptive SDS product that offers surgical decisions based on an AI database grounded in surgical decision making.  The data-focused product leaves the surgeon fully responsible, while the decision based product makes it questionable who is fully responsible if the surgeon followed an AI-based decision and there was a poor outcome. Another benefit of focusing on descriptive technologies is there is a much smaller technology adoption hurdle for the surgeon when faced with trusting descriptive statistics compared to an AI-based prescriptive decision support tool.

An ML-based descriptive low-hanging fruit could be data-driven surgical reporting and documentation. Surgical procedures are currently documented as one to two pages of text. While a six to eight hour video will not serve as a report in itself, SDS may help extract relevant information from this video by automatically documenting important steps in the procedure.
Here, computer vision algorithms for recognition of surgical phases and instruments may be used to extract metainformation from videos~\citep{MasacagniAlapatt2021_EndoDigest}. 

An ML-based predictive low-hanging fruit could lie in the optimization of OR logistics. Prediction of procedure time either preoperatively or utilizing intraoperative sensor data may not improve patient outcome, but could provide value to hospital managers if it helps cut down costs in the OR by optimizing patient volume \citep{aksamentov_deep_2017, bodenstedt_prediction_2019, twinanda_rsdnet_2019}.
This, too, harbors low risk for patients and has a low barrier for market entry. Furthermore, the reference information, i.e., time between incision and suture, is already documented in most hospitals and no laborious annotation by surgical experts is necessary to train the respective ML algorithms. Since OR management tools already exist, SDS applications could even yield success stories within existing tools without having to establish entirely new software tools.
 Improvements in patient safety may already result from a simple tool that combines SDS algorithms for object recognition in laparoscopic video (e.g. gauze, specimen bag or suture needle) with a warning for surgeons and scrub nurses if these objects are introduced into the patient's abdomen but not removed afterwards. Since such an SDS application warns clinical staff but does not perform an action on the patient itself, the risk for the patient is inherently low. Here, a combination of surgical knowledge (which objects are at what time introduced into the patient’s body?) with SDS algorithms (which objects can robustly be detected?) and an unobtrusive user interface with a low false alarm rate may result in a low-hanging fruit. Along these lines, automation of the surgical checklist \citep{conley_effective_2011} would be a technically feasible SDS application with high clinical value.

The impending success of next-generation surgical ro\-bot\-ics in the OR may bring further opportunities to the clinical translation of SDS. The da Vinci\textsuperscript{\textregistered} surgical system (Intuitive Surgical Inc., Sunnyvale, CA, USA)  and its upcoming competitors lay the foundation for systematic data capture as well as surgical guidance by information augmentation in the OR. A relatively low-hanging fruit with benefit to the surgeon in the domain of surgical robotics may be an automated camera guidance system, as suggested by Wagner et al.~\citep{wagner_learning_2021}. On the one hand, the risk of poor camera positioning for the patient is low compared to that of invasive tasks such as suturing. On the other hand, correcting the camera position is currently a highly disruptive task to the surgeon. The first products for autonomous endoscopic camera control are now emerging in robotic surgery, such as the FDA-approved system from \citeauthor{transenterix_transenterix_nodate} (Morrisville, NC, USA). %

\subsection{Current challenges and next steps}

As highlighted in several previous publications~\citep{maier-hein_surgical_2017, maier-hein_surgical_2018, hager_chapter_2020}, clinical applications for SDS are manifold, ranging from pre- and intraoperative decision support to context-aware assistance and surgical skills training. The clinical translation-related goals generated by the consortium as part of the Delphi process
are provided in Tab.~\ref{tab:goalsMission4}. 
\begin{table}
\caption{Mission statement corresponding to clinical translation (Sec.~\ref{sec:clinicalTranslation}) along with corresponding goals. The median priority for each goal is symbolized by a vertical priority bar representing the priorities "Not a priority" (one square filled), "Low priority" (two squares filled), "Medium priority" (three squares filled), "High priority" (four squares filled), "Essential priority" (five squares filled).}
\begin{tcolorbox}[title= Mission IV: Clinical translation, colback=white, halign=left, subtitle style={boxrule=0.4pt,colback=gray}]
    %
    \textbf{Promote a cultural shift toward SDS-driven clinical practice.}
    \tcbsubtitle{Goals}
    \hspace{-0.31cm}
    \begin{tabular}[c]{ m{0.9\textwidth} m{0.1\textwidth} }
        \begin{description}
            \item[Goal 4.1] Catalyze clinical translation by generating visible and impactful SDS success stories
        \end{description} & \includegraphics[width=5pt]{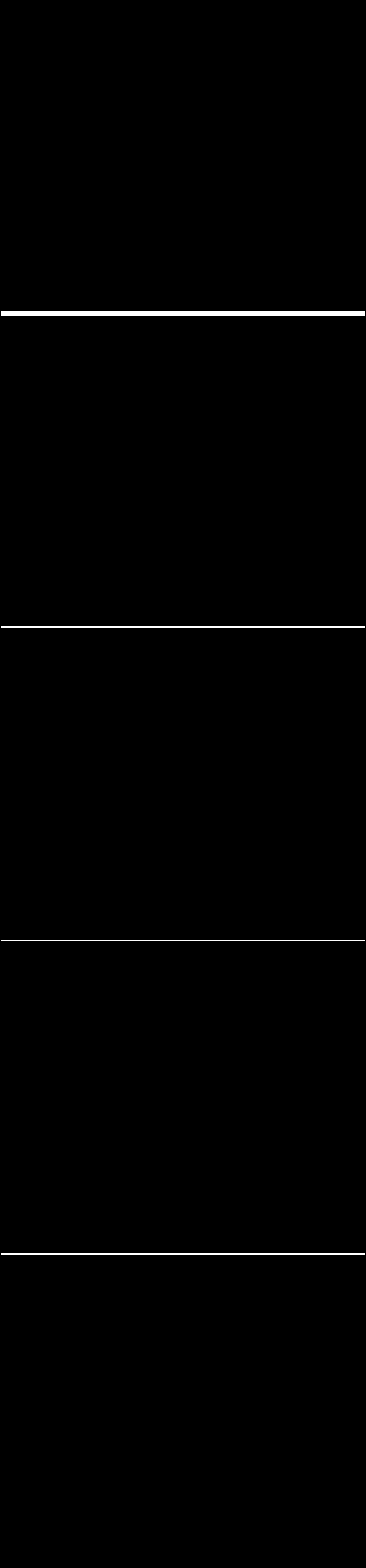} \newline \\
        \vspace{-0.2cm}
        \begin{description}
            \item[Goal 4.2] Build public trust in SDS
        \end{description} & \includegraphics[width=5pt]{figures/Priority4} \newline \\
        \vspace{-0.2cm}
        \begin{description}
            \item[Goal 4.3] Improve transfer of knowledge among different SDS stakeholders
        \end{description} & \includegraphics[width=5pt]{figures/Priority4} \newline \\
        \vspace{-0.2cm}
        \begin{description}
            \item[Goal 4.4] Establish SDS as a career path in healthcare institutions
        \end{description} & \includegraphics[width=5pt]{figures/Priority3} \newline \\
        \vspace{-0.2cm}
        \begin{description}
            \item[Goal 4.5] Ensure high-quality external validation of SDS applications
        \end{description} & \includegraphics[width=5pt]{figures/Priority4} \newline \\
        \vspace{-0.2cm}
        \begin{description}
            \item[Goal 4.6] Develop new performance metrics for AI algorithms that measure clinically relevant parameters not accounted for by existing metrics
        \end{description} & \includegraphics[width=5pt]{figures/Priority4} \newline \\
        \vspace{-0.2cm}
        \begin{description}
            \item[Goal 4.7] Allocate economic, infrastructural and personnel resources within healthcare institutions for SDS
        \end{description} & \includegraphics[width=5pt]{figures/Priority4} \newline \\
        \vspace{-0.2cm}
        \begin{description}
            \item[Goal 4.8] Develop widely accepted frameworks for obtaining patient consent to data sharing
        \end{description} & \includegraphics[width=5pt]{figures/Priority4} \newline \\
        \vspace{-0.2cm}
        \begin{description}
            \item[Goal 4.9] Establish liability and medical insurance regulations for data-driven clinical practice
        \end{description} & \includegraphics[width=5pt]{figures/Priority4} \newline \\
    \end{tabular}
\end{tcolorbox}
\label{tab:goalsMission4}
\end{table}
 The following aspects deserve particular attention:\\

\textbf{How to catalyze clinical translation of SDS?} (goals 4.1/4.2)  Clinical data is recognized as \enquote{the resource most central to healthcare progress} \citep{institute_of_medicine_us_roundtable_on_value__science-driven_health_care_clinical_2010}. What is needed is thus a cultural shift towards data acquisition, annotation and analysis within a well-defined data governance framework as a primary clinical 
task~\citep{august_value_2021}. 
The allocation of economic, infrastructural and personnel resources within hospitals for this appears as a non-negotiable requirement for the purpose.
The need for creating value from large amounts of representative data, both for de novo development/validation and external validation studies, further necessitates multi-institutional collaborations. Researchers in other domains have achieved such collaborations, for example in genomics and bioinformatics; SDS would benefit from adopting relevant aspects of these domains' research cultures. In addition, enabling explicit academic recognition for developing rigorously annotated data sets can facilitate data resources for research in SDS, as discussed in Sec.~\ref{sec:dataAnnotation}. Paving the way for short-term clinical success stories as well as long-term clinical translation further requires SDS applications to be integrated into clinical workflows. In fact, the sparsity of studies on SDS solutions for intraoperative care illustrate the challenge of conducting multi-disciplinary research while prioritizing the patient. Therefore, research on SDS products should consider the impact on workflow early in product development and closely engage relevant stakeholders (see Tab.~\ref{tab:stakeholders}). Impactful success stories could then be generated by focusing on low-hanging fruit presented in the previous section. These, in turn, would contribute to building public trust in SDS and boost public enthusiasm to spark patient demand.\\



\textbf{How to improve knowledge transfer among different stakeholders?} (goal 4.3)  The creation of interdisciplinary networks involving the different stakeholders and the regular organization of SDS events in conjunction with both technical and medical conferences is key to improving knowledge transfer between the groups. Such events should, in part, be dedicated to specific questions, such as annotation guidelines, data structures or good practices with respect to external validation.  As a means for actively disseminating, discussing, and promoting new insights in the field of SDS, a well-curated community web platform should be established as the central information hub.
One could even go further and offer e.g. a prize for clinical trials demonstrating SDS success.
A good means for public outreach could be the hosting of public days focused on a particular topic at major conferences in the field, as a way of creating awareness for that topic, or campaigns e.g. in the vein of "Stop the Bleed"~\citep{acs_committee_on_trauma_stop_nodate}.\\

\textbf{How to train key SDS personnel?} (goal 4.4)
In order to facilitate clinical translation of SDS in the long term, it will further be crucial to promote the transdisciplinary training of future surgical data scientists and thereby establish SDS as a career path. Computer scientists will have to enter operating rooms on a regular basis to understand real clinical problems and to get an impression of the obstacles in clinical translation. Similarly, surgeons will have to  understand the basic principles, capabilities and limits of data science techniques to identify solvable clinical problems and proper applications for SDS. A viable path to improve knowledge transfer would be to establish SDS as a commonly respected career path in hospitals. In this context, both technical and clinical disciplines should be complemented by knowledge and expertise in clinical research methodology, i.e., epidemiology and biostatistics. Moreover, human factors engineering and human computer interaction researchers should be integrated into the community. Setting up such an SDS career path should also involve the definition of specifics and skills an 'AI-ready' clinician should meet. A curriculum should put a specific focus on medical statistics covering confounding variables, risk correction and data biases, as well as on regulatory issues (e.g. SaMD). On top of the research-oriented positions, we should further seek to establish SDS-related jobs for data acquisition, management and annotation, specifically in university hospitals. \\

\textbf{How to ensure high-quality external validation of SDS applications?} (goal 4.5-4.7) A critical pitfall with clinical prediction models, which include models for diagnosis and prognosis, is unbridled proliferation of de novo development and validation studies, but scant external validation studies \citep{adibi_validation_2020}. Research to support regulatory approval of SDS products, i.e., in order to market these products, would typically address external validation. However, advances in clinical care are not restricted to marketed products. Therefore, it is necessary for the research community to not only conduct de novo development and validation studies but also well designed external validation studies. Past experience with clinical prediction models shows the need for creative solutions. While some solutions, such as \enquote{living registries}, have been proposed \citep{adibi_validation_2020}, proactive effort by the SDS community to develop effective solutions that allow for consistent and uniform external validation can be a transformative contribution. The status quo, summarized in a review of existing literature in AI-based intraoperative decision-making, shows that the SDS community has not addressed the pitfall of inadequate external validation studies \citep{navarrete-welton_current_2020}. This challenge is systematically addressed when the end-goal for the translational research is regulatory approval to market a SDS product; the regulatory agency serves as a steward in this case. Similar stewardship may benefit translational research in SDS that is not intended to support regulatory approval. Finally, it is important to develop new performance metrics for AI algorithms that quantify clinically relevant parameters currently not accounted for in outcome validation studies. ML techniques offer the possibility to capture data patterns that may serve as potential surrogate measures of long-term outcomes. For example, many established metrics, such as 5-year-survival, are not immediately available after a surgical intervention for cancer. Surgical video or motion data localized to anatomy through imaging studies may be used to identify activities or events that increase the risk of cancer cell seeding (and subsequent metastasis).\\


\textbf{How to ensure ethical and legal guidance?} (goals 4.8/ 4.9) With the face of data-driven clinical practice about to change in a vast manner, unprecedented ethical and legal questions pertaining to both the regulation of medical AI as well as its practical use will be raised. Moving forward, liability and medical negligence/insurance regulations need to be adapted for data-driven clinical practice. A recent survey among Dutch surgeons revealed privacy and liability concerns as significant grounds for objection to video and audio recording of surgical procedures \citep{van_de_graaf_current_2020}, reinforcing the importance of clear regulatory frameworks towards better clinical acceptance. New regulations will have to go much further than these current considerations, with a particular focus to be placed on cases of AI failure, human rejection of AI recommendations, or potentially the omission of AI \citep{directorate-general_for_parliamentary_research_services_european_parliament_ethics_2020}. Notably, the FDA recently put forth an \textit{Artificial Intelligence and Machine Learning (AI/ML) Software as a Medical Device Action Plan} \citep{health_artificial_2021}. These regulatory issues strongly interconnect with previously raised issues of trust in as well as transparency and explainability of AI models, which have also been raised in the very recent WHO report \textit{Ethics \& Governance of Artificial Intelligence for Health} \citep{health_ethics__governance_ethics_2021}. An ethical and human rights-based framework intended to guide the development and use of AI was further proposed  by \cite{fjeld2020principled}, taking eight key themes such as privacy, accountability, safety/security, transparency/explainability, fairness and non-discrimination, human control of technology, professional responsibility, and promotion of human values into account.
Moreover, ethical and moral considerations regarding the democratization of data and/or AI model access will be necessary. 
In the specific context of surgery, first guidance on the ethical implications of integrating AI algorithms into surgical training workflows has recently become available \citep{collins_ethical_2021}. Similarly, new concepts for obtaining patient consent to data sharing that take into account the dynamics and unforeseeability of data usage in future SDS applications need to be established. One way to go might be the introduction of a data donor card, analogously to organ donor cards, as suggested in Sec.~\ref{sec:dataAnnotation_currentChallengesAndNextSteps}. Both patient- and healthcare professional-centric ethical and legal considerations are likely to have a large impact on the public perception of and trust in SDS, which needs to be boosted for higher patient demand. Above all, patient safety must be supported by the development of contemplative regulatory frameworks.


In summary, a multi-pronged approach to address challenges that can catalyze rapid advances in SDS and to develop solutions to problems considered low-hanging fruit will be crucial to the future of SDS as a scientific field. The introduction of initial features that provide clear benefits can facilitate advanced changes. To this end, a compositional approach may be pursued wherein complex SDS products reuse simpler AI models that have been previously approved and adopted in clinical care. Once a number of high value applications are established and there is hospital buy-in, a virtuous circle of SDS can be expected to begin, enabling more applications, higher volume data collection, stronger models, streamlined regulation, and better acceptance.

\section{Discussion}
\label{sec:discussion}

15 years have passed since the vision of the operating room of the future was sketched for the year 2020 \citep{cleary_or2020_2004}. A central goal of the SDS 2019 workshop was to revisit the paper and report produced by \citet{cleary_or_2005} and \citet{mun_operating_2005} and investigate where we stand, what has hindered us to achieve some of the goals envisioned and what are new trends that had not been considered at the time.

When asked: \enquote{What has really changed when you are entering the OR of today as compared to the setting in 2004?}, participants came to the conclusion that they do not perceive any disruptive changes. Improvements were stated to be of rather incremental nature including advances in visualization (e.g. 3D visualization and 4K video imaging \citep{ceccarelli_evolving_2018, dunkin_3d_2015, rigante_preliminary_2017}) and improvements in tissue dissection, which is now safer, easier and faster to perform due to ultrasound scissors and impedance controlled electrosurgery, for example. None of these innovations includes a relevant AI or ML aspect. And some developments did not even come with the envisioned benefits. For instance, staplers of today are by far more sophisticated than 10 years ago, but the problem of anastomotic leakage is still relevant \citep{stamos_anastomotic_2018}. The following paragraphs put the main (six) topics of the 2004 workshop into today’s perspective.

\textbf{Operational efficiency and workflow:} Core problems identified in 2004 were the \enquote{absence of a standard, computerized medical record for patients that documents their histories and their needs} as well as \enquote{multiple and disparate systems for tracking related work processes}. While these problems have remained until today (see Sec.~\ref{sec:technicalInfrastructure}), the challenge of integrating the different information sources related to the entire patient pathway has meanwhile been widely acknowledged. Emerging standards like HL7 FHIR and the maturing efforts of IHE form a solid base for future developments. However, standards alone are not sufficient to solve the problem; hospitals need to make data acquisition, exchange and accessibility a requirement. HIT that enables fast deployment of tools for data acquisition, annotation and processing should be seen as a core service to enable cutting edge research. By centralizing such efforts, data pools can be maintained over the scope of many projects instead of creating isolated databases. This brings with it the need to standardize regulatory workflows. Getting access to data for research is often highly challenging. By outlining clear guidelines and codes of conduct, time spent on formalities can be cut while reducing uncertainties regarding what is the right or wrong way to handle sensitive data. Finally, the prevalence of unstructured data needs to be decreased in order to increase data accessibility. At this point, this also seems to be a matter of user interfaces - by providing clinicians with tools to rapidly create structured reports, reliance on free text can be reduced. This, however, requires training and acceptance by clinical personnel - which could be increased through education in data science topics.

\textbf{Systems integration and technical standards:} OR integration was the aim of multiple international initiatives, such as OR.NET, the Smart Cyber Operating Theater (SCOT) project \citep{iseki_scot_2012} and the Medical Device \enquote{Plug-and-Play} (MD PnP) Interoperability Program. Despite these ongoing efforts we are, however, still far from an OR in which \enquote{all machines and imaging modalities can talk to each other}, as postulated in 2004. Again, interoperability with intraoperative devices should be viewed as a prerequisite by clinical management, and as an investment in future workflow and cost optimization. Emerging standards like SDC provide a means to enable data exchange; however, more work needs to be invested in the creation of platforms that enable dynamic reactions to events and complex interactions.

\textbf{Telecollaboration:} While the OR of the twenty-first century connects many different individuals from various disciplines, telecollaboration has only slightly evolved during the last one and a half decades, and a genuine breakthrough has not yet been achieved \citep{choi_telesurgery_2018}. Many of the impediments can be seen in missing technical developments (e.g. regarding data compression and latency), coordination issues and knowledge gaps on the part of the prospective users as well as the aforementioned lack of data standardization \citep{mun_operating_2005}. It is to be hoped that coming improvements in  intelligent telecommunication networks (e.g. 5G) might trigger future progress in telecollaboration.

\textbf{Robotics and surgical instrumentation:} In 2020, numerous surgical procedures, including major surgery on the esophagus, pancreas or rectum, are feasible to be performed using surgical robots. In striking contrast, the actual use of surgical robotics is still marginal. A number of high-quality controlled trials failed to prove superiority, making the use of surgical robotics in many cases difficult to justify \citep{roh_robot-assisted_2018}. Another reason for the poor progress may lie in the lack of competition in hardware. Since the discontinuation of the development of the ZEUS device in 2003, the field has been clearly dominated by the da Vinci system. Only in recent times, truly competitive systems such as the Senhance\texttrademark{} (TransEnterix) or the Versius\textsuperscript{\textregistered} (Cambridge Medical Robotics Ltd., Cambridge, UK) system have begun to emerge \citep{peters_review_2018}. It will be exciting to see whether a broader range of technical solutions, along with, perhaps, a stronger interlocking with next-generation intraoperative imaging, will stimulate this particular aspect of the next OR.

\textbf{Intraoperative diagnosis and imaging:} While intraoperative imaging appeared very promising in 2004, the modest successes that have been made in that area are mostly related to mobile X-Ray based devices and drop-in devices in robotics \citep{diana_prospective_2017, goyal_new_2018}. The pivotal problem of matching pre- and intraoperative images still remains, as does the unsolved issue of adaptive real-time visualization during intraoperative deformation of soft tissue. One emerging and very promising field is the field of biophotonics (see Sec.~\ref{sec:technicalInfrastructure}). Benefiting from a lack of ionizing radiation, low hardware complexity and easy integrability into the surgical workflow, biophotonics has yielded an increasing number of success stories in intraoperative imaging \citep{bruins_vascular_2020, neuschler_pivotal_2017}.

\textbf{Surgical informatics:} In 2004, the term SDS had not been invented. At that time, surgical informatics was defined as the collection, storage/organization, retrieval, sharing, and rendering of biomedical information that is relevant to the care of the surgical patient, with an aim to provide comprehensive support to the entire healthcare team \citep{mun_operating_2005}. Since the beginnings of the field of computer-aided surgery, however, artificial intelligence and in particular ML have arisen as new enabling techniques that were not in the focus 15 years ago. While these techniques have begun revolutionizing other areas of medicine, in particular radiology \citep{kickingereder_automated_2019, shen_deep_2017}, SDS still suffers from a notable absence of success stories. This can be attributed to a number of various challenges, specifically related to high quality and high volume data annotation, as well as intraoperative data acquisition and analysis and surgical workflow integration, as detailed in Sec.~\ref{sec:technicalInfrastructure}-~\ref{sec:clinicalTranslation}.\\

Overall, the comparison between the workshop topics discussed in 2004 and 2019 revealed that the most fundamental perceived difference is related to how the future of surgery is envisioned by experts in the field. While discussions in 2004 were mainly centered around devices, AI is now seen as a key enabling technique for the future OR. This article has therefore been centered around technical challenges related to applying AI/ML techniques to surgery.

A core challenge now is to put the vision of SDS into clinical practice, but the hurdles appear high. A recent query among surgeons clearly demonstrated that \enquote{digitalization} in general and typical SDS topics in particular are still far removed from their focus of interest \citep{wilhelm_digitalization_2020}. Accordingly, their readiness to get involved in scientific contributions to SDS is still low. As evidenced in a recent systematic review, even in the context of medical image diagnosis, only few prospective deep learning studies and randomized clinical trials exist, with the vast majority of the latter at a high risk of bias and suffering from reproducibility issues, small human comparator groups as well as deviation from existing reporting standards \citep{nagendran_artificial_2020}. Striving for more and higher quality studies on the impact of AI thus becomes imperative for the advancement of the field towards clinical applicability as well as for precluding the rise of false expectations or unattainable promises for the future.

Last but not least, in order to position SDS as a thriving research field in the 21st century, it is of the utmost importance to recognize and enhance the interdisciplinary nature of the field. The establishment of SDS as an independent transdisciplinary career path between the fields of surgery and informatics, which is currently being initiated, is one step on a potential roadmap towards restructuring SDS. Further steps will include enhancement of visibility and clearer communication of challenges to the computer vision and ML communities. Since many SDS applications could be beyond the capabilities of today's ML models (especially time-series data), it becomes crucial to enlist and bring together as many and highly qualified AI researchers of different backgrounds as possible. In a similar vein, clinical researchers who have worked independently on surgical process modeling and objective skill assessment need to be featured more prominently in the development of SDS applications. Clinicians need to become actively involved in driving applications that will be of value to them, and common goals between clinicians and researchers need to be exploited, for instance in the shape of shared task forces. A concrete example for this can be seen in the integration of the SDS community with the video-based assessment program SAGES, where clinicians and researchers alike benefit from shared goals regarding data collection and annotation for surgical education \citep{feldman_sages_2020}.

SDS is a multidisciplinary field that requires motivation, discipline and involvement from a number of different stakeholders and scientific or clinical backgrounds. Together, it becomes viable to strive for fast and efficient development of better models that accurately represent the complexities of SDS issues, thereby moving closer to clinical translation.

\section*{Acknowledgments}
\textcolor{black}{Many thanks to Annika Reinke (DKFZ, Germany) for designing Fig. 1}. We further thank all participants of the workshop, in particular those who filled out the questionnaire including Max Allan (Intuitive Surgical Inc., United States), Mark Asselin (Queen’s University, Canada), Steven Bishop (CMR Surgical Ltd., United Kingdom), Sebastian Bodenstedt (National Center for Tumor Diseases (NCT), Germany), Harold Jay Bolingot (Kyushu Institute of Technology, Japan), Elvis Chen (Robarts Research Institute, Canada), Bijan Dastgheib (International Centre for Surgical Safety ICSS, Canada), Roger Daglius Dias (Brigham Health / Harvard Medical School, United States), Luc Duong (Ecole de technologie superieure, Canada), Ulrich Eck (Technical University of Munich, Germany), Isabel Funke (National Center for Tumor Diseases (NCT) Dresden, Germany), Cong Gao (Johns Hopkins University, United States), Pablo Garcia Kilroy (Verb Surgical Inc., United States), Matthias Grimm (Technische Universität München, Germany), Tamas Haidegger (Obuda University, Hungary), Georges Hattab (National Center for Tumor Diseases (NCT), Germany), Changyan He (Johns Hopkins University, United States), Enes Hosgor (surgical.ai, United States), Hassan Ismail Fawaz (Université Haute-Alsace, France), Anthony Jarc (Intuitive Surgical, United States), Leo Joskowicz (The Hebrew University of Jerusalem, Israel), Ertugrul Karademir (German Aerospace Center (DLR), Germany), Tae Soo Kim (Johns Hopkins University, United States), Kirsten Klein (KARL STORZ SE \& Co. KG, Germany), Michael Kranzfelder (Klinikum rechts der Isar, TU München, Germany), Shlomi Laufer (Technion, Israel), Greg Nelson (Aesculap AG, Germany), Chinedu Nwoye (University of Strasbourg, France), Molly O'Brien (Johns Hopkins University, United States), Daniel Ostler (Klinikum rechts der Isar of Technical University of Munich, Germany), Micha Pfeiffer (National Center for Tumor Diseases Dresden (NCT), Germany), Mohammad Rahbari (University Hospital Mannheim, Medical Faculty Mannheim of the University of Heidelberg, Germany), Wolfgang Reiter (Wintegral GmbH, Germany), Nicola Rieke (NVIDIA, Germany), Tobias Roß (German Cancer Research Center (DKFZ), Germany), Roozbeh Shams (École Polytechnique de Montreal, Canada), Amber Simpson (Memorial Sloan Kettering Cancer Center, United States), Vinkle Srivastav (University of Strasbourg, France), Sarina Thomas (German Cancer Research Center (DKFZ), Germany), Liset Vazquez Romaguera (Polytechnique Montreal, Canada), Tong Yu (University of Strasbourg, France). We further thank Tim Rädsch (German Cancer Research Center (DKFZ), Germany) for his contribution to the data annotation section \textcolor{black}{and Alexander Jenke (NCT Dresden, Germany) for his contribution to the data repository table}.

This work was supported by the European Research Council (ERC) starting grant COMBIOSCOPY under the New Horizon Framework Programme grant agreement [ERC-2015-StG-637960]; the Helmholtz Imaging Platform (HIP); the NCT Heidelberg; BPI France (project CONDOR); the Johns Hopkins Science of Learning Institute Research Grant; the National Institutes of Health [NIDCR R01 DE025265, P41 EB015902, P41 EB015898, R01 CA235589]; the National Center For Tumor Diseases (NCT) Surgical Oncology Program; KARL STORZ SE \& Co. KG; the Royal Society (UF140290) and NIHR Imperial BRC (Biomedical Research Centre); the ERC - H2020 Autonomous Robotic Surgery (ARS) grant agreement [ERC-2016-ADG-742671]; the Surgical Metrics Project - American College of Surgeons (National Society Contract); the Quantified Physician - 7-SIGMA Simulation Systems, Minnesota, MN (Industry Contract); the Tourniquet Master Training - DOD SBIR Phase IIb – Continuation Award [W81XWH-13-C-0021]; the Ontology for Human Motion and Psychomotor Performance - Stanford University Media-X; Motion Analysis for Microvascular Anastomosis - University of Wisconsin (Academic Contract); the Precision Learning Initiative - American Medical Association (National Society Grant); Quantifying the Metrics of Surgical Mastery: An Exploration in Data Science (NIH) [R01DK123445]; the Wellcome/EPSRC Centre for Interventional and Surgical Sciences (WEISS) [203145Z/16/Z]; Engineering and Physical Sciences Research Council (EPSRC) [EP/P027938/1, EP/R004080/1, EP/P012841/1]; the Royal Academy of Engineering Chair in Emerging Technologies; the St. Michael’s Hospital; the University of Toronto; the Grant-in-Aid for Scientific Research on Innovative Area from MEXT, Japan; the National Cancer Data Ecosystem, contract number 19X037Q under Task Order HHSN26100071 from NCI; the project ProteCT [BMBF 16SV8568]; the German Research Foundation (DFG, Deutsche Forschungsgemeinschaft) as part of Germany’s Excellence Strategy - EXC 2050/1 - Project ID 390696704 - Cluster of Excellence \enquote{Centre for Tactile Internet with Human-in-the-Loop} (CeTI); the Canada Research Chair in Computer Integrated Surgery, Natural Sciences and Engineering Research Council of Canada; \textcolor{black}{the ANR with grants ANR-16-CE33-0009 (project DeepSurg), ANR-10-IAHU-02 (IHU Strasbourg) and ANR-20-CHIA-0029-01 (Chair AI4ORSafety); and the Federal Ministry of Economics and Energy (grant number BMWI 01MT17001C) within the OP 4.1 project.}

\section*{Conflicts of interest}
Anand Malpani is a future employee at Mimic Technologies Inc. (Seattle, WA, US). Johannes Fallert and Lars Mündermann are employed at KARL STORZ SE \& Co. KG (Tuttlingen, Germany). Hirenkumar Nakawala is employed at CMR Surgical Ltd (Cambridge, UK). \textcolor{black}{Nicolas Padoy is a scientific advisor of Caresyntax (Berlin, Germany). Daniel A. Hashimoto is a consultant for Johnson \& Johnson (New Brunswick, NJ, USA), Verily Life Sciences (San Francisco, CA, USA), and Activ Surgical (Boston, MA, USA). He has received research support from Olympus Corporation and the Intuitive Foundation.} Carla Pugh is the founder of 10 Newtons Inc. (Madison, WI, US). Danail Stoyanov is employed at Digital Surgery Ltd (London, UK) and Odin Vision Ltd (London, UK). Teodor Grantcharov is the founder of Surgical Safety Technologies Inc. (Toronto, Ontario, Canada). All other authors do not declare any conflicts of interest.


\bibliographystyle{cas-model2-names.bst}
\bibliography{refs}

\onecolumn


\begin{appendices}



\newpage
\renewcommand{\arraystretch}{1.2}
\renewcommand{\appendixtitle}{Publicly accessible and annotated surgical data repositories}
\section*{\appendixtitleFull}
\label{app:publicDataRepositories}

\vspace{\baselineskip}

\begin{center}
\tablefirsthead{%
  \toprule
  \multicolumn{1}{l}{Source} &
  \multicolumn{1}{l}{Procedure(s)/Activity(ies)} &
  \multicolumn{1}{l}{Data Source} &
  \multicolumn{1}{l}{Data Type} &
  \multicolumn{1}{l}{Reference/Annotation} &
  \multicolumn{1}{l}{Year}
  \\
  \midrule }
\tablehead{%
  \midrule
  \multicolumn{6}{l}{\small\sl continued from previous page} \\
  \toprule
  \multicolumn{1}{l}{Source} &
  \multicolumn{1}{l}{Procedure(s)/Activity(ies)} &
  \multicolumn{1}{l}{Data Source} &
  \multicolumn{1}{l}{Data Type} &
  \multicolumn{1}{l}{Reference/Annotation} &
  \multicolumn{1}{l}{Year}
  \\
  \midrule }
\tabletail{%
  \midrule
  \multicolumn{6}{r}{\small\sl continued on next page} \\
  \midrule }
\tablelasttail{}
\topcaption{List of publicly accessible and annotated surgical data repositories, assigned to the categories (1) robotic minimally-invasive surgery, (2) laparoscopic surgery, (3) endoscopy, (4) microscopic surgery, and (5) surgery in sensor-enhanced OR, (6) other. Note that each repository occurs only once in the table although some categories overlap. }
\begin{supertabular}{p{3.5cm} p{3.6cm} p{2.7cm} p{1.5cm} p{2.8cm} p{0.5cm}}
\multicolumn{6}{l}{ROBOTIC MINIMALLY-INVASIVE SURGERY}\\
\midrule
  \citeappendix{appendix:endovis-surgvisdom_endovis_nodate}  &
  needle-driving, knot tying, dissection in training setting &
  virtual/ex-vivo
phantom/porcine &
  video &
  activity &
  2020 \\
 \citeappendix{appendix:endovis-misaw_micro-surgical_nodate}  &
  micro-surgical anastomosis (suturing, knot-tying) in training setting &
  ex-vivo, phantom &
  kinematics, video &
  phase, step, activity &
  2020 \\
  EndoVis-Scared \citepappendix{appendix:allan_stereo_2021}  &
  exploration of abdominal organs &
  ex-vivo, porcine &
  video &
  depth maps, calibration &
  2019 \\
  \citeappendix{appendix:endovis-robseg_endovis_nodate}  &
  nephrectomy &
  in-vivo, porcine &
  video &
  segmentation of instrument parts, objects, anatomy/tissue &
  2018 \\
   EndoVis-RobInstrument \citepappendix{appendix:allan_2017_2019} &
  different porcine procedures &
  in-vivo, porcine &
  video &
  segmentation of instrument parts, instrument type & 
  2017
  \\
  \citeappendix{appendix:EndoVis-SimSurgSkill_nodate} & 
  multiple training tasks &
  virtual &
  video &
  bounding boxes tool, skill &
  2021\\
  \citeappendix{appendix:EndoVis-PETRAW_nodate} & 
  multiple training tasks &
  virtual &
  video, kinematics &
  segmentation of instruments/pegs/blocks, phase, steps, activity &
  2021\\
  ATLAS Dione \newline \citepappendix{appendix:sarikaya_detection_2017} &
  ball placement, ring peg transfer, suture pass, suture and knot tie, urethrovesical anastomosis &
  ex-vivo, phantom &
  video &
  activity, skill, instrument bounding box &
  2017
  \\
  Nephrec9 \citepappendix{appendix:nakawala_nephrec9_2017} &
  partial nephrectomy &
  in-vivo, human &
  video &
  phase &
  2017 \\
  EndoVis-Kidney
  \citepappendix{appendix:hattab2020kidney} &
  partial nephrectomy &
  in-vivo, porcine &
  video &
  kidney boundary &
  2017 \\
   Hamlyn Centre Laparoscopic /Endoscopic Video data sets  \citepappendix{appendix:stoyanov_soft-tissue_2005, appendix:lerotic_dynamic_2008, appendix:mountney_three-dimensional_2010, appendix:pratt_dynamic_2010, appendix:stoyanov_real-time_2010, appendix:giannarou_probabilistic_2013, 
  appendix:ye_self-supervised_2017} &
  diverse  procedures, e.g. partial nephrectomy, totally endoscopic coronary artery bypass graft, intra-abdominal exploration &
  in-vivo/ex-vivo, human/porcine/phantom &
  video &
  depth maps, calibration, 3D surface reconstruction &
  2010 - 2017 \\
  EndoAbS \newline \citepappendix{appendix:veronica_penza_endoabs_2016} &
  exploration abdominal organs &
  ex-vivo, phantom &
  images &
  3D surface reconstruction, calibration &
   2016 \\
  JIGSAWS \citepappendix{appendix:gao_jhu-isi_2014} &
  suturing,  knot-tying,  needle passing in training setting &
  ex-vivo, phantom &
  kinematics, video &
  activity, skill &
  2014 \\
  SARAS-ESAD \newline \citepappendix{appendix:bawa2021saras} & 
  prostatectomy &
  in-vivo, human &
  video &
  action bounding boxes &
  2020\\
  \citeappendix{appendix:SARAS-MESAD_nodate} &
  prostatectomy &
  in-vivo/ex-vivo, human/phantom &
  video &
  action bounding boxes &
  2021\\
  \midrule
  \multicolumn{6}{l}{LAPAROSCOPIC SURGERY}\\
\midrule
  EndoVis-ROBUST-MIS \citepappendix{appendix:ross2021comparative}  &
  laparoscopic rectal resection, proctocolectomy &
  in-vivo, human &
  video &
  multi-instance segmentation of instruments &
  2019 \\
 \citeappendix{appendix:endovis-workflowandskill_endovissub-workflowandskill_nodate} &
  cholecystectomy &
  in-vivo, human &
  video &
  phase, action, instrument type, skill &
  2019 \\
Cholec80  \citepappendix{appendix:twinanda_endonet_2017} &
  cholecystectomy &
  in-vivo, human &
  video &
  phase, instrument type &
  2017 \\
\citeappendix{appendix:endovis-_endovissub-workflow_nodate}  &
  laparoscopic colorectal surgery &
  in-vivo, human &
  video, device signals &
  phase, instrument type &
  2017 \\
  TrackVes \newline \citepappendix{appendix:penza_trackves_2017} &
  exploration of abdominal organs &
  in-vivo/ex-vivo, 
human/porcine/goat &
  video &
  2D polygon around area of interest, attributes of area &
  2017 \\
     m2cai16-tool \citepappendix{appendix:twinanda_endonet_2017} &
  cholecystectomy &
  in-vivo, human &
  video &
  instrument type &
   2016 \\
m2cai16-workflow  \citepappendix{appendix:twinanda_endonet_2017, appendix:stauder_tum_2017} &
  cholecystectomy &
  in-vivo, human &
  video &
  phase &
  2016 \\
m2cai16-tool-locations \citepappendix{appendix:jin2018tool} &
  cholecystectomy &
  in-vivo, human &
  video &
  instrument bounding box &
  2018 \\
\citeappendix{appendix:endovis-instrument_endovissub-instrument_nodate} &
  laparoscopic colorectal surgery, robotic minimally invasive surgery &
  in-vivo/ex-vivo, 
  human/porcine &
  video, \newline images &
  segmentation of instrument parts and center, 2D pose &
  2015 \\
  TMI Dataset
  \citepappendix{appendix:maier-hein_can_2014}  &
  exploration of abdominal organs &
  ex-vivo, porcine &
  images &
  3D surface reconstruction, calibration &
  2014 \\
  Crowd-Instrument \citepappendix{appendix:maier-hein_can_2014} &
  laparoscopic adrenalectomy, pancreas resection &
  in-vivo, human &
  images &
  segmentation of instruments &
   2014 \\
     Laparoscopy Instrument Sequence  \citepappendix{appendix:sznitman_data-driven_2012} &
 cholecystectomy &
 in-vivo, human &
  video &
  instrument center, scale &
  2012 \\
LapGyn4 \citepappendix{appendix:leibetseder2018lapgyn4} &
  gynecologic laparoscopic surgeries  &
  in-vivo, human &
  images, video &
  {actions, anatomy, instrument count} & 
  2018 \\
GLENDA \citepappendix{appendix:leibetseder2020glenda} &
  laparoscopic gynecology  &
  in-vivo, human &
  images, video &
  {segmentation of pathol. endometriosis categories, pathology type} & 
  2020 \\
\citeappendix{appendix:EndoVis-CholecTriplet2021_nodate} &
  cholecystectomy  &
  in-vivo, human &
  images &
  instrument, verb, target & 
  2021 \\
\citeappendix{appendix:EndoVis-HeiSurf_nodate} &
  cholecystectomy &
  in-vivo, human &
  images, video &
  segmentation of 23 different structures, phase, action, tool & 
  2021 \\
\midrule
\multicolumn{6}{l}{MICROSCOPIC SURGERY}\\
\midrule
\citeappendix{appendix:endovis-cataracts-semseg_endovis_nodate}  &
  cataract surgery &
  in-vivo, human &
  images &
  segmentation of anatom\-i\-cal structures and in\-stru\-ments &
  2020 \\
  \citeappendix{appendix:endovis-cataracts-workflow_endovis_nodate}  &
  cataract surgery &
  in-vivo, human &
  video &
  phase &
  2020 \\
\citeappendix{appendix:endovis-cataracts_endovis_nodate} &
  cataract surgery and surgical tray &
  in-vivo, human &
  video &
  instrument type &
  2018 \\
NeuroSurgicalTools data set \citepappendix{appendix:neurosurgicaltools_dataset_medicis} &
  neurosurgery &
  in-vivo, human &
  images &
  instrument bounding polygon, instrument type &
  2015 \\
Retinal Microsurgery Instrument Tracking (RMIT) \citepappendix{appendix:sznitman_data-driven_2012} &
  retinal surgery &
  in-vivo, human &
  video &
  instrument center, scale &
  2012 \\
Cataract-101 \citepappendix{appendix:schoeffmann2018cataract} &
  cataract surgery  &
  in-vivo, human &
  video &
  phase, experience level of surgeon & 
  2018 \\
  
\midrule
  \multicolumn{6}{l}{ENDOSCOPY}\\
\midrule
\citeappendix{appendix:endoscopy_disease_detection_and_segmentation_edd_edd2020_nodate}  &
  gastroscopy, gastro-esophageal, colonoscopy &
  in-vivo, human &
  video &
  bounding boxes and segmentation of multi-class disease regions &
  2020 \\
  Endoscopic Artefact Detection (EAD)
  \citeappendix{appendix:ali2020objective}  &
  gastroscopy, cystoscopy, gastro-esophageal, colonoscopy &
  in-vivo, human &
  video &
  bounding box and segmentation of multi-class artefacts &
  2019 \\
\citeappendix{appendix:EndoCV21_nodate} &
  colonoscopy  &
  in-vivo, human &
  video &
  bounding box and pixel-wise segmentation of polyps & 
  2021 \\
  
  \citeappendix{appendix:endovis-giana_gastrointestinal_nodate}  \citepappendix{appendix:bernal2017comparative}  &
  colonoscopy, wireless capsule endoscopy &
  in-vivo, human &
  video, \newline images &
  segmentation and classification of polyps / angiodysplasia / bowel lesions &
  2015 - 2018 \\
   NBI-InfFrames  \citepappendix{appendix:moccia_nbi-infframes_2018} &
  laryngeal endoscopy &
  in-vivo, human &
  video &
  informative frames &
  2018 \\
\citeappendix{appendix:EndoVis-GIANA21_nodate} &
  colonoscopy  &
  in-vivo, human &
  images, video &
  polyp masks, classification of polyps & 
  2021 \\
Laryngeal data set  \citepappendix{appendix:sara_moccia_laryngeal_2017} &
  laryngeal endoscopy &
  in-vivo, human &
  images &
  patches healthy/cancerous laryngeal tissues &
  2017 \\
  Hamlyn Centre Laparoscopic / Endoscopic Video data sets \citepappendix{appendix:ye_online_2016} &
  gastrointestinal \newline endoscopy &
  in-vivo, human &
  video &
  bounding box of optical biopsy sites & 
  2016 \\
\citeappendix{appendix:AIDA_nodate} &
  gastrointestinal confocal endoscopy, gastric chromoendoscopy, esophagus microendoscopy  &
  in-vivo, human &
  images &
  bounding box of abnormalities & 
  2017 \\
KVASIR \citepappendix{appendix:pogorelov2017kvasir} &
  gastro- and colonoscopy  &
  in-vivo, human &
  images &
  anatomical landmarks, pathologies & 
  2016 \\
Hyper-KVASIR \citepappendix{appendix:borgli2020hyperkvasir} &
  gastro- and colonoscopy  &
  in-vivo, human &
  images, video &
  anatomical landmarks, pathologies, partially segmentation mask and bounding boxes & 
  2020 \\
KVASIR-Capsule \citepappendix{appendix:smedsrud2020kvasir} &
  capsule endoscopy  &
  in-vivo, human &
  images, video &
  anatomical landmarks, quality of mucosal view and pathological findings & 
  2021 \\
KID \citepappendix{appendix:koulaouzidis2017kid} &
  capsule endoscopy &
  in-vivo, human &
  images, video &
  abnormalities & 
  2017 \\
Sinus-Surgery-Endoscopic-Image-Datasets \newline \citepappendix{appendix:qin2020towards} &
  endoscopic sinus surgery  &
  ex-vivo/in-vivo, human &
  images &
  segmentation of instruments & 
  2020 \\
\citeappendix{appendix:AdaptOR_nodate} &
  endoscopic heart surgery  &
  ex-vivo/in-vivo, phantom/human &
  images &
  landmarks in phantoms & 
  2021 \\
EndoSLAM \newline \citepappendix{appendix:ozyoruk2021endoslam} &
  standard/capsule \newline endoscopy  &
  ex-vivo/synthetic, porcine/phantom &
  images &
  6 DoF pose, 3D map ground truth & 
  2021 \\
\citeappendix{appendix:EndoVis-FetReg_nodate} & 
  fetoscopy &
  in-vivo, humanl &
  images, video &
  segmentation of vessel/tool/fetus, phase, steps, activity &
  2021\\
\midrule
\multicolumn{6}{l}{SURGERY IN SENSOR-ENHANCED OR}\\
\midrule
Multi-View Operating Room (MVOR)  \citepappendix{appendix:srivastav_mvor_2019} &
  vertebroplasty, lung biopsy &
  in-vivo, human &
  RGB-D &
  human bounding boxes, 2D/3D human body pose key points &
  2018 \\
xawAR16 \citepappendix{appendix:loy_rodas_see_2017} &
   experimental setting for radiation awareness in hybrid operating room &
  ex-vivo, phantom &
  RGB-D &
  poses of the moving camera &
  2016 \\
  \midrule
\multicolumn{6}{l}{OTHER}\\
\midrule
\citeappendix{appendix:CuRIOUS_nodate} &
  neurosurgery  &
  in-vivo, human &
  images &
  MRI images and intra-operative ultrasound with labeled anat. landmarks & 
  2019 \\
DeepFluoroLabeling-IPCAI2020 \newline \citepappendix{appendix:grupp2020automatic} &
  fluoroscopy  &
  ex-vivo, human &
  images &
  segmentation of hip in CT and fluoroscopy, anat. landmarks & 2020 \\
\bottomrule
\end{supertabular}
\label{tab:publicDataRepositories}
\end{center}


\newpage
\renewcommand{\arraystretch}{1.5}
\renewcommand{\appendixtitle}{Surgical Data Science standards \& tools}
\section*{\appendixtitleFull}
\label{app:standardsAndTools}

\vspace{\baselineskip}

\begin{center}
\tablefirsthead{%
  \toprule
  \multicolumn{1}{l}{Standard} &
  \multicolumn{1}{l}{Organization} &
  \multicolumn{1}{l}{Stage of} &
  \multicolumn{1}{l}{Acceptance } &
  \multicolumn{1}{l}{Purpose} \\
  \multicolumn{1}{l}{} &
  \multicolumn{1}{l}{} &
  \multicolumn{1}{l}{interoperability} &
  \multicolumn{1}{l}{in / outside healthcare} &
  \multicolumn{1}{l}{} \\
  \midrule }
\tablehead{%
  \midrule
  \multicolumn{5}{l}{\small\sl continued from previous page} \\
  \toprule
  \multicolumn{1}{l}{Standard} &
  \multicolumn{1}{l}{Organization} &
  \multicolumn{1}{l}{Stage of} &
  \multicolumn{1}{l}{Acceptance } &
  \multicolumn{1}{l}{Purpose} \\
  \multicolumn{1}{l}{} &
  \multicolumn{1}{l}{} &
  \multicolumn{1}{l}{interoperability} &
  \multicolumn{1}{l}{in / outside healthcare} &
  \multicolumn{1}{l}{} \\
  \midrule }
\tabletail{%
  \midrule
  \multicolumn{5}{r}{\small\sl continued on next page} \\
  \midrule }
\tablelasttail{}
\topcaption{Selected standards relevant to data acquisition, access, storage and communication in SDS.}
\begin{supertabular}{p{2cm} p{3cm} p{2cm} p{3.5cm} p{4cm}}
AVRO &
  Apache Software   Foundation &
  syntactic &
  rare / widespread &
  Data   serialization format, especially for Apache Hadoop \\
DICOM &
  National   Electrical Manufacturers Association &
  syntactic &
  quasi-universal   / none &
  Defines usage of   medical imaging information \\
HDF5 &
  HDF Group &
  syntactic &
  rare /   occasional &
  Data format \\
HL7 FHIR &
  Health Level   Seven International (HL7) &
  syntactic / \newline semantic &
  widespread /   none &
  Focuses on   interoperability of electronic health information in healthcare \\
HL7 Version 2 \& 3 (including CDA) &
  Health Level Seven International (HL7) &
  syntactic / \newline semantic &
  widespread / none &
  Defines   exchange, integration, distribution and retrieval of electronic health   information \\
IEEE 11073 SDC standard family &
  IEEE, OR.NET e.V. &
  IEEE 11073-20702   syntactic interoperability, IEEE 11073-20701 binding standard &
  IEEE 11073-20702   is based on industry standard DPWS, other substandards occasional / rare &
  Communication   protocol for service-oriented medical devices and IT systems \\
IoT &
  Public consensus &
  syntactic &
  rare /   widespread &
  Collective term   describing the interconnection of various systems and actors through the   internet with the purpose of providing intelligent services. \\
JSON &
  Ecma &
  syntactic &
  occasional /   widespread &
  Format for data   exchange and serialization, especially in REST-APIs \\
LOINC &
  Regenstrief   Institute at Indiana University School of Medicine (IUSM) in Indianapolis +   Community &
  semantic &
  widespread /   rare &
  Terminology   standard for laboratory and clinical measurements, observations and documents \\
OpenIGTLink &
  primarily   supported by the U.S. National Institutes of Health (NIH R01EB020667, PI:   Junichi Tokuda) &
  syntactic &
  occasional /   rare &
  Enables   communication between various systems and devices in the operating room for   image-guided therapy \\
openEHR &
  openEHR International &
  syntactic / \newline semantic &
  widespread / none &
  Architecture used for modelling patient-centric health data and management of electronic health records with a query language and an open API \\
Protobuf   (Protocol buffers) &
  Google &
  syntactic &
  rare / occasional &
  Data format \\
RDF &
  RDF Working   group from the World Wide Web Consortium (W3C) &
  semantic &
  occasional /   widespread &
  Data model for   describing resources and their relationship to each other \\
REST &
  Public consensus &
  syntactic &
  occasional /   widespread &
  Set of   principles for web services \\
XML &
  XML Working group from the World Wide Web Consortium (W3C), derived from SGML (ISO 8879) &
  syntactic / \newline semantic &
  widespread / universal &
  Data   serialization format for textual information. \\
\bottomrule
\end{supertabular}
\label{tab:standards}
\end{center}

\newpage

\begin{center}
\tablefirsthead{%
  \toprule
  \multicolumn{1}{l}{Tool} &
  \multicolumn{1}{l}{Organization} &
  \multicolumn{1}{l}{Acceptance } &
  \multicolumn{1}{l}{Purpose} \\
  \multicolumn{1}{l}{} &
  \multicolumn{1}{l}{} &
  \multicolumn{1}{l}{in / outside healthcare} &
  \multicolumn{1}{l}{} \\
  \midrule }
\tablehead{%
  \midrule
  \multicolumn{4}{l}{\small\sl continued from previous page} \\
  \toprule
  \multicolumn{1}{l}{Tool} &
  \multicolumn{1}{l}{Organization} &
  \multicolumn{1}{l}{Acceptance } &
  \multicolumn{1}{l}{Purpose} \\
  \multicolumn{1}{l}{} &
  \multicolumn{1}{l}{} &
  \multicolumn{1}{l}{in / outside healthcare} &
  \multicolumn{1}{l}{} \\
  \midrule }
\tabletail{%
  \midrule
  \multicolumn{4}{r}{\small\sl continued on next page} \\
  \midrule }
\tablelasttail{}
\topcaption{Selected tools relevant to data acquisition, access, storage and communication in SDS.}
\begin{supertabular}{p{3cm} p{4cm} p{3.5cm} p{5cm}}
Amazon AWS &
  Amazon Web   Services Inc.,  Amazon &
  occasional /   widespread &
  Cloud Computing \\
Apache   Kafka &
  Apache Software   Foundation &
  rare /   widespread &
  Streaming   platform for message distribution \\
Docker\textsuperscript{\textregistered} &
  Docker Inc. &
  rare /   widespread &
  Tool for   building software packages, called containers \\
Docker\textsuperscript{\textregistered} Swarm &
  Google Inc. &
  rare /   widespread &
  Orchestration   tool for Docker containers \\
Elasticsearch &
  Elastic &
  rare /   occasional &
  Search and   Analytics Engine \\
Google Cloud Platform\texttrademark &
  Google Inc. &
  occasional /   widespread &
  Cloud Computing \\
Hadoop\textsuperscript{\textregistered} &
  Apache Software   Foundation &
  occasional /   occasional &
  Framework for   distributed computing \\
Kibana &
  Elastic &
  rare /   occasional &
  Dashboard for   data visualization \\
Kubernetes\textsuperscript{\textregistered} &
  Cloud Native   Computing Foundation &
  rare / widespread &
  Orchestration   tool for Docker containers \\
LevelDB &
  Google Inc. &
  rare / occasional &
  Key-value storage \\
Microsoft Azure &
  Microsoft Corporation &
  occasional / widespread &
  Cloud Computing \\
RabbitMQ\textsuperscript{\textregistered} &
  Pivotal Software &
  rare / widespread &
  Message broker \\
ROS &
  Community &
  occasional / occasional &
  Framework with a set of libraries and tools for robot applications \\
\bottomrule
\end{supertabular}
\label{tab:tools}
\end{center}

\noindent RabbitMQ is a trademark of VMware, Inc. in the U.S. and other countries. Elasticsearch is a trademark of Elasticsearch BV, registered in the U.S. and in other countries. Kibana is a trademark of Elasticsearch BV, registered in the U.S. and in other countries.

\newpage

\begin{center}
\tablefirsthead{%
  \toprule
  \multicolumn{1}{l}{Discipline} &
  \multicolumn{1}{l}{Representative software tools} \\
  \midrule }
\tablehead{%
  \midrule
  \multicolumn{2}{l}{\small\sl continued from previous page} \\
  \toprule
  \multicolumn{1}{l}{Discipline} &
  \multicolumn{1}{l}{Representative software tools} \\
  \midrule }
\tabletail{%
  \midrule
  \multicolumn{2}{r}{\sffamily\small\sl continued on next page} \\
  \midrule }
\tablelasttail{}
\topcaption{Disciplines that intersect with SDS and representative software tools that are commonly used in each discipline.}
\begin{supertabular}{p{6cm} p{10.5cm}}

Classical statistics &
  \citeappendix{appendix:r_r_nodate}, \citeappendix{appendix:python_scipystats_statistical_nodate}, \citeappendix{appendix:python_statsmodels_introduction_nodate}, \citeappendix{appendix:matlab_statistics_and_machine_learning_toolbox_statistics_nodate} \\
General machine learning &
  \citeappendix{appendix:scikit-learn_scikit-learn_nodate}, \citeappendix{appendix:python_statsmodels_introduction_nodate}, \citeappendix{appendix:matlab_statistics_and_machine_learning_toolbox_statistics_nodate} \\
Deep learning &
  Frameworks: \citeappendix{appendix:tensorflow_tensorflow_nodate} (including \citeappendix{appendix:keras_keras_nodate}), \citeappendix{appendix:pytorch_pytorch_nodate}, \citeappendix{appendix:caffe_caffe_nodate}, \citeappendix{appendix:apache_mxnet_apache_nodate}, \citeappendix{appendix:microsoft_cognitive_toolkit_cntk_microsoft_nodate}, \citeappendix{appendix:deep_learning_toolbox_deep_nodate}, \citeappendix{appendix:opencv_opencv_nodate}, \citeappendix{appendix:nvidia_clara_nvidia_2019}, \citeappendix{appendix:dltk_dltk_nodate}, \citeappendix{appendix:niftynet_niftynet_nodate} \newline \newline
  Pre-trained model repositories:
  \citeappendix{appendix:model_zoo_model_nodate}, \citeappendix{appendix:onnxmodels_onnxmodels_nodate}, \citeappendix{appendix:tensorflow_model_garden_tensorflowmodels_nodate}, \citeappendix{appendix:torchvision_models_torchvisionmodels_nodate} \\
Data visualization &
  \citeappendix{appendix:python_matplotlib_matplotlib_nodate}, \citeappendix{appendix:python_seaborn_seaborn_nodate}, \citeappendix{appendix:matlab_matlab_nodate} \\
Medical image processing and visualization &
  \citeappendix{appendix:vtk_vtk_nodate}, \citeappendix{appendix:itk_itk_nodate}, \citeappendix{appendix:itk-snap_itk-snap_nodate}, \citeappendix{appendix:3d_slicer_3d_nodate}, \citeappendix{appendix:mitk_medical_nodate} \newline
  Visualization tools survey: \citeappendix{appendix:haak_survey_2016} \\
Classical computer vision &
  \citeappendix{appendix:opencv_opencv_nodate}, \citeappendix{appendix:pcl_point_nodate}, \citeappendix{appendix:vlfeat_vlfeat_nodate}, \citeappendix{appendix:matlab_computer_vision_toolbox_computer_nodate} \\
Natural language processing &
  \citeappendix{appendix:python_nltk_natural_nodate} and \citeappendix{appendix:spacy_spacy_nodate}, \citeappendix{appendix:pytorch-nlp_petrochukmpytorch-nlp_nodate}, \citeappendix{appendix:google_cloud_natural_language_cloud_nodate}, \citeappendix{appendix:amazon_comprehend_amazon_nodate} \\
Signal processing &
  \citeappendix{appendix:python_scipysignal_signal_nodate}, \citeappendix{appendix:matlab_signal_processing_toolbox_signal_nodate} \\
Surgical simulation &
  \citeappendix{appendix:sofa_sofa_nodate}, \citeappendix{appendix:imstk_imstk_nodate}, \citeappendix{appendix:opensurgsim_opensurgsim_nodate} \\
Surgery navigation / Augmented Reality &
  \citeappendix{appendix:slicerigt_slicerigt_nodate}, \citeappendix{appendix:imfusion_suite_imfusion_nodate} \\
Robotics &
  \citeappendix{appendix:ros_robot_nodate} \\
Software engineering &
  \citeappendix{appendix:git_git_nodate}, \citeappendix{appendix:docker_empowering_nodate}, \citeappendix{appendix:jupyter_notebook_project_nodate}, \citeappendix{appendix:data_version_control_dvc_data_nodate} \\
\bottomrule
\end{supertabular}
\label{tab:commonTools}
\end{center}

\newpage
\renewcommand{\appendixtitle}{Surgical Data Science annotation tools \& services}
\section*{\appendixtitleFull}
\label{app:annotationTools}

\vspace{\baselineskip}

\begin{center}
\tablefirsthead{%
  \toprule
  \multicolumn{1}{l}{Tool} &
  \multicolumn{1}{l}{Data type} &
  \multicolumn{1}{l}{Ontology} &
  \multicolumn{1}{l}{Automatic annotation tools} \\
  \multicolumn{1}{l}{} &
  \multicolumn{1}{l}{} &
  \multicolumn{1}{l}{integration} &
  \multicolumn{1}{l}{} \\
  \midrule }
\tablehead{%
  \midrule
  \multicolumn{4}{l}{\small\sl continued from previous page} \\
  \toprule
  \multicolumn{1}{l}{Tool} &
  \multicolumn{1}{l}{Data type} &
  \multicolumn{1}{l}{Ontology} &
  \multicolumn{1}{l}{Automatic annotation tools} \\
  \multicolumn{1}{l}{} &
  \multicolumn{1}{l}{} &
  \multicolumn{1}{l}{integration} &
  \multicolumn{1}{l}{} \\
  \midrule }
\tabletail{%
  \midrule
  \multicolumn{4}{r}{\small\sl continued on next page} \\
  \midrule }
\tablelasttail{}
\topcaption{Selection of annotation tools for spatial, spatio-temporal and temporal annotations.}
\begin{supertabular}{p{5.2cm} p{3.5cm} p{1cm} p{5.2cm}}
\multicolumn{4}{l}{Spatial annotation} \\
\midrule
\citeappendix{appendix:3d_slicer_3d_nodate} &
  Images &
  - &
  Plugins for AI-assisted annotation \\
\citeappendix{appendix:deeplabel_jveitchmichaelisdeeplabel_2020} &
  Images &
  - &
  Automatic tagging \\
\citeappendix{appendix:labelme_labelme_nodate} &
  Images &
  - &
  - \\
\citeappendix{appendix:make_sense_make_nodate} &
  Images &
  - &
  Semi-automatic bounding box annotation, detection \\
\citeappendix{appendix:mitk_medical_nodate} &
  Images &
  - &
  Plugins for AI-assisted annotation \\
\citeappendix{appendix:nvidia_clara_imaging_nvidia_nodate} &
  Images &
  - &
  Semi-automatic segmentation + interactive mode \\
\citeappendix{appendix:pixel_annotation_tool_github_nodate} &
  Images &
  - &
  Watershed segmentation \\
\citeappendix{appendix:semantic_segmentation_editor_hitachi-automotive-and-industry-labsemantic-segmentation-editor_nodate} &
  Images, point clouds &
  - &
  Polygon (automatic option) \\
  EXACT \citep{marzahl2021exact} &
  Images &
  - &
  Version control system \\
\midrule
\multicolumn{4}{l}{Spatio-temporal annotation} \\
\midrule
\citeappendix{appendix:amazon_sagemaker_ground_truth_amazon_nodate} &
  Images, videos, \newline 3D point clouds, text &
  - &
  Interactive mode, semi-automated labeling \\
\citeappendix{appendix:cvat_opencvcvat_nodate} &
  Images, videos &
  - &
  Semi-automatic segmentation, detection \\
\citeappendix{appendix:superannotate_desktop_superannotate_nodate} &
  Images, videos &
  - &
  Active learning, interactive mode \\
\citeappendix{appendix:ultimatelabeling_alexandre01ultimatelabeling_nodate} &
  Videos &
  - &
  Semi-automatic detection + tracking \\
\citeappendix{appendix:vatic_video_nodate} &
  Videos &
  - &
  Optical flow, crowdsourcing \\
\citeappendix{appendix:vott_microsoftvott_nodate} &
  Images, videos &
  - &
  Automatic object detection \\
\midrule
\multicolumn{4}{l}{Temporal annotation} \\
\midrule
\citeappendix{appendix:anvil_anvil_nodate} &
  Videos, audio &
  - &
  - \\
\citeappendix{appendix:bcom_surgery_workflow_toolbox_bcom_nodate} &
  Videos &
  yes &
  - \\
\citeappendix{appendix:observer_xt_behavioral_nodate} &
  Multimodal &
  - &
  - \\
\citeappendix{appendix:swan_sw_nodate} &
  Videos &
  yes &
  - \\
\bottomrule
\end{supertabular}
\label{tab:annotationTools}
\end{center}

\newpage

\begin{center}
\tablefirsthead{%
  \toprule
  \multicolumn{1}{l}{Company} &
  \multicolumn{1}{l}{Domain} \\
  \midrule }
\tablehead{%
  \midrule
  \multicolumn{2}{l}{\small\sl continued from previous page} \\
  \toprule
  \multicolumn{1}{l}{Company} &
  \multicolumn{1}{l}{Domain} \\
  \midrule }
\tabletail{%
  \midrule
  \multicolumn{2}{r}{\small\sl continued on next page} \\
  \midrule }
\tablelasttail{}
\topcaption{Leading companies providing data set annotations with managed human workforces.}
\begin{supertabular}{p{9cm} p{7cm}}
\citeappendix{appendix:alegion_austin_tx_us_alegion_nodate} & General computer vision \\
\citeappendix{appendix:appen_ltd_chatswood_nsw_australia_confidence_nodate} & General computer vision \\
\citeappendix{appendix:cloudfactory_ltd_richmond_uk_data_nodate} & General computer vision \\
\citeappendix{appendix:cogito_new_york_ny_us_training_nodate} & General computer vision \\
\citeappendix{appendix:general_blockchain_inc_san_jose_ca_us_image_nodate} & General computer vision \\
\citeappendix{appendix:samasource_impact_solutions_inc_san_francisco_ca_us_samasource_nodate} & General computer vision \\
\citeappendix{appendix:scale_ai_inc_san_francisco_ca_us_scale_nodate} & General computer vision \\
\citeappendix{appendix:capestart_inc_cambridge_ma_us_medical_nodate} & Medical imaging \\

\citeappendix{appendix:edgecase_ai_llc_hingham_ma_us_data_nodate} & Specialized computer vision \& medical imaging \\
\citeappendix{appendix:imerit_kolkata_west_bengal_india_imerit_nodate} & Specialized computer vision \& medical imaging \\
\citeappendix{appendix:infolks_ptv_ltd_mannarkkad_kerala_india_image_nodate} & Specialized computer vision \& medical imaging \\
\citeappendix{appendix:labelbox_inc_san_francisco_ca_us_labelbox_nodate} & Specialized computer vision \& medical imaging \\
\citeappendix{appendix:steldia_services_ltd_limassol_agios_athanasios_cyprus_outsourcing_nodate} & Specialized computer vision \& medical imaging \\
\citeappendix{appendix:superannotate_llc_sunnyvale_ca_us_superannotate_nodate} & Specialized computer vision \& medical imaging \\
\citeappendix{appendix:telus_international_ai_nodate} & Specialized computer vision \& medical imaging \\
\bottomrule
\end{supertabular}
\label{tab:annotationServices}
\end{center}

\newpage
\renewcommand{\arraystretch}{1.28}
\renewcommand{\appendixtitle}{Published SDS clinical studies - perioperative}
\section*{\appendixtitleFull}
\label{app:clinicalStudies-peri-op}

\vspace{\baselineskip}

\begin{center}
\tablefirsthead{%
  \toprule
  \multicolumn{1}{l}{Publication} &
  \multicolumn{1}{l}{Subject} &
  \multicolumn{1}{l}{Type of study} &
  \multicolumn{1}{l}{Study size} \\
  \multicolumn{1}{l}{} &
  \multicolumn{1}{l}{} &
  \multicolumn{1}{l}{} &
  \multicolumn{1}{l}{(\# patients)} \\
  \midrule }
\tablehead{%
  \midrule
  \multicolumn{4}{l}{\small\sl continued from previous page} \\
  \toprule
  \multicolumn{1}{l}{Publication} &
  \multicolumn{1}{l}{Subject} &
  \multicolumn{1}{l}{Type of study} &
  \multicolumn{1}{l}{Study size} \\
  \multicolumn{1}{l}{} &
  \multicolumn{1}{l}{} &
  \multicolumn{1}{l}{} &
  \multicolumn{1}{l}{(\# patients)} \\
  \midrule }
\tabletail{%
  \midrule
  \multicolumn{4}{r}{\small\sl continued on next page} \\
  \midrule }
\tablelasttail{}
\topcaption{Selection of perioperative SDS clinical studies. Searches were performed in June 2021  using {[machine learning]} AND {[surgery]} or {[deep learning]} AND {[surgery]} or {[artificial intelligence]} AND {[surgery]} or {[decision support]} AND {[surgery]} or {[surgical data science]} AND {[clinical]} in PubMed and Google. Search results were manually evaluated and all studies that analyzed a perioperative SDS system with a machine learning (ML)-based component were included.}
\centering
\begin{supertabular}{p{3.3cm} p{8cm} p{2.2cm} p{1.9cm}}
\citet{bahl_high-risk_2017} &
ML-based prediction of pathological upgrade of high-risk breast lesions and reduction of unnecessary surgical excision based on data such as histologic results and text features from pathologic reports & 
Retrospective cohort & 986\\
\citet{corey_development_2018} &
  ML-based prediction of postoperative complication risk in surgical patients based on electronic health record data &
  Prospective \newline cohort & 66,370\\
\citet{de_silva_spinecloud_2020} &
  ML-based prediction models for postoperative outcomes of lumbar spine surgery based on image features and patient characteristics &
  Retrospective cohort & 64\\
\citet{duke_university_pilot_2016} &
  ML-based clinical analytical platform for predicting risk of surgical complications and improving surgical outcomes based on patient care parameters &
  Prospective \newline cohort & 200\\
\citet{futoma_improved_2017} &
  ML-based sepsis prediction based on clinical patient data over time &
  Prospective \newline cohort & 51,697\\
\citet{hyland_early_2020} &
  ML-based early prediction of circulatory failure in the intensive care unit based on physiological (clinical and laboratory) measurements from multiple organ systems &
  Prospective \newline cohort & 36,098\\
\citet{komorowski_artificial_2018} &
  ML-based identification of optimal treatment strategies for sepsis in intensive care based on laboratory and clinical patient data &
  Prospective \newline cohort & 17,083\\
\citet{mai_artificial_2020} &
ML-based preoperative prediction of severe liver failure after hemihepatectomy in hepatocellular carcinoma patients based on laboratory and clinical parameters & 
Prospective \newline cohort & 353\\
\citet{marcus_improved_2020} &
ML-based prediction of surgical resectability in patients with glioblastoma based on preoperative MRI imaging & 
Retrospective cohort & 135\\
\citet{MasacagniAlapatt2021_EndoDigest} &
  ML-based detection of critical moments in laparoscopic cholecystectomy videos for selective video documentation & 
  Cross-sectional &
  n/a \newline (155 videos)\\
\citet{meyer_machine_2018} &
  ML-based real-time prediction of severe complications in post-cardiosurgical critical care based on electronic health record data &
  Prospective \newline cohort & 42,007\\
\citet{tomasev_clinically_2019} &
  ML-based prediction of future acute kidney injury based on electronic health records &
  Prospective \newline cohort & 703,782\\
\citet{vijayan_automatic_2019} &
ML-based automatic pedicle screw planning in cone-beam guided spine surgery based on CT imaging data & 
Cross-sectional & 40\\
\bottomrule
\end{supertabular}
\end{center}

\newpage
\renewcommand{\appendixtitle}{Registered SDS clinical studies}
\section*{\appendixtitleFull}
\label{app:registeredClinicalStudies}

\vspace{\baselineskip}

\begin{center}
\tablefirsthead{%
  \toprule
  \multicolumn{1}{l}{Study summary} &
  \multicolumn{1}{l}{Patient data} &
  \multicolumn{1}{l}{Study type} &
  \multicolumn{1}{l}{Period} &
  \multicolumn{1}{l}{\# Participants} &
  \multicolumn{1}{l}{Locations} \\
  \midrule }
\tablehead{%
  \midrule
  \multicolumn{6}{l}{\small\sl continued from previous page} \\
  \toprule
  \multicolumn{1}{l}{Study summary} &
  \multicolumn{1}{l}{Patient data} &
  \multicolumn{1}{l}{Study type} &
  \multicolumn{1}{l}{Period} &
  \multicolumn{1}{l}{\# Participants} &
  \multicolumn{1}{l}{Locations} \\
  \midrule }
\tabletail{%
  \midrule
  \multicolumn{6}{r}{\small\sl continued on next page} \\
  \midrule }
\tablelasttail{}
\topcaption{Registered SDS clinical studies at ClinicalTrials.gov as of October 2020. Searches were performed using the following keywords: {[machine learning]} AND {[surgery]} or {[deep learning]} AND {[surgery]} or {[artificial intelligence]} AND {[surgery]} or {[decision support]} AND {[surgery]} or {[data science]} AND {[surgery]} or {[surgical data science]}. Search results were manually evaluated and all studies were included that either test an SDS system or component, or collect data to create and test an SDS system or component. ID is the ClinicalTrials.gov identifier.}
\begin{supertabular}{p{3cm} p{3cm} p{2.4cm} p{1.65cm} p{1.2cm} p{2.8cm}}
\multicolumn{6}{l}{PREOPERATIVE APPLICATIONS}\\
\midrule
Evaluation of an ML-based CDSS to help decide if a patient should undergo hip or knee replacement surgery based on functional and health related quality of life (HRQoL) changes. \newline ID: NCT04332055 &
  Preoperative patient questionnaire &
  Interventional, \newline randomized, \newline single-center &
  Oct. 2020 - \newline Oct. 2025 &
  600 &
  Northern Orthopaedic Division, Clinic Farsø, Aalborg University Hospital, Farsø, Northern Jutland, Denmark \\
Evaluation of an ML-based CDSS (IBM Watson) for hepatocellular carcinoma treatment, prognosis and assessment of  surgical resection risk with radiomics. \newline ID: NCT03917017 &
  Preoperative abdominal images and radiomic parameters &
  Interventional, \newline non-randomized, \newline single-center &
  Jan. 2019 - \newline Dec. 2024 &
  100 &
  Zhujiang Hospital of Southern Medical University, Guangzhou, Guangdong, China \\
Evaluation of an ML-based CDSS to predict ST-segment elevation myocardial infarction (STEMI). \newline ID: NCT03317691 &
  Preoperative ECG &
  Observational, \newline retrospective, \newline single-center &
  Oct. 2017 - \newline Oct. 2018 &
  2,000 &
  Shanghai Tenth People's Hospital, Shanghai, China \\
Evaluation of an ML-based CDSS to help assess risk of refractive eye surgery complications from corneal ectasia. \newline ID: NCT04313387 &
  Preoperative corneal tomography parameters &
  Observational, \newline retrospective, \newline single-center &
  Jan. 2012 - \newline Jan. 2018 &
  558 &
  Visum Eye Center, São José do Rio Preto Medical School, São José do Rio Preto, Brazil \\
Data collection and creation of an ML-based CDSS to detect if a patient has an airway that increases risk of anesthesia related injury. \newline ID: NCT04458220 &
  Preoperative 3D face scans in different positions and from different angles &
  Observational, \newline retrospective, \newline single-center &
  Jul. 2020 - \newline May 2023 &
  4,000 &
  The Ninth People's Hospital of Shanghai Jiaotong University School of Medicine, Shanghai, China \\
Data collection and creation of an ML-based CDSS to predict total knee arthroplasty (TKA) surgery outcome. \newline ID: NCT03894514 &
  Demographic, psychosocial and preoperative clinical parameters from the EHR &
  Observational, \newline prospective, \newline single-center &
  May 2019 - \newline May 2020 &
  150 &
  The University of Valencia, Valencia, Spain \\
Data collection and creation of an ML-based CDSS to assess risk and treatment strategy of patients with acute coronary syndromes in emergency departments. \newline ID: NCT03286491 &
  Unspecified &
  Observational, \newline prospective, \newline single-center &
  Aug. 2017 - \newline Feb. 2018 &
  400 &
  Izmir Bozyaka Training and Research Hospital, Izmir, Turkey \\
Data collection and creation of an ML-based CDSS to detect if a patient has an airway that increases risk of anesthesia related injury. \newline ID: NCT03125837 &
  Preoperative digital photographs in different positions and from different angles &
  Observational, \newline prospective, \newline single-center &
  May 2017 - \newline May 2022 &
  50,000 &
  School of Medicine, Zhejiang University, Hangzhou, China \\
Data collection and creation of an ML-based CDSS to predict pain response, opioid response and morphine usage requirements in pediatric patients requiring surgery, using electronic health record and genetic data. \newline ID: NCT01140724 &
  Genetic &
  Observational, \newline prospective, \newline multi-center &
  Apr. 2008 - \newline Aug. 2021 &
  1,200 &
  Children's Hospital Medical Center, Cincinnati, Ohio, United States \\
Data collection and creation of an ML-based CDSS to assess patient risk of elective heart valve surgery. \newline ID: NCT03724123 &
  Demographic and preoperative clinical parameters from the EHR &
  Observational, \newline retrospective, \newline single-center &
  Jan. 2008 - \newline Dec. 2014 &
  2,229 &
  Kepler University Hospital, Linz, Austria \\
\midrule
\multicolumn{6}{l}{INTRAOPERATIVE APPLICATIONS}\\
\midrule
Evaluation of an ML-based CDSS \citepappendix{appendix:edwards_hemosphere_platform_hemosphere_nodate} to detect and prevent arterial hypotension during abdominal surgery with the Hypotension Prediction Index (HPI) using the  \citeappendix{appendix:flotrac_system_flotrac_nodate}.\newline ID: NCT04301102 &
  Intraoperative hemodynamic parameters &
  Interventional, \newline randomized, \newline multi-center &
  Sep. 2020 - \newline May 2021 &
  80 &
  Hospital de Jerez de la Frontera, Cádiz, Spain \\
Evaluation of an ML-based CDSS \citepappendix{appendix:edwards_hemosphere_platform_hemosphere_nodate} to detect and prevent arterial hypotension during lung surgery with the Hypotension Prediction Index (HPI) using the\citeappendix{appendix:flotrac_system_flotrac_nodate}. \newline ID: NCT04149314 &
  Intraoperative hemodynamic parameters &
  Interventional, \newline randomized, \newline single-center &
  Nov. 2019 - \newline Dec. 2022 &
  150 &
  University of Giessen, Giessen, Germany \\
Evaluation of an ML-based CDSS \citepappendix{appendix:alertwatch_anesthesia_control_tower_home_nodate} to support risk assessment for the anesthesiology team. \newline ID: NCT03923699 &
  Physiological parameters, EHR, anesthesia machine parameters, laboratory results &
  Interventional, \newline randomized, \newline single-center &
  Jul. 2019 - \newline Jul. 2024 &
  40,000 &
  Washington University School of Medicine, Saint Louis, Missouri, United States \\
Evaluation of an ML-based CDSS to detect intraoperative hypertension, using blood pressure \citepappendix{appendix:nexfin_finger_cuff_bmeye_nodate}. \newline ID: NCT03533205 &
  Intraoperative hemodynamic parameters (blood pressure) &
  Observational, \newline prospective, \newline single-center &
  Apr. 2015 - \newline Apr. 2018 &
  507 &
  The Academic Medical Center, The University of Amsterdam, Amsterdam, Netherlands \\
Data collection and creation of an ML-based CDSS to recognize healthy and abnormal tissue characteristics in abdominal surgery. \newline ID: NCT04589884 &
  Intraoperative hyperspectral images (HSI) &
  Observational, \newline prospective, \newline single-center &
  Sep. 2020 - \newline Oct. 2024 &
  600 &
  The Digestive and endocrine surgery service, NHC, Strasbourg, France \\
Multi-objective data collection of colorectal cancer surgery videos and biopsy samples for developing ML-based systems. \newline ID: NCT04220242 &
  Colorectal surgery videos and tissue microsections &
  Observational, \newline prospective and retrospective, multi-center &
  Dec. 2019 - \newline Dec. 2022 &
  250 &
  The Mater Misericordiae University Hospital, Dublin, Ireland \\
Data collection and creation of an ML-based CDSS to detect cerebral ischemia and reperfusion during cardiac surgery. \newline ID: NCT03919370 &
  Intraoperative hemodynamic and cerebral oxygenation parameters &
  Observational, \newline prospective, \newline single-center &
  Dec. 2019 - \newline Dec. 2022 &
  10 &
  Sahlgrenska University Hospital, Gothenburg, Sweden \\
Data collection and creation of an ML-based CDSS to predict postoperative outcomes (mortality, morbidity, Intensive Care Unit admission, length of hospital stay, and hospital readmission). \newline ID: NCT04014010 &
  Intraoperative hemodynamic parameters (blood pressure, heart rate), oxygen level, carbon dioxide level and hemodynamic medication records &
  Observational, \newline retrospective, \newline single-center &
  Jan. 2013 - \newline Dec. 2017 &
  35,000 &
  Nova Scotia Health Authority Queen Elizabeth II hospitals, Halifax, Canada \\
\midrule
\multicolumn{6}{l}{POSTOPERATIVE APPLICATIONS}\\
\midrule
Evaluation of an ML-based CDSS for real-time vasoactive and inotropic support de-escalation in pediatric patients following cardiac surgery. \newline ID: NCT04600700 &
  Postoperative blood oxygenation parameters (the inadequate oxygen delivery index) &
  Observational, \newline retrospective, \newline single-center &
  Jan. 2021 - \newline Mar. 2022 &
  250 &
  Boston Children’s Hospital, Boston, United States \\
Evaluation of a gait monitoring system with ML components \citepappendix{appendix:gaitsmart_digital_nodate} to detect  gait deficiencies after total hip or knee replacement surgery, and detect differences from different rehabilitation programs. \newline ID: NCT04289025 &
  Postoperative gait parameters from inertial motion units (IMUs) &
  Interventional, \newline randomized, \newline single-center &
  Jan. 2021 - \newline Mar. 2021 &
  100 &
  Norfolk and Norwich University Hospital, Norwich, Norfolk, United Kingdom \\
Evaluation of an ML-based CDSS \citepappendix{appendix:alertwatch_anesthesia_control_tower_home_nodate} for risk forecasting immediately after surgery with telemedicine notifications. \newline ID: NCT03974828 &
  Physiological parameters, EHR, anesthesia machine parameters, laboratory results &
  Interventional, \newline randomized, \newline single-center &
  Nov. 2020 - \newline Jan. 2024 &
  3,375 &
  Washington University School of Medicine, St. Louis, Missouri, United States \\
Evaluation of an ML-based system \citepappendix{appendix:caption_healthcaption_ai_products_nodate} to improve cardiac ultrasound image standardization and quality after surgery (step down unit). \newline ID: NCT04203251 &
  Postoperative cardiac ultrasound &
  Observational, \newline prospective, \newline single-center &
  Mar. 2020 - \newline May 2020 &
  100 &
  University of California San Francisco, San Francisco, California, United States \\
Evaluation of an at-home ML-based postoperative monitoring system \citepappendix{appendix:smart_angel_support_2020} to reduce unplanned recourse. \newline ID: NCT04068584 &
  Postoperative hemodynamic, blood oxygenation and well-being parameters (pain, nausea, vomiting, comfort) &
  Interventional, \newline randomized, \newline multi-center &
  Feb. 2020 - \newline Aug. 2021 &
  1,260 &
  Nīmes University Hospital Centre, Nīmes, France \\
Evaluation of an ML-based CDSS (CALYPSO) that creates personalized risk predictions to reduce postoperative complications. \newline ID: NCT02828475 &
  Unspecified &
  Observational, \newline prospective, \newline single-center &
  Jun. 2016 - \newline Jan. 2017 &
  200 &
  Duke University Medical Center, Durham, North Carolina, United States \\
Evaluation of an ML-based CDSS to help manage postoperative cataract surgery patients. \newline ID: NCT04138771 &
  Postoperative visual acuity parameters, intraocular pressure parameters and slit-lamp images &
  Interventional, \newline single-center &
  Jan. 2013 - \newline Mar. 2020 &
  300 &
  Zhongshan Ophthalmic Center, Sun Yat-sen University, Guangzhou, Guangdong, China \\
Data collection and creation of an ML-based CDSS to predict postoperative respiratory failure within 7 days. \newline ID: NCT04527094 &
  Pre- and intraoperative EHR &
  Observational, \newline prospective, \newline single-center &
  Nov. 2020 - \newline Aug. 2021 &
  8,000 &
  Seoul National University Hospital, Seoul, Republic of Korea \\
Data collection and creation of an ML-based CDSS to predict postoperative outcomes after vascular stent placement using data from a wearable device (ECG bracelet). \newline ID: NCT04455568 &
  Postoperative ECG &
  Observational, \newline prospective, \newline multi-center &
  Jul. 2020 - \newline Jul. 2024 &
  400 &
  Taipei Medical University Shuang Ho Hospital, New Taipei City, Taiwan \\
Data collection and creation of an ML-based system to compute continuous blood pressure of patients in surgical intensive care non-invasively, using a wearable blood pressure measuring device and  a patient monitor \citepappendix{appendix:intellivue_mx700_philips_intellivue_nodate}. \newline ID: NCT04261062 &
  Postoperative hemodynamics (blood pressure) &
  Observational, \newline prospective, \newline single-center &
  May 2020 - \newline Jan. 2022 &
  220 &
  Yonsei University College of Medicine, Department of Anesthesiology and Pain Medicine, Seoul, Republic of Korea \\
Data collection and creation of an ML-based CDSS to detect and predict opioid induced respiratory compromise (OIRC) events in postoperative pain management. \newline ID: NCT03968094 &
  EHR and postoperative blood oxygenation, ventilation and transcutaneous PCO2 parameters &
  Observational, \newline prospective, \newline single-center &
  Jun. 2019 - \newline Mar. 2020 &
  50 &
  Buffalo General Medical Center, Buffalo, New York, United States \\
Data collection and creation of an ML-based CDSS to assess postoperative glioblastoma surgery images to distinguish progression from pseudo-progression. \newline ID: NCT04359745 &
  Preoperative and postoperative MRI &
  Observational, \newline prospective, \newline multi-center &
  Mar. 2019 - \newline May 2023 &
  500 &
  Guy's and St Thomas' NHS Foundation Trust and King’s College, London, United Kingdom \\
Data collection and creation of an ML-based CDSS to predict kidney injury after hyperthermic intraperitoneal chemotherapy (HIPEC). \newline ID: NCT03895606 &
  Preoperative and intraoperative physiological parameters including hemodynamics, blood oxygenation, body temperature, cardiac index and stroke volume variation &
  Observational, \newline prospective, \newline single-center &
  Mar. 2019 - \newline Mar. 2020 &
  57 &
  Gangnam Severance Hospital, Seoul, Republic of Korea \\
Data collection and creation of an ML-based CDSS to predict risk of readmission following discharge after cardiovascular surgery, using data from a wearable device \citepappendix{appendix:snap40_monitor_full-service_nodate}. \newline ID: NCT03800329 &
  Postoperative hemodynamic, blood oxygenation, respiration, body temperature and movement parameters &
  Interventional, \newline randomized, \newline single-center &
  Mar. 2018 - \newline Mar. 2021 &
  100 &
  Mayo Clinic in Rochester, Rochester, Minnesota, United States \\
\midrule
\multicolumn{6}{l}{MULTI-STAGE/OTHER APPLICATIONS}\\
\midrule
Evaluation of an CDSS (Digital Surgery GoSurgery) with ML components for OR workflow assistance and analytics. \newline ID: NCT03955614 &
  Surgery workflow and OR video &
  Interventional, \newline non-randomized, \newline multi-center &
  Oct. 2019 - \newline Oct. 2020 &
  150 &
  Imperial College Hospitals NHS Trust, London, United Kingdom \\
Evaluation of an ML-based CDSS to predict motor response after subthalamic nucleus deep brain stimulation (STN DBS) therapy in Parkinson patients. \newline ID: NCT04093908 \newline &
  Demographic, clinical and postoperative UPDRS variables &
  Observational, \newline retrospective, \newline multi-center &
  Aug. 2019 - \newline Dec. 2019 &
  322 &
  Maastricht UMC, Maastricht, Limburg, Netherlands \\
Evaluation of an ML-based CDSS \citepappendix{appendix:kia_mews_2020} to predict if a hospitalized patient requires care escalation within 6 hours. \newline ID: NCT04026555 \newline &
  Admission discharge transfer (ADT) events, structured clinical assessments (e.g. nursing notes), physiological parameters, ECG and laboratory results &
  Interventional, \newline non-randomized, \newline single-center &
  Jun. 2019 - \newline Mar. 2020 &
  2,915 &
  Mount Sinai Hospital, New York, New York, United States \\
Evaluation of an ML-based CDSS to help report and monitor patients before and after total knee arthroplasty (TKA), using data from a wearable device (unspecified). \newline ID: NCT03406455 &
  Preoperative and postoperative physical activity parameters including step counting and knee range-of-motion &
  Observational, \newline prospective, \newline single-center &
  Jul. 2018 - \newline May 2019 &
  25 &
  Cleveland Clinic, Cleveland, Ohio, United States \\
Evaluation of a deep brain stimulation surgery navigation system \citepappendix{appendix:surgical_information_sciences_surgical_nodate} with ML components for enhanced image visualization. \newline ID: NCT02902328 &
  Preoperative MRI &
  Observational, \newline prospective, \newline single-center &
  Mar. 2016 - \newline Sep. 2016 &
  30 &
  Surgical Information Sciences Inc., Minneapolis, Minnesota, United States \\
Data collection and creation of an ML-based system for early sepsis detection for patients in ICUs including surgical ICUs. \newline ID: NCT04130789 &
  ICU device parameters, microbiology parameters and laboratory results &
  Observational, \newline prospective, \newline multi-center &
  Nov. 2019 - \newline Jun. 2023 &
  17,500 &
  Clinical Microbiology, University Hospital Basel, Basel, Switzerland \\
Data collection and creation of an ML-based CDSS to predict liver transplant (LT) complication risk using microbial flora data at pre-LT, early post-LT and late post-LT timepoints. \newline ID: NCT03666312 &
  Preoperative and intraoperative microbial flora parameters &
  Observational, \newline prospective, \newline multi-center &
  Sep. 2019 - \newline Aug. 2021 &
  275 &
  IRCCS San Raffaele, Milan, Italy \\
Multi-objective data collection to create and evaluate ML-based systems for liver volume assessment before and after surgery, and liver lesion detection. \newline ID: NCT03960710 &
  Preoperative and postoperative CT images &
  Observational, \newline retrospective, \newline single-center &
  Apr. 2019 - \newline Sep. 2019 &
  120 &
  Radiology service, Imaging research unit, Edouard Herriot Hospital, Lyon, France \\
Data collection and creation of an ML-based CDSS to predict risk of postoperative cognitive complications. ID: NCT03175302 &
  Preoperative digital cognititive testing data &
  Observational, \newline prospective, \newline single-center &
  Jun. 2018 - \newline Aug. 2021 &
  25,240 &
  University of Florida, Gainesville, Florida, United States \\
Data collection and creation of an ML-based CDSS to predict risk of  postoperative complications (Clavien-Dindo score). \newline ID: NCT04092933 &
  Patient Data Management System (PDMS) data including physiological parameters (vitals and respiratory), medication, intraoperative events and times &
  Observational, \newline retrospective, \newline single-center &
  May 2014 - \newline Feb. 2022 &
  109,000 &
  The Technical University of Munich, Munich, Germany \\
Data collection and creation of an ML-based CDSS to predict postoperative acute renal failure after liver resection. \newline ID: NCT01318798 &
  Preoperative and intraoperative physiological data (unspecified) &
  Observational, \newline retrospective, \newline single-center &
  Jan. 2010 - \newline Apr. 2012 &
  549 &
  University Hospital of Zurich, Department of Visceral and Transplantation Surgery, Zurich, Switzerland \\
\bottomrule
\end{supertabular}
\label{tab:registeredClinicalStudies}
\end{center}

\newpage
\renewcommand{\appendixtitle}{Stakeholder importance}
\section*{\appendixtitleFull}
\label{app:stakeholderImportance}

Importance of stakeholders as determined in the Delphi process.

\begin{center}
    \includegraphics[height=0.93\textheight]{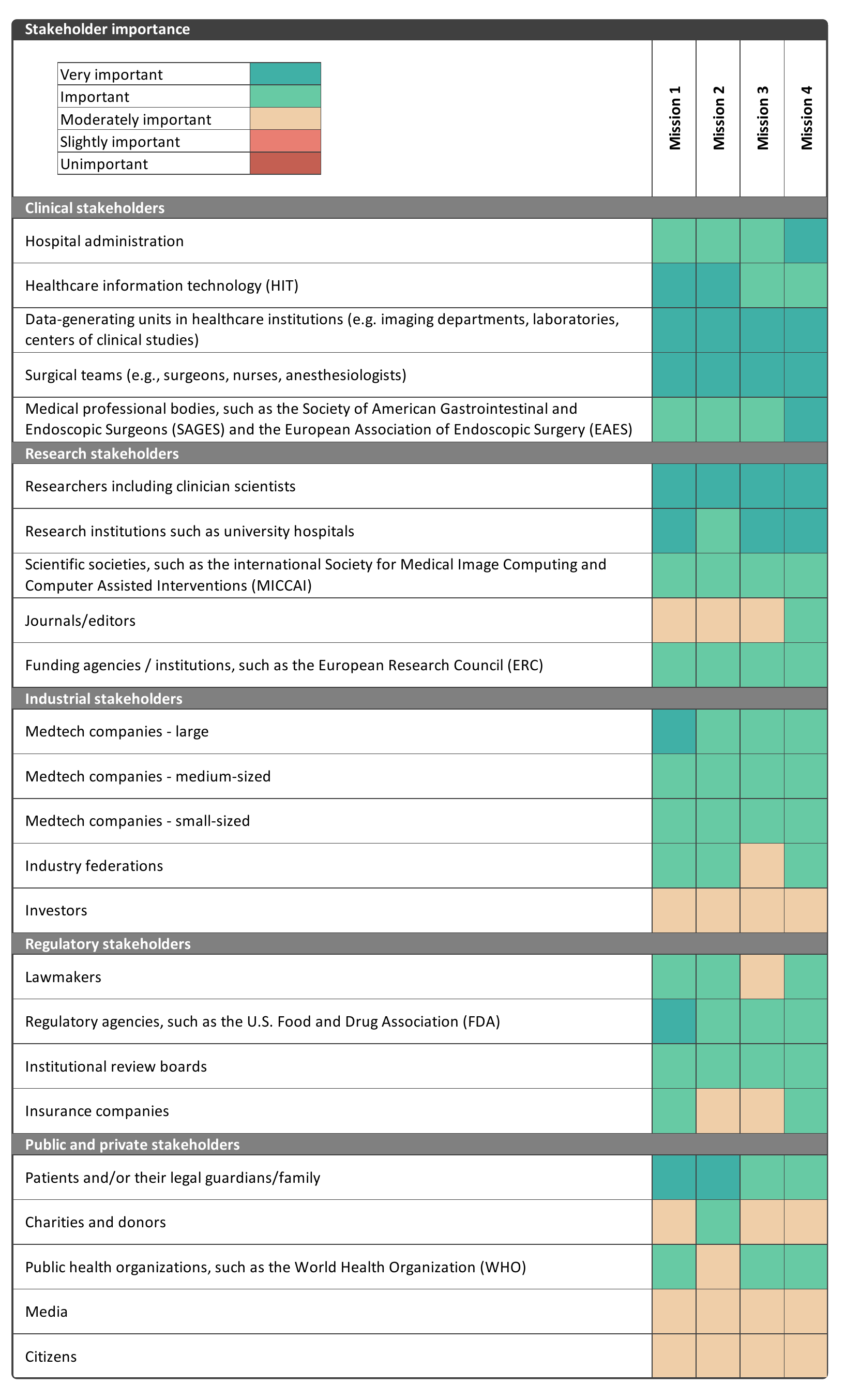}
\end{center}


\newpage

\end{appendices}

\newpage
\bibliographystyleappendix{cas-model2-names.bst}
\bibliographyappendix{refsAppendix}

\end{document}